\documentclass[a4paper,11pt]{article}
\pdfoutput=1 

\usepackage{jcappub} 
\usepackage{subcaption,graphicx,makecell,mathtools}
\usepackage[T1]{fontenc} 

\title{Free-streaming and Coupled Dark Radiation Isocurvature Perturbations: Constraints and Application to the Hubble Tension}

\author[a]{Subhajit Ghosh,}
\author[b,c]{Soubhik Kumar,}
\author[a]{Yuhsin Tsai}

\affiliation[a]{Department of Physics, University of Notre Dame, IN 46556, USA}
\affiliation[b]{Berkeley Center for Theoretical Physics, Department of Physics,
University of California, Berkeley, CA 94720, USA}
\affiliation[c]{Theoretical Physics Group, Lawrence Berkeley National Laboratory, Berkeley, CA 94720, USA}
\emailAdd{sghosh5@nd.edu}
\emailAdd{soubhik@berkeley.edu}
\emailAdd{ytsai3@nd.edu}

\newcommand{\powab}{\mathcal{P}_{ab}(k)}

\newcommand{\pow}{\mathcal{P}}

\newcommand{\poab}{\mathcal{P}_{ab}^{(1)}}
\newcommand{\ptab}{\mathcal{P}_{ab}^{(2)}}

\newcommand{\lcdm}{$\Lambda$CDM{}}
\newcommand{\mpcinv}{{\rm Mpc}^{-1}}
\newcommand{\neff}{$N_{\rm tot}$}
\newcommand{\ndr}{$N_{\rm dr}$}
\newcommand{\phpiia}{$N_{\rm dr}^2\mathcal{P}_\mathcal{II}^{(1)}$}
\newcommand{\phpiib}{$N_{\rm dr}^2\mathcal{P}_\mathcal{II}^{(2)}$}

\newcommand{\powrr}{\mathcal{P}_\mathcal{RR}}
\newcommand{\powii}{\mathcal{P}_\mathcal{II}}
\newcommand{\powri}{\mathcal{P}_\mathcal{RI}}

\newcommand{\iso}{{\rm iso}}
\newcommand{\fracdr}{f_{\rm dr}}
\newcommand{\sdr}{\mathcal{S}_{\rm DR}}
\newcommand{\gsim}{\lower.7ex\hbox{$\;\stackrel{\textstyle>}{\sim}\;$}}
\newcommand{\lsim}{\lower.7ex\hbox{$\;\stackrel{\textstyle<}{\sim}\;$}}

\abstract{Dark radiation (DR) appears as a new physics candidate in various scenarios beyond the Standard Model. While it is often assumed that perturbations in DR are adiabatic, they can easily have an isocurvature component if more than one field was present during inflation, and whose decay products did not all thermalize with each other. 
By implementing the appropriate isocurvature initial conditions (IC), we derive the constraints on both uncorrelated and correlated DR density isocurvature perturbations from the full Planck 2018 data alone, and also in combination with other cosmological data sets.
Our study on free-streaming DR (FDR) updates and generalizes the existing bound on neutrino density isocurvature perturbations by including a varying number of relativistic degrees of freedom, and for coupled DR (CDR) isocurvature, we derive the first bound. We also show that for CDR 
qualitatively new physical effects arise compared to FDR. One such effect is that for isocurvature IC, FDR gives rise to larger CMB anisotropies compared to CDR---contrary to the adiabatic case.
More generally, we find that a blue-tilt of DR isocurvature spectrum is preferred. This gives rise to a larger value of the Hubble constant $H_0$ compared to the standard $\Lambda$CDM+$\Delta N_{\rm eff}$ cosmology with adiabatic spectra and relaxes the $H_0$ tension.}

\begin{document} 
\maketitle
\flushbottom

\section{Introduction}
The precision cosmological data from Cosmic Microwave Background (CMB) and Large Scale Structure (LSS) are crucial probes of physics beyond the Standard Model (BSM). In particular, CMB can be very sensitive to new relativistic degrees of freedom present around recombination~\cite{Aghanim:2018eyx}. Many BSM scenarios contain ``dark'' radiation (DR) that are such ultralight degrees of freedom, very weakly coupled to the Standard Model (SM) (see~\cite{Green:2019glg} and the references therein). The energy density $\rho_{\rm DR}$ in such DR is usually compared to the energy density in one SM neutrino species, $\rho_{1\nu}$ and is parametrized by $\Delta N_{\rm tot}\equiv\rho_{\rm DR}/\rho_{1\nu}$~\cite{1969ZhPmR...9..315S,1977PhLB...66..202S}. Depending on the microphysics, DR can be free-streaming (FDR) or coupled (CDR). Given their feeble couplings to the SM, both kinds of DR can be extremely difficult to probe at collider or direct detection experiments. However, even for gravitational coupling of the DR to the SM, CMB can still be sensitive to the energy density in DR, and CMB observations can be the first to discover the new physics associated with DR. More importantly, we will see that beyond the existence of DR, precision CMB observations are also sensitive to the origin of DR, and hence to the inflationary and reheating history of the Universe~\cite{Planck:2018jri}. 
 
FDR that carries an adiabatic perturbation modifies the CMB power spectrum mainly via two effects. First, the radiation energy increases the rate of cosmic expansion and thereby decreases the sound horizon at recombination. For a fixed value of the angular size of the sound horizon $\theta_s$, this increases the angular size of the damping scale $\theta_d$, implying more Silk damping at high multipole moments $\ell$~\cite{Hu:1996vq, Bashinsky:2003tk, 2013PhRvD..87h3008H}. Moreover, the supersonic propagation of free-streaming radiation perturbation induces a phase shift in the sound waves in the photon-baryon plasma, resulting in a phase shift of the CMB and baryon acoustic oscillations from the $\Lambda$CDM result~\cite{1973ApJ...180....1P,2013PhRvD..87h3008H,Bashinsky:2003tk,Baumann:2015rya,Chacko:2015noa}. The Planck measurements are sensitive enough to probe both of these effects~\cite{Baumann:2015rya,Follin:2015hya}.

On the other hand, for CDR, where the coupling can come from self-interacting particles~\cite{Jeong:2013eza,Buen-Abad:2015ova,Kreisch:2019yzn,Forastieri:2019cuf} or from scatterings between radiation and non-relativistic particles~\cite{Chacko:2016kgg,Buen-Abad:2017gxg,DiValentino:2017oaw,Ghosh:2019tab}, the radiation can behave as an ideal relativistic fluid. The ideal fluid perturbations no longer propagate supersonically and they generate a different phase shift in photon perturbation equations than the free-streaming radiation. 
The scattering also forbids the diffusion damping of the DR perturbation and enhances the initial metric perturbation compared to the free-streaming radiation~\cite{Ma:1995ey,Hu:1995en,Bashinsky:2003tk}. 
This latter effect increases the CMB temperature fluctuation, and the Planck measurements are again sensitive to both of these effects. The $\Delta N_{\rm tot}$ constraints therefore take different values between the FDR and CDR even with adiabatic perturbations~\cite{Baumann:2015rya,Brust:2017nmv,Blinov:2020hmc}.

In the discussion so far we assumed that all the primordial perturbations in the Universe come from the quantum fluctuations of a single field, such as the inflaton. In this case, both FDR and CDR carry the same \emph{adiabatic} perturbations as the SM neutrinos and photons, before being further processed by the subhorizon physics~\cite{Dodelson:2003ft}. However, the inflationary and the reheating history of the Universe can easily be different from this simplest setup. As a simple example, DR can originate from the decay of a {\em curvaton} or an {\em axion} field $\chi$ that can obtain its own quantum fluctuations uncorrelated with the inflaton~\cite{Enqvist:2001zp,Lyth:2001nq,Moroi:2001ct,Kawasaki:2011rc}. In this case, FDR and CDR inherit the \emph{isocurvature} perturbations (see e.g.~\cite{Linde:1985yf, Polarski:1992dq, Polarski:1994rz, Langlois:1999dw, Gordon:2000hv} for inflationary models generating isocurvature perturbations) of $\chi$, and generate different corrections to the CMB spectrum compared to the adiabatic DR perturbations discussed above.

To characterize these differences, we first derive the appropriate initial conditions for DR isocurvature density perturbations~(DRID) for both FDR and CDR. Implementing these, we then update and generalize the constraints on DRID perturbations using the current CMB power spectrum and the baryon acoustic oscillation (BAO) data~\cite{Alam:2016hwk}. As we will show, depending on the size and the spectral tilt of the DRID, current data sets different bounds on $\Delta N_{\rm tot}$ compared to the adiabatic DR perturbations.

In fact, for the case of FDR isocurvature, both the initial conditions and the Boltzmann equations of the perturbations are identical to the scenario of neutrino density isocurvature (NDI)~\cite{Bucher:1999re,Planck:2018jri,Kawasaki:2011rc,Adshead:2020htj}. Planck has presented constraints on NDI in Ref.~\cite{Planck:2018jri} by fixing the neutrino number to be the SM prediction $N_{\rm tot}=3.046$. We repeat such an analysis. However, to
make the bound more applicable to different BSM scenarios, we also present bounds on FDR isocurvature with \textit{varying} $\Delta N_{\rm tot}$, i.e., varying the amount of both SM neutrinos and DR, using Planck temperature, polarization and lensing data. Ref.~\cite{Kawasaki:2011rc} also studied the constraint on the FDR isocurvature with varying $\Delta N_{\rm tot}$ using the 7-year WMAP data. As we will show, the bound improves significantly when the Planck data is taken into account. 

Beyond studying NDI and the related FDR isocurvature perturbations, we further consider the isocurvature perturbations in CDR that can easily arise in various BSM scenarios. Since in this case DR behaves as an ideal fluid, it has a vanishing anisotropic stress. As we will show later, this fact leads to a sign difference in the total anisotropic stress of the system (for fixed choices of initial isocurvature density perturbations) between FDR and CDR. This then gravitationally affect the photon perturbations, and in particular, for isocurvature initial conditions the CMB power spectrum becomes enhanced in the case of FDR compared to CDR. This feature is exactly opposite to the case of adiabatic perturbations in DR, for which CDR enhances the CMB spectrum further compared to FDR. We will provide both numerical results and an analytical explanation of this intriguing phenomena. As in the case of FDR, we present results for two cases, one where the energy density in SM neutrinos is kept fixed, and the other where it is varied.

Furthermore, using the CMB data we show that the addition of DRID helps to accommodate a larger value of the Hubble constant $H_0$ compared to the $\Lambda$CDM fit. In particular, the changes induced by a blue tilted isocurvature spectrum can be compensated by a higher value of $\Delta N_{\rm tot}$ which increases the value of $H_0$. Therefore, DR isocurvature spectrum relaxes the tension between the CMB and the local measurement of $H_0$ by the SH0ES collaboration~\cite{Riess:2019cxk} to a better extent, compared to scenarios with adiabatic perturbations in either FDR or CDR.\footnote{There are other local $H_0$ measurements, for instance, based on the TRGB distance ladder measurement~\cite{Freedman:2021ahq} or measurements using strong gravitational lensing systems~\cite{Suyu:2016qxx,Birrer:2020tax}. The obtained $H_0$ varies between these different measurements. To show the application of the DRID to the $H_0$ fit, we will focus on the SH0ES result that has the largest deviation from the Planck result.
}\footnote{For recent reviews and other proposed solutions to the $H_0$ tension see, e.g.,~\cite{Knox:2019rjx, DiValentino:2021izs,Dainotti:2021pqg}.} 

In Table~\ref{tab:h0olympic}, we summarize our results for $H_0$ for different scenarios, and quantify the improvement compared to $\Lambda$CDM following the procedure in Ref.~\cite{Schoneberg:2021qvd}. Here `GT' refers to Gaussian Tension which is a simple measure of disagreement between two measurements,
\begin{align}
\rm{GT} \equiv \frac{H_{0,\mathcal{D}} - H_{0,\rm{SH0ES}}}{\sqrt{\sigma^2_\mathcal{D} + \sigma^2_{\rm{SH0ES}}}},
\end{align}
where $H_{0,\mathcal{D}}$ and $\sigma_\mathcal{D}$ are the central value and 68\% CL error, obtained from our simulations with Planck TTTEEE+low E+ lensing data.\footnote{For our simulations, the upper and lower errorbars are asymmetric around the best fit value for $H_0$ (see e.g.~Table~\ref{tab:param-fdr-fn}) and we use the smaller of the two errors to be conservative.} $H_{0,\rm{SH0ES}}$ and $\sigma_{\rm SH0ES}$ are obtained from~\cite{Riess:2021jrx} and are given by $73.04$ and $1.04$ in units of km/s/Mpc. The second criterion $\sqrt{\Delta\chi^2}$ quantifies how much the inclusion of the SH0ES data affects the best fit in the context of a given model, with a larger value signifying a larger tension.\footnote{For our numerical results, we use two collections of dataset. In the first one, we include only Planck data.
Since the $H_0$ determined using BAO dataset is in agreement with the Planck~\cite{Aghanim:2018eyx}, inclusion of the BAO data would not give
any additional inconsistency.
Therefore, in the second collection we include BAO and SH0ES at the same time. This way we can still make statements about how the $H_0$ values are changed in the absence or presence of SH0ES data.} This is defined via,
\begin{align}
\Delta\chi^2 \equiv \chi^2_{\rm min, \mathcal{D}+BAO+SH0ES} - \chi^2_{\rm min, \mathcal{D}},
\end{align}
where $\mathcal{D}$ includes full Planck data. The last criterion `$\Delta$AIC' measures how well a model M performs compared to $\Lambda$CDM when SH0ES data is included,
\begin{align}
\Delta\rm{AIC} = \chi^2_{\rm{min}, M} - \chi^2_{\rm{min}, \Lambda\rm{CDM}} + 2(N_M - N_{\Lambda\rm{CDM}}).
\end{align}
Due to the last factor, models with more parameters $\rm{N}_M$, but without associated improvement in $\chi^2$, are less favored according to this criterion. As will be explained later, the number of extra parameters for `FN' scenario is 3 (the energy density in DR, the amplitude of DR perturbation and its tilt). For `VN' this number is 4 since we also let the energy density in SM neutrino to vary.

\renewcommand{\arraystretch}{1.2}
\begin{table}
\centering
\begin{tabular}{c|c|c|c|c}
\hline
\hline
Scenario & $H_0$(km/s/Mpc) & GT & $\sqrt{\Delta\chi^2}$ & $\Delta$AIC \\
\hline
FN, FDR & $ 69.69^{+0.82}_{-1.3}$  & 2.5$\sigma$ & 4.0 & -9.4\\
FN, CDR & $69.57^{+0.88}_{-1.5}$ & 2.5$\sigma$ & 4.1 & -5.1\\
VN, FDR & $68.8^{+1.6}_{-1.6}$ & 2.2$\sigma$ & 4.4 & -6.1\\
VN, CDR & $69.2^{+1.6}_{-1.8}$ & 2.0$\sigma$ & 4.1 & -3.7\\
\hline
\end{tabular}
\caption{Quantification of the tension using different criteria following~\cite{Schoneberg:2021qvd} (see text for a description of various criteria). `FN' or `VN' refer to the case where neutrino energy density is kept fixed or varied, respectively. Here we also assume no correlation between isocurvature and adiabatic perturbations. We use the latest SH0ES data~\cite{Riess:2021jrx} which gives $H_0 = 73.04^{+1.04}_{-1.04}$km/s/Mpc.}
\label{tab:h0olympic}
\end{table}

As one example, when we allow isocurvature perturbations in CDR in `VN' scenario, the fit with the Planck data gives $H_0 = 69.2^{+1.6}_{-1.8}$~km/s/Mpc. The Gaussian tension with the \emph{latest} SH0ES measurement~\cite{Riess:2021jrx} $73.04\pm 1.04$~km/s/Mpc is then reduced to approximately $2.0\sigma$ compared to the $4.8\sigma$ ($3.1\sigma$) Gaussian tension between the SH0ES measurement and the $\Lambda$CDM~\cite{Aghanim:2018eyx} (plus adiabatic CDR studied in~\cite{Blinov:2020hmc}). When further including the SH0ES data in our analysis, the CDR-DRID `VN' scenario gives $H_0 = 71.46\pm 0.87$~km/s/Mpc, and the discrepancy is reduced to $1.2\sigma$. On the other hand, for `FN' scenario with FDR, the $\Delta$AIC result is promising and passes the requirement $\Delta \rm{AIC} < -6.91$ described in Ref.~\cite{Schoneberg:2021qvd}.




The outline of the paper is as follows. In the next section we discuss the definition of DR isocurvature, and present the isocurvature initial conditions for both FDR and CDR. We also write down a curvaton model that relates the cosmological observables used in the data analysis to the primordial fluctuations of the curvaton and the inflaton fields. This model serves as a simple example of how DR isocurvature can arise. Then
in Sec.~\ref{sec.sig} we derive model-independent results from the Markov-Chain Monte-Carlo (MCMC) study of the FDR and CDR scenarios using the Planck~2018 and BAO data, assuming negligible correlation between isocurvature and adiabatic perturbations. We show how DRID relaxes the tension between the CMB and the SH0ES measurements of $H_0$, and present bounds on DR isocurvature. While this updates the previous study of FDR isocurvature~\cite{Kawasaki:2011rc} by incorporating the Planck 2018 data, to the best of our knowledge, phenomenology of and bounds on CDR isocurvature are derived here for the first time. In Sec.~\ref{sec.CDRFDR} we explain the difference of the FDR and CDR isocurvature using analytical arguments, and explain why for isocurvature initial conditions FDR gives rise to larger CMB anisotropies than CDR, opposite to the case of adiabatic initial conditions. Our conclusions are in Sec.~\ref{sec.con}. In Appendix~\ref{sec.corr} we show the MCMC results when correlations between adiabatic and isocurvature perturbations are taken into account. In Appendix~\ref{sec:app:tri} and \ref{sec:app:tri-wcor} we include more detailed triangle plots for the analysis without and with correlation, respectively. 

\paragraph{Notations and conventions.} Here we summarize some of the notations often used throughout this paper. When discussing the model involving a curvaton, we use $R_{i}=\bar{\rho}_i/(\bar{\rho}_\gamma+\bar{\rho}_\nu+\bar{\rho}_{\rm DR})$ for $i=(\gamma,\nu,\text{DR})$ to denote the fractional homogeneous energy density in radiation species $i$. For our numerical simulations, $N_{\rm ur}$ and $N_{\rm dr}$ denote the effective number of degrees of freedom in SM neutrinos ($\nu$) and DR respectively, with $N_{\rm ur}=3.046$ being the standard $\Lambda$CDM choice. We define
 $N_{\rm tot}=N_{\rm ur}+N_{\rm dr}$ and $f_{\rm dr}=N_{\rm dr}/N_{\rm tot}$.  Curvature perturbation on uniform density hypersurfaces is denoted by $\zeta$~\cite{Malik:2008im} and the corresponding quantity for any individual species $i$ (e.g., $\gamma$, $\nu$ etc.) is denoted by $\zeta_i$. Isocurvature perturbations are defined with respect to photon perturbations. For example, DR isocurvature perturbations are defined via, $\sdr\equiv3(\zeta_{\rm DR}-\zeta_{\gamma})$. Given an isocurvature perturbation $\mathcal{S}_i$, we use $f_{\rm iso}\equiv A_{\rm iso}/A_{s}$ to denote the ratio of the square of the primordial power spectra of the isocurvature ($A_{\rm iso}$) and adiabatic ($A_{s}$) contributions. We also use $n_{\rm iso}$ to denote the tilt of the power spectrum of $\mathcal{S}_i$.

\section{Dark radiation isocurvature perturbations}\label{sec.DRiso}
If the Universe obtains all its density perturbations from a single source of quantum fluctuations, as in single-field inflation, then the density perturbations are necessarily adiabatic. In such a scenario, all of the Universe undergo the exact same expansion history albeit with different time delays, $\delta t(t,{\bf x})$ that varies as a function of $t$ and ${\bf x}$. In particular, we can use this time delay $\delta t(t,{\bf x})$ to describe the perturbations in all the components of the Universe,
\begin{equation}
    \delta t=\frac{\delta\rho_{\rm rad}}{\dot{\bar\rho}_{\rm rad}}=\frac{\delta\rho_c}{\dot{\bar\rho}_c}=\frac{\delta\rho_b}{\dot{\bar\rho}_b}\,.
\end{equation}
Here $\bar{\rho}_i$'s are the homogeneous energy densities of various components: $c$ and $b$ represent the cold dark matter (CDM) and baryon respectively, and the radiation `rad' can be photon $\gamma$, neutrino $\nu$, or DR. The overdot here denotes a derivative with respect to physical time $t$. From the continuity equation $\dot{\bar\rho}=-3H(\bar{\rho}+\bar{p})$, the perturbations are then seen to follow the \emph{adiabaticity} condition
\begin{equation}
    \frac{3}{4}\frac{\delta\rho_{\rm rad}}{\bar\rho_{\rm rad}}=\frac{\delta\rho_c}{\bar\rho_c}=\frac{\delta\rho_b}{\bar\rho_b}\,.
\end{equation}

However, there is no prior reason for all the perturbations to originate from a single source of fluctuations. For example, if there is another fluctuating scalar field $\sigma$ during the inflation which later reheats into DR with energy density $\rho_{\rm DR}$, and if DR only couples to the $\Lambda$CDM components through gravity, then the perturbations of $\rho_{\rm DR}$ need no longer respect the adiabaticity condition. Instead, a non-zero \emph{isocurvature} perturbation can arise~\footnote{In terms of the individual perturbations~$\zeta_i$ for species $i$~\cite{Wands:2000dp}, $\mathcal{S}_{\rm DR} \equiv3(\zeta_{\rm DR}-\zeta_{\gamma})$.}
\begin{equation}\label{eq:sdr}
    \sdr =\frac{3}{4}\left(\frac{\delta\rho_{\rm DR}}{\bar\rho_{\rm DR}}-\frac{\delta\rho_\gamma}{\bar\rho_\gamma}\right).
\end{equation}
Compared to the adiabatic perturbations, isocurvature perturbations in DR contribute differently to the coupled evolution of different species, and give rise to modifications of the CMB spectra beyond the standard $\Lambda\text{CDM}+\Delta N_{\rm tot}$-only result with all perturbations being adiabatic. 

While the scenario in which DR is free-streaming (FDR) and has isocurvature perturbation, is similar to the well-studied case of neutrino density isocurvature (NDI), when DR is a coupled fluid (CDR), the physics is qualitatively different. Such situations can arise, for example, if DR consists of massless gauge bosons from a deconfined non-abelian gauge theory, or massless dark photons scattering with dark electrons. In these cases, the corrections to the CMB spectra is also different compared to just having isocurvature perturbations in freely streaming extra neutrinos. To capture this difference in physics, we first need to derive the initial conditions for isocurvature perturbations for both CDR and FDR. Following this, we give a simple toy model of the $\chi$ field that can produce isocurvature perturbations in DR.  After discussing the constraints on FDR and CDR from different datasets in Sec.~\ref{sec.sig}, we come back in Sec.~\ref{sec.CDRFDR} to get an analytical understanding for the difference between FDR and CDR.

\subsection{Isocurvature initial conditions}
To derive the isocurvature initial conditions\footnote{See, e.g.,~\cite{Doran:2003xq,Valiviita:2008iv,Majerotto:2009np} for discussions of isocurvature initial conditions in different cosmological models.}, we work in the synchronous gauge parametrized by (see e.g.~\cite{Ma:1995ey}),
\begin{align}\label{eq.synch1}
ds^2=a^2(\tau)\left(-d\tau^2+(\delta_{ij}+h_{ij})dx^i dx^j\right)    
\end{align}
with,
\begin{align}\label{eq.synch2}
h_{ij}(\tau,\Vec{x})=\int d^3k e^{i\Vec{k}\cdot\Vec{x}}\left(\hat{k}_i\hat{k}_j h(\tau,\Vec{k})+\left(\hat{k}_i\hat{k}_j-\frac{1}{3}\delta_{ij}\right)6\eta(\tau,\Vec{k})\right).    
\end{align}
We also use the standard notation $\delta_i\equiv \delta\rho_i/\bar{\rho}_i$ to denote density perturbations in species $i$, and write the conformal Hubble rate in the presence of radiation and matter as,
\begin{align}
    \mathcal{H}(\tau)\equiv \frac{da/d\tau}{\tau}=\frac{1}{\tau}\frac{1+\frac{1}{2}\omega \tau}{1+\frac{1}{4}\omega \tau},
\end{align}
where $\omega\equiv a(\tau_i)\bar{\rho}_m(\tau_i)/(\sqrt{3\bar{\rho}_{r}(\tau_i)}M_{\rm pl})$ is determined in terms matter and radiation energy densities at initial time $\tau_i$.
For our choices of initial time, $\omega\tau\ll 1$ at early enough times and it will serve as an expansion parameter. Using these relations, we now sketch our derivation of the initial conditions that have a non-zero $\sdr$. The derived initial conditions can then be used along with the standard Boltzmann and Einstein equations to obtain the late time perturbations. In practice, we do this with our modified version of $\texttt{CLASS}$~\cite{lesgourgues2011cosmic} where we encode the new initial conditions.


As with isocurvature perturbations in other components, such as baryons, CDM or neutrinos, isocurvature in DR correspond to vanishing metric perturbations in the superhorizon limit, $k\tau\rightarrow 0$, where $k$ is a comoving momentum and $\tau$ is the conformal time.  As the modes renter the horizon at $k\tau=1$, DR perturbations gravitationally source the metric perturbations which in turn modify the perturbations in other components as well, eventually contributing to $C_{\ell}^{\rm TT,TE,EE}$ spectra.

Following the convention for the neutrino density isocurvature (NDI), e.g.,~\cite{Bucher:1999re}, we define the DR isocurvature initial condition by requiring the sum of the density perturbations in radiation to vanish
\begin{equation}\label{eq.sumrad}
    \delta\rho_{\rm DR}+\delta\rho_{\gamma}+\delta\rho_{\nu}=0\,,
\end{equation}
so that in the radiation dominated Universe at $k\tau\to 0$,  the radiation fluid is homogeneous and there is no curvature perturbation at all. Therefore, this initial condition does not affect the adiabatic curvature perturbations, as required. Furthermore, we also do not want any isocurvature perturbations for neutrinos, and therefore we require $\delta_\nu=\delta_\gamma$, and also choose a normalization, $\delta_{\rm DR}=1$.\footnote{Note, this normalization choice ignores the physical size of the isocurvature perturbations, i.e., the reader should imagine $\delta_\gamma$, for example, is multiplied by a factor $\sim\mathcal{O}(10^{-5})$ to get its physical size. This size will be accounted for precisely in our MCMC runs.} These two requirements, along with eq.~\eqref{eq.sumrad}, then completely fix the non-zero perturbations in the $k\tau\rightarrow 0$ limit,
\begin{align}
\delta_{\rm DR}=1,~~\delta_\gamma=\delta_\nu=-\frac{R_{\mathrm{DR}}}{1-R_{\mathrm{DR}}},
\end{align}
with all other perturbations vanishing at least as fast as $\mathcal{O}(k\tau)$.  Here we have defined energy fraction 
\begin{equation}
    R_{i}\equiv\frac{\bar{\rho}_i}{\bar{\rho}_{\gamma}+\bar{\rho}_{\nu}+\bar{\rho}_{\rm DR}}.
\end{equation}
To determine the rest of the perturbations, such as velocity divergence $\theta_i$, stress perturbation $\sigma_i$ of various components and the metric perturbations $h,\eta$, we proceed analytically order by order in $k\tau$ and $\omega\tau$ to obtain a power series solution of the coupled Boltzmann-Einstein equations in the synchronous gauge. These equations are as in Ma-Bertschinger~\cite{Ma:1995ey} with appropriate modifications to the Einstein equations to take into account the presence of DR.

All the above considerations apply to both FDR and CDR. Their difference however lies in the fact that CDR behaves like a fluid with negligible stress $\sigma_{\rm DR}\approx 0$, whereas FDR free streams and develops a non-zero stress $\sigma_{\rm DR}$, just like the neutrinos.
\subsubsection{Free-streaming DR (FDR)}
In this case, the equation of motion for FDR is identical to that of the neutrinos, namely,
\begin{align}
\dot{\delta}_{\rm DR}=&-\frac{4}{3}\theta_{\rm DR}-\frac{2}{3}\dot{h},\nonumber\\
\dot{\theta}_{\rm DR}=&k^2\left(\frac{1}{4}\delta_{\rm DR}-\sigma_{\rm DR}\right),\nonumber\\
\dot{\sigma}_{\rm DR}=&\frac{4}{15}\theta_{\rm DR}+\frac{2}{15}\dot{h}+\frac{4}{5}\dot{\eta},
\end{align}
where we have ignored moments higher than the quadrupole. To derive the initial conditions, we solve these above equations along with the rest of the Einstein and Boltzmann equations, order by order in $k\tau$ and $\omega\tau$. The resulting power series solutions are shown in Table~\ref{tab:fdric} up to $\mathcal{O}((k\tau)^2)$ or $\mathcal{O}((\omega\tau)(k\tau)^2)$ depending on the specific perturbations.

\renewcommand{\arraystretch}{1.2}
\begin{table}[h]
    \begin{center}
    \begin{tabular}{|c|c|c|c|c|c|c|}
    \hline
    \text{variable} & $\mathcal{O}(0)$ & $\mathcal{O}(k\tau)$ & $\mathcal{O}((k\tau)^2)$  & $\mathcal{O}(\omega k^2 \tau^3)$\\
    \hline
    \hline
    $\delta_{\gamma}$ & $-\frac{R_{\mathrm{DR}}}{1-R_{\mathrm{DR}}}$ & 0 & $\frac{R_{\mathrm{DR}}}{6(1-R_{\mathrm{DR}})}$ &  \\
    \hline
    $\theta_\gamma/k$ & 0 & $-\frac{R_{\mathrm{DR}}}{4(1-R_{\mathrm{DR}})}$ & 0 &    \\
    \hline
    $\delta_\nu$ & $-\frac{R_{\mathrm{DR}}}{1-R_{\mathrm{DR}}}$ & 0 & $\frac{R_{\mathrm{DR}}}{6(1-R_{\mathrm{DR}})}$ &   \\
    \hline
    $\theta_\nu/k$ & 0 & $-\frac{R_{\mathrm{DR}}}{4(1-R_{\mathrm{DR}})}$ & 0 &   \\
    \hline
    $\sigma_\nu$ & 0 & 0 & $-\frac{19 R_{\mathrm{DR}}}{30(1-R_{\mathrm{DR}})(15+4 R_{\mathrm{DR}} +4 R_\nu)}$ &   \\
    \hline
    $\delta_{\rm DR}$ & 1 & 0 & $-\frac{1}{6}$ &   \\
    \hline
    $\theta_{\rm DR}/k$ & 0 & $\frac{1}{4}$ & 0 &  \\
    \hline
    $\sigma_{\rm DR}$ & 0 & 0 & $\frac{15-15R_{\mathrm{DR}}+4 R_\nu}{30(1-R_{\mathrm{DR}})(15+4R_{\mathrm{DR}}+4 R_\nu)}$ &  \\
    \hline
    $\eta$ & 0 & 0 & $\frac{-R_{\mathrm{DR}}+R_{\mathrm{DR}}^2+R_{\mathrm{DR}} R_\nu}{6(1-R_{\mathrm{DR}})(15+4 R_{\mathrm{DR}}+4R_\nu)}$ &  \\
    \hline
    $h$ & 0 & 0 & 0 & $\frac{R_{\mathrm{DR}} R_b}{40(1-R_{\mathrm{DR}})}$ \\
    \hline
    $\delta_b$ & 0 & 0 & $\frac{R_{\mathrm{DR}}}{8(1-R_{\mathrm{DR}})}$ &   \\
    \hline
    $\delta_c$ & 0 & 0 & 0 & $-\frac{R_{\mathrm{DR}} R_b}{80(1-R_{\mathrm{DR}})}$ \\
    \hline
    \end{tabular}
    \caption{Isocurvature initial conditions for free-streaming DR (FDR) with the normalization $\delta_{\rm DR}=1$. For the contributions at $\mathcal{O}(\omega k^2 \tau^3)$, we only show the results of $h$ and $\delta_c$ since this is the order at which they are first non-zero.}
    \label{tab:fdric}
    \end{center}
\end{table}

\paragraph{Comparison with NDI.}
Due to their free-streaming nature, isocurvature initial conditions (IC) in FDR should be equivalent to that in neutrinos. To see this, we can make the replacement in FDR IC in the following order: $R_\nu\rightarrow 0; R_{\mathrm{DR}}\rightarrow R_\nu; \{\delta,\theta,\sigma\}_{\rm DR}\rightarrow \{\delta,\theta,\sigma\}_\nu$. Upon doing that, we see that the results in Table~\ref{tab:fdric} matches with the standard NDI initial condition result~\cite{Bucher:1999re}.\footnote{To exactly match with~\cite{Bucher:1999re} we need to make the identification $\Omega_{b,0}\equiv \frac{1}{4}R_b\omega$.} Therefore, we recover the expected result that FDR has the same physical effect as NDI with varying $\Delta N_{\rm tot}$ as noted, for example, in Ref.~\cite{Kawasaki:2011rc,Adshead:2020htj}.

\subsubsection{Coupled DR (CDR)}
The procedure for deriving initial conditions in this case is identical to the case of FDR, except we drop the DR Boltzmann hierarchy equations involving moments higher than the dipole, $\theta_{\rm DR}$, and set $\sigma_{\rm DR}=0$. The results are shown in Table~\ref{tab:cdric}. 

\begin{table}[h]
    \begin{center}
    \begin{tabular}{|c|c|c|c|c|c|c|}
    \hline
    \text{variable} & $\mathcal{O}(0)$ & $\mathcal{O}(k\tau)$ & $\mathcal{O}((k\tau)^2)$ &  $\mathcal{O}(\omega k^2 \tau^3)$ \\
    \hline
    \hline
    $\delta_{\gamma}$ & $-\frac{R_{\mathrm{DR}}}{1-R_{\mathrm{DR}}}$ & 0 & $\frac{R_{\mathrm{DR}}}{6(1-R_{\mathrm{DR}})}$ &   \\
    \hline
    $\theta_\gamma/k$ & 0 & $-\frac{R_{\mathrm{DR}}}{4(1-R_{\mathrm{DR}})}$ & 0 &    \\
    \hline
    $\delta_\nu$ & $-\frac{R_{\mathrm{DR}}}{1-R_{\mathrm{DR}}}$ & 0 & $\frac{R_{\mathrm{DR}}}{6(1-R_{\mathrm{DR}})}$ &   \\
    \hline
    $\theta_\nu/k$ & 0 & $-\frac{R_{\mathrm{DR}}}{4(1-R_{\mathrm{DR}})}$ & 0 &   \\
    \hline
    $\sigma_\nu$ & 0 & 0 & $-\frac{R_{\mathrm{DR}}}{2(1-R_{\mathrm{DR}})(15+4 R_\nu)}$ &   \\
    \hline
    $\delta_{\rm DR}$ & 1 & 0 & $-\frac{1}{6}$ &   \\
    \hline
    $\theta_{\rm DR}/k$ & 0 & $\frac{1}{4}$ & 0 &  \\
    \hline
    $\eta$ & 0 & 0 & $\frac{R_{\mathrm{DR}} R_\nu}{6(1-R_{\mathrm{DR}})(15+4R_\nu)}$  &  \\
    \hline
    $h$ & 0 & 0 & 0 & $\frac{R_{\mathrm{DR}} R_b}{40(1-R_{\mathrm{DR}})}$ \\
    \hline
    $\delta_b$ & 0 & 0 & $\frac{R_{\mathrm{DR}}}{8(1-R_{\mathrm{DR}})}$ &   \\
    \hline
    $\delta_c$ & 0 & 0 & 0 & $-\frac{R_{\mathrm{DR}} R_b}{80(1-R_{\mathrm{DR}})}$ \\
    \hline
    \end{tabular}
    \caption{Isocurvature initial conditions for coupled DR (CDR) in synchronous gauge and otherwise same as Table~\ref{tab:fdric}.}
    \label{tab:cdric}
    \end{center}
\end{table}

Importantly, these initial conditions can not be recasted as NDI initial conditions. We see from Tables~\ref{tab:fdric}~and~\ref{tab:cdric}, that the primary difference among the common quantities arise for $\eta$ and $\sigma$. The reason and the physical implication of this will be discussed in detail in Sec.~\ref{sec.CDRFDR}.

\subsection{A curvaton model}
Here we discuss a simple model involving a light axion or curvaton-like field $\chi$ that acquires isocurvature fluctuations during inflation. We assume that both during and after inflation, the energy density in $\chi$ is subdominant compared to the inflaton and its decay products.  For masses $m_\chi\ll H_{\rm inf}$, the inflationary Hubble scale,  $\chi$ remains frozen to a `misaligned' value $\chi_*$ during inflation.  Subsequently, when the post-inflationary Hubble scale falls below its mass, it starts oscillating coherently around the minimum of its potential. Eventually, $\chi$ decays into DR which inherits the isocurvature fluctuations in $\chi$. 

To write an expression for DR isocurvature perturbation, we
assume a simple quadratic potential for $\chi$,
\begin{equation}
    V(\chi)=\frac{1}{2}m_{\chi}^2\chi^2\,,
\end{equation}
along with the fact that $\chi$ contributes negligibly to the energy density during inflation. In that case, the isocurvature fluctuation in $\chi$ is given by (see e.g.~\cite{Langlois:2004nn})
\begin{align}
\mathcal{S}_\chi=\frac{2\delta\chi}{\chi}\bigg\rvert_*,
\end{align}
where $*$ denotes the fact that the RHS is evaluated during the horizon exit and $\delta\chi$ denotes the quantum fluctuation of $\chi$.  To see this, one can go to the uniform density gauge at the onset of $\chi$ oscillations, determined by $H\approx2m_\chi/3$, to write 
\begin{align}
\mathcal{S}_\chi\equiv3(\zeta_\chi-\zeta_\phi)=&3(\zeta_\chi-\zeta_{\rm SM})\nonumber \\
=&\frac{\delta\rho_{\chi}}{\bar{\rho}_{\chi}}-\frac{3}{4}\frac{\delta\rho_{\rm SM}}{\bar{\rho}_{\rm SM}}\approx \frac{\delta\rho_{\chi}}{\bar{\rho}_{\chi}}.
\end{align}
In the above, we have used that the SM radiation bath comes from the inflaton decay $\zeta_\phi=\zeta_{\rm SM}$, and since the SM radiation bath dominates the energy density,  on the uniform density slice we have $\delta\rho_{\rm SM}\approx\delta\rho=0$.  Finally, we can use the fact $\delta\rho_\chi/\bar{\rho}_\chi\approx 2\delta\chi/\chi$ remains constant with time on superhorizon scales~\cite{Polarski:1992dq}. Therefore, we can write the power spectrum of the $\chi$ perturbation in a form similar to the adiabatic case,
\begin{equation}
\mathcal{P}_{\delta\chi} \simeq \left(\frac{H}{\pi\chi_*}\right)^2\left(\frac{k}{aH}\right)^{2\eta_\chi - 2\epsilon}\,,
\end{equation}
where $\eta_\chi=m_\chi^2/(3H_{\rm inf}^2)$ and $\epsilon=-\dot{H}_{\rm inf}/H_{\rm inf}^2$. The tilt of the spectrum relates to the spectral index $n_{\rm iso}=1+2\eta_\chi-2\epsilon$ in Eq.~(\ref{eq:fiso}).

To relate the fluctuations in DR, $\zeta_{\rm DR}$ to the fluctuations of $\chi$ $\zeta_\chi$, we use the sudden decay approximation and go to the uniform density hypersurface at the time of $\chi$ decay, determined by $\Gamma_\chi=H$~\cite{Sasaki:2006kq}. On this hypersurface, the various curvature perturbations to first order in fluctuations are given by,
\begin{align}
\zeta=-\psi,~~
\zeta_{\chi}=-\psi+\frac{1}{3}\frac{\delta\rho_\chi}{\bar{\rho}_\chi},\\
\zeta_{\rm DR}=-\psi+\frac{1}{4}\frac{\delta\rho_{\rm DR}}{\bar{\rho}_{\rm DR}},~~
\zeta_{\rm SM}=&-\psi+\frac{1}{4}\frac{\delta\rho_{\rm SM}}{\bar{\rho}_{\rm SM}}.
\end{align}
Since DR originates from $\chi$-decay, we can relate their energy densities on the above hypersurface,
\begin{align}
\bar{\rho}_{\rm DR}=\bar{\rho}_\chi,~~\delta\rho_{\rm DR}=\delta\rho_\chi,
\end{align}
which implies,
\begin{align}
\zeta_{\rm DR}=\frac{3}{4}\zeta_\chi+\frac{1}{4}\zeta.   
\end{align}
Finally, since we are on the uniform density hypersurface, we have $\delta\rho_{\chi}+\delta\rho_{\rm SM}=0$ which implies,
\begin{align}\label{eq.totzeta}
\zeta=\frac{4(1-f_\chi)\zeta_{\rm SM}+3 f_\chi \zeta_\chi}{4(1-f_\chi)+3 f_\chi},    
\end{align}
where $f_\chi=\bar{\rho}_\chi/(\bar{\rho}_\chi+\bar{\rho}_{\rm SM})$ is the energy density fraction in $\chi$ at the time of its decay.
Therefore we can write the final DR density perturbation as,
\begin{align}
\zeta_{\rm DR}=\frac{3}{4-f_\chi}\zeta_\chi+\frac{1-f_\chi}{4-f_\chi}\zeta_{\phi}.
\end{align}
Using the isocurvature perturbation due to $\chi$, $\mathcal{S}_\chi=3(\zeta_\chi-\zeta_\phi)$ we can now write the expression for DR isocurvature perturbation,
\begin{align}
\mathcal{S}_{\rm DR}=\frac{3}{4-f_\chi}\mathcal{S}_\chi.    
\end{align}
Therefore, the power spectrum of DR isocurvature perturbation is given by,
\begin{align}
    \mathcal{P}_{\rm DR} =\left(\frac{3}{4-f_\chi}\right)^2
    \mathcal{P}_{\delta\chi}.
\end{align}
Now to get the correlation of $S_{\rm DR}$ with primordial curvature perturbation after curvaton decay, we write eq.~\eqref{eq.totzeta} as,
\begin{align}
\zeta=\zeta_{\rm SM}+\frac{f_\chi}{3}\mathcal{S}_{\rm DR},
\end{align}
implying a correlation between primordial curvature perturbation after $\chi$ decay and DR isocurvature perturbation,
\begin{align}
\cos\Delta=\frac{\langle\zeta \mathcal{S}_{\rm DR}\rangle}{\langle\zeta\zeta\rangle^{1/2}\langle \mathcal{S}_{\rm DR}\mathcal{S}_{\rm DR}\rangle^{1/2}}\propto f_\chi.
\end{align}
Here we have used that $\langle\zeta_{\rm SM}\mathcal{S}_{\rm DR}\rangle\propto \langle\delta\phi\delta\chi\rangle=0$. This implies for scenarios where DR contribution to the radiation energy density is subdominant $f_\chi
\ll 1$, as we will be interested in this work, the correlation $\cos\Delta\ll 1$. In our numerical study, the ratio of the curvature perturbation in Eq.~(\ref{eq:fiso}) is $f_{\rm iso}\simeq (\mathcal{S}_{\rm DR}/\zeta)^2$. While the above discussion serves as an example, where correlation between curvature perturbation and isocurvature perturbation is small, we now focus on constraining DR isocurvature using CMB and other data sets in a model-\textit{independent} manner. In the main text, we will give the results assuming zero correlation between curvature and isocurvature perturbations, while leaving the more general results with correlation for Appendix~\ref{sec.corr}. To this end, we will also use the comoving curvature perturbation $\mathcal{R}$, as opposed to $\zeta$, to describe our constraints following Planck~\cite{Planck:2018jri}. This difference will not be important for the initial conditions we use, since on superhorizon scales, $\mathcal{R}\approx\zeta$, see e.g.~\cite{Malik:2008im,Gordon:2000hv}. Also for notational similarity with Planck, we will use $\mathcal{I}\equiv\mathcal{S}_{\rm DR}$ to denote DR isocurvature.

\section{CMB signals of DR isocurvature}\label{sec.sig}

In this section, we describe our Bayesian analysis of cosmologies with (mixed) FDR and CDR isocurvature initial conditions, with current cosmological datasets using Markov Chain Monte Carlo (MCMC) sampler \texttt{MontePython}~\cite{Audren:2012wb,Brinckmann:2018cvx}. We used the latest Planck 2018 CMB temperature, polarization and lensing power spectra~\cite{Aghanim:2019ame}. In addition, we have also used Baryon acoustic oscillation (BAO) measurements and local measurement of the Hubble constant to constrain the parameter space. The plots are generated using the python package \texttt{GetDist}~\cite{Lewis:2019xzd}.

\paragraph{Datasets:} The dataset combination `P18-TTTEEE+lowE+lensing' denotes the combination of low-$\ell ~(\ell < 30)$ TT, low-$\ell ~ (\ell < 30)$ EE, high-$\ell ~(\ell \ge 30)$ TTTEEE and lensing likelihoods. Thus, `P18-TTTEEE+lowE+lensing' contains the full information of the temperature, polarization and lensing power spectra measurements from Planck. For BAO data we use the 6DF Galaxy survey~\cite{2011MNRAS.416.3017B}, SDSS-DR7 MGS data~\cite{Ross:2014qpa}, and the BOSS measurement of BAO scale and $f\sigma_8$ from DR12 galaxy sample~\cite{Alam:2016hwk}. For a local measurement of the Hubble constant, we use the latest measurement $H_0 = 73.04 \pm 1.04$~km/s/Mpc by the SH0ES collaboration~\cite{Riess:2021jrx}, and denote it by `SH0ES(L)'. We use the following likelihood combination for our analysis:`P18-TTTEEE+lowE+lensing' and `P18-TTTEEE+lowE+lensing+BAO+SH0ES(L)'. 

\subsection{Parameters of DR isocurvature}

Following the Planck analysis of isocurvature perturbation~\cite{Planck:2013jfk,Ade:2015lrj,Planck:2018jri}, we use the `two-scale' parametrization for the primordial power spectra of the dark radiation isocurvature perturbations. Here, we briefly describe the notations and derive the relation with the conventional amplitude and spectral index parametrization. A generalised power spectrum $\pow_{ab}(k)$, having power law dependence of $k$, can be parametrized with its value specified at two scales $k=k_1$ and $k=k_2$ as~\cite{Planck:2013jfk}
\begin{equation}\label{eq:twoscaledef}
\powab =\exp\left[ \ln\poab {\ln k - \ln k_2 \over \ln k_1 - \ln k_2 } + \ln\ptab {\ln k - \ln k_1 \over \ln k_2 - \ln k_1 }\right]\;,
\end{equation} 
with $\poab \equiv \pow_{ab}(k_1)$ and $\ptab \equiv \pow_{ab}(k_2)$ being the corresponding amplitudes. Here $ a,b = \mathcal{R} , \mathcal{I} $ where $\mathcal{R}$ and $\mathcal{I}$ stand for adiabatic and isocurvature (DRID) perturbations, respectively. In accordance with the Planck analysis~\cite{Planck:2018jri}, we choose $k_1 = 0.002 \ {\rm Mpc}^{-1}$ and $k_2 = 0.1 \ {\rm Mpc}^{-1}$ so that the range $[k_1,k_2]$ spans most of the modes constrained by the Planck data.
The spectral index and the amplitude of the primordial adiabatic perturbations can be derived from the above parametrization as,
\begin{equation}\label{eq:nsAs}
n_s = 1+  {\ln \mathcal{P}_\mathcal{RR}^{(1)} -\ln \mathcal{P}_\mathcal{RR}^{(2)} \over \ln k_1 -\ln k_2}\;\quad\text{and}\quad A_s =  \mathcal{P}_\mathcal{RR}^{(1)} \exp\left[(n_s -1)\ln\left(k_\ast \over k_1\right)\right]\;.
\end{equation}
Here $k_\ast$ is the pivot scale where the amplitude is defined. Similarly, for isocurvature perturbation we have
\begin{equation}\label{eq:niAi}
    n_\text{iso} = 1+  {\ln \mathcal{P}_\mathcal{II}^{(1)} -\ln \mathcal{P}_\mathcal{II}^{(2)}  \over \ln k_1 -\ln k_2}\quad\text{and}\quad A_{\rm iso} =  \mathcal{P}_\mathcal{II}^{(1)} \exp\left[(n_{\rm iso} -1)\ln\left(k_\ast \over k_1\right)\right]\;.
\end{equation}
For this analysis, we choose $k_\ast = 0.05~{\mpcinv}$. To compare the strength of the isocurvature perturbation with respect to the adiabatic perturbation, it is often useful to defined isocurvature perturbation fraction $(f_{\rm iso})$ as,\
\begin{equation}\label{eq:fiso}
f_\text{iso} \equiv {A_\text{iso} \over A_s} = {\mathcal{P}_\mathcal{II}^{(1)}  \over \mathcal{P}_\mathcal{RR}^{(1)}}\left(k_\ast \over k_1\right)^{n_\text{iso} -n_s}.
\end{equation}

In the main text, we focus on the \emph{uncorrelated} DRID perturbations. Therefore, we ignore the correlation between DRID and the adiabatic perturbation by setting $\pow_\mathcal{RI}^{(1)} = \pow_\mathcal{RI}^{(2)} = 0$. Later in Appendix~\ref{sec.corr} and \ref{sec:app:tri-wcor} we will relax this assumption by allowing non-zero correlation between these perturbations. However, as shown later, uncorrelated DRID perturbation scenario can successfully capture all the essential physical effects of DRID perturbations relevant for cosmology.

Therefore, the simplest extension of the \lcdm{} cosmology to study DRID perturbation would include three new parameters: the energy density of dark radiation which is introduced in terms of the effective number of degrees of freedom $N_{\rm dr}$, the amplitudes of DRID perturbations $\mathcal{P}_\mathcal{II}^{(1)}$ and $\mathcal{P}_\mathcal{II}^{(2)}$ at the scales $k_1$ and $k_2$, respectively. However, this vanilla setup is not suitable for the analysis with cosmological data and have issues with convergence, as we now explain.

As shown in Tables~\ref{tab:fdric}~and~\ref{tab:cdric}, in the presence of DRID perturbation, the initial photon perturbations and the metric perturbations at leading order are proportional to the fractional energy density of the dark radiation:
\begin{equation}
    \delta_\gamma, \theta_\gamma, \eta, h \propto \frac{R_{\mathrm{DR}}}{1-R_{\mathrm{DR}}} \approx R_{\mathrm{DR}} \propto N_{\rm dr}\;.
\end{equation}
In the last equation, we have used the fact that the energy fraction of dark radiation $R_{\mathrm{DR}}$ is expected to be small  since \lcdm{} is an excellent description of the observed universe. This scaling of photon perturbations also holds for sub-horizon evolution described by linearized Boltzmann equations, i.e.,
\begin{equation}
    F_{\gamma \ell}(k) \propto N_{\rm dr}\;~\text{for}~N_{\rm dr}\ll 1,
\end{equation}
where $F_{\gamma \ell}(k)$ is the $\ell$-th multipole of photon transfer function.
Therefore, the CMB spectrum induced by DRID perturbation has the (approximate) degeneracy in the following two parameters,
\begin{equation}\label{eq:drid-dege}
    C_{\ell,{\rm DRID}} \propto A_{\rm iso}N_{\rm dr}^2\;.
\end{equation}
Since, $N_{\rm dr}$ can in principle be very small, $A_{\rm iso}$ can take very large value in those cases due to the degeneracy. Thus, for a fixed magnitude of $C_{\ell,{\rm DRID}}$, the DRID perturbation amplitude  $A_{\rm iso} $ (equivalently $\mathcal{P}_\mathcal{II}^{(1)}$ and $\mathcal{P}_\mathcal{II}^{(2)}$) varies across a wide range of scales depending on the value of $N_{\rm dr}$. Therefore, due to the large variation of $\mathcal{P}_\mathcal{II}^{(1)}$ and $\mathcal{P}_\mathcal{II}^{(2)}$ across several orders of magnitude, the convergence of the MCMC runs with these two variable as primary cosmological parameters is rather poor.

\renewcommand{\arraystretch}{1.4}
\begin{table}[h!]
    \centering
    \begin{tabular}{|l|c| c| }
\hline
FDR (FN \& NC)& \makecell[c]{P18-TTTEEE\\+lowE+lensing}& \makecell[c]{P18-TTTEEE+lowE+ \\lensing+BAO+SH0ES(L)}\\
\hline
$10^2 \omega_{b}$& $ 2.264^{+0.018}_{-0.021}$& $ 2.281\pm 0.015$\\
\hline
$\omega_{cdm }$& $ 0.1217^{+0.0017}_{-0.0025}$& $ 0.1245\pm 0.0025$\\
\hline
$100\theta_{s }$& $ 1.04219\pm 0.00045$& $ 1.04193\pm 0.00047$\\
\hline
$\tau_{reio }$& $ 0.0563^{+0.0070}_{-0.0079}$& $ 0.0561\pm 0.0072$\\
\hline
$10^{10}\mathcal{P}_\mathcal{RR}^{(1)}$& $ 23.11\pm 0.49$& $ 22.83\pm 0.47$\\
\hline
$10^{10}\mathcal{P}_\mathcal{RR}^{(2)}$& $ 20.55^{+0.35}_{-0.41}$& $ 20.72\pm 0.36$\\
\hline
$ 10^{10} N_{\rm dr}^2 \mathcal{P}_\mathcal{II}^{(1)}$& $< 17.9$& $ 14.8^{+2.1}_{-15}$\\
\hline
$10^{10} N_{\rm dr}^2\mathcal{P}_\mathcal{II}^{(2)}$& $ 106^{+40}_{-70}$& $ 129^{+50}_{-60}$\\
\hline
$N_{\rm dr}$& $< 0.216$& $ 0.36\pm 0.13$\\
\hline
\hline$H_0 ({\rm km/s/Mpc})$& $ 69.69^{+0.82}_{-1.3}$& $ 70.94\pm 0.80$\\
\hline
$\sigma_8$& $ 0.8249^{+0.0075}_{-0.0087}$& $ 0.8313\pm 0.0085$\\
\hline
$10^9 A_s$& $ 2.098^{+0.031}_{-0.035}$& $ 2.107\pm 0.032$\\
\hline
$n_{s }$& $ 0.9700^{+0.0062}_{-0.0074}$& $ 0.9752\pm 0.0062$\\
\hline
$n_{\rm iso}$& $ 1.54^{+0.34}_{-0.29}$& $ 1.61^{+0.32}_{-0.27}$\\
\hline
$f_{\rm iso}$& $< 18.7$& $< 6.52$\\
\hline
$N_{\rm tot}$& $< 3.26$& $ 3.41\pm 0.13$\\
\hline
$f_{\rm dr}$& $ 0.052^{+0.022}_{-0.047}$& $ 0.104^{+0.036}_{-0.032}$\\
\hline
\hline
$ \chi^2 - \chi^2_{\Lambda \rm{CDM}}$& $-1.94$& $-15.4$\\
\hline
\end{tabular}
    \caption{Mean and $1\sigma$ error of parameters for FDR-DRID (uncorrelated and fixed $N_{\rm ur}$) for the corresponding datasets. The limits are at 68\% C.L. The constraints on the primary parameters and the derived parameters are shown in two separate blocks. The $\chi^2$ difference with respect to the \lcdm{} (fixed \neff{}) model for the corresponding data-set is shown on the last line.}
    \label{tab:param-fdr-fn}
\end{table}

\begin{table}[htb!]
    \centering
    \begin{tabular}{|l|c| c| }
\hline
CDR (FN \& NC)& \makecell[c]{P18-TTTEEE\\+lowE+lensing}& \makecell[c]{P18-TTTEEE+lowE+ \\lensing+BAO+SH0ES(L)}\\
\hline
$10^2 \omega_{b}$& $ 2.262^{+0.019}_{-0.024}$& $ 2.286\pm 0.016$\\
\hline
$\omega_{cdm }$& $ 0.1228^{+0.0018}_{-0.0030}$& $ 0.1268\pm 0.0028$\\
\hline
$100\theta_{s }$& $ 1.04230^{+0.00034}_{-0.00038}$& $ 1.04260\pm 0.00034$\\
\hline
$\tau_{reio }$& $ 0.0562^{+0.0070}_{-0.0080}$& $ 0.0568^{+0.0066}_{-0.0075}$\\
\hline
$10^{10}\mathcal{P}_\mathcal{RR}^{(1)}$& $ 23.32\pm 0.47$& $ 23.19\pm 0.47$\\
\hline
$10^{10}\mathcal{P}_\mathcal{RR}^{(2)}$& $ 20.33\pm 0.35$& $ 20.22\pm 0.36$\\
\hline
$10^{10} N_{\rm dr}^2 \mathcal{P}_\mathcal{II}^{(1)}$& $< 20.6$& $< 15.8$\\
\hline
$10^{10} N_{\rm dr}^2 \mathcal{P}_\mathcal{II}^{(2)}$& $ 241^{+70}_{-200}$& $ 339^{+100}_{-300}$\\
\hline
$N_{\rm dr}$& $< 0.244$& $ 0.43\pm 0.13$\\
\hline
\hline$H_0 ({\rm km/s/Mpc})$& $ 69.57^{+0.88}_{-1.5}$& $ 71.28\pm 0.85$\\
\hline
$\sigma_8$& $ 0.8237\pm 0.0069$& $ 0.8270\pm 0.0069$\\
\hline
$10^9 A_s$& $ 2.083\pm 0.032$& $ 2.072\pm 0.032$\\
\hline
$n_{s }$& $ 0.9649^{+0.0062}_{-0.0056}$& $ 0.9650^{+0.0071}_{-0.0055}$\\
\hline
$n_{\rm tot}$& $ 1.69^{+0.40}_{-0.34}$& $ 1.86^{+0.42}_{-0.31}$\\
\hline
$f_{\rm tot}$& $< 22.4$& $ 7.1^{+1.7}_{-3.2}$\\
\hline
$N_{\rm tot}$& $< 3.29$& $ 3.48\pm 0.13$\\
\hline
$f_{\rm dr}$& $ 0.059^{+0.024}_{-0.052}$& $ 0.123\pm 0.034$\\
\hline
\hline
$ \chi^2 - \chi^2_{\Lambda \rm{CDM}}$& $1.34$& $-11.06$\\
\hline
\end{tabular}
    \caption{Mean and $1\sigma$ error of parameters for CDR-DRID (uncorrelated and fixed $N_{\rm ur}$)  for the corresponding datasets. The limits are at 68\% C.L. The constraints on the primary parameters and the derived parameters are shown in two separate blocks. The $\chi^2$ difference with respect to the \lcdm{} (fixed \neff{}) model for the corresponding datasets is shown on the last line.}
    \label{tab:param-cdr-fn}
\end{table}

To circumvent the convergence issue at $N_{\rm dr}\ll1$, we decided to vary the composite isocurvature amplitude parameters $N_{\rm dr}^2\mathcal{P}_\mathcal{II}^{(1)}$ and $N_{\rm dr}^2\mathcal{P}_\mathcal{II}^{(2)}$ as primary parameters for the MCMC runs. These parameter combinations are more physical since they directly influence the observed $C_\ell$. Thus, for the analysis we augment the \lcdm{} cosmology with three new parameters: $N_{\rm dr}$, $N_{\rm dr}^2\mathcal{P}_\mathcal{II}^{(1)}$ and $N_{\rm dr}^2\mathcal{P}_\mathcal{II}^{(2)}$.
We perform two separate sets of analysis where we treat the neutrino contribution to the total energy density differently.
In the first scenario (dubbed as `FN'), we keep the effective number of degrees of freedom in neutrinos: $N_{\rm ur}$ fixed to the $\Lambda$CDM value $3.046$. In the other scenario (dubbed as `VN'), we let $N_{\rm ur}$ vary as a cosmological parameter.
In both these model, the total effective neutrino degrees of freedom is given by, $N_{\rm tot} = N_{\rm ur} + N_{\rm dr}$ and the DR fraction is defined as $\fracdr = N_{\rm dr}/N_{\rm tot}$.
In our analysis we took neutrinos and the dark radiation to be massless. For incorporating adiabatic perturbation, we use the `two-scale' amplitudes $\mathcal{P}_\mathcal{RR}^{(1)}$ and $\mathcal{P}_\mathcal{RR}^{(2)}$ whose relation with the familiar $A_s$ and $n_s$ are given in Eq.~\eqref{eq:nsAs}. We have used Metropolis Hastings algorithm to perform the Bayesian analysis. We choose flat prior for all four additional parameters (compared to \lcdm{}) with no hard prior upper boundary. The Gelman-Rubin convergence criterion~\cite{Gelman:1992zz} $R-1<0.01$ was satisfied by all the MCMC chains.


\subsection{Constraint on DR isocurvature}

\subsubsection{with fixed $N_{\rm ur} = 3.046$ (FN)}
\begin{figure}[htb!]
	\centering
		\includegraphics[width=0.85\linewidth]{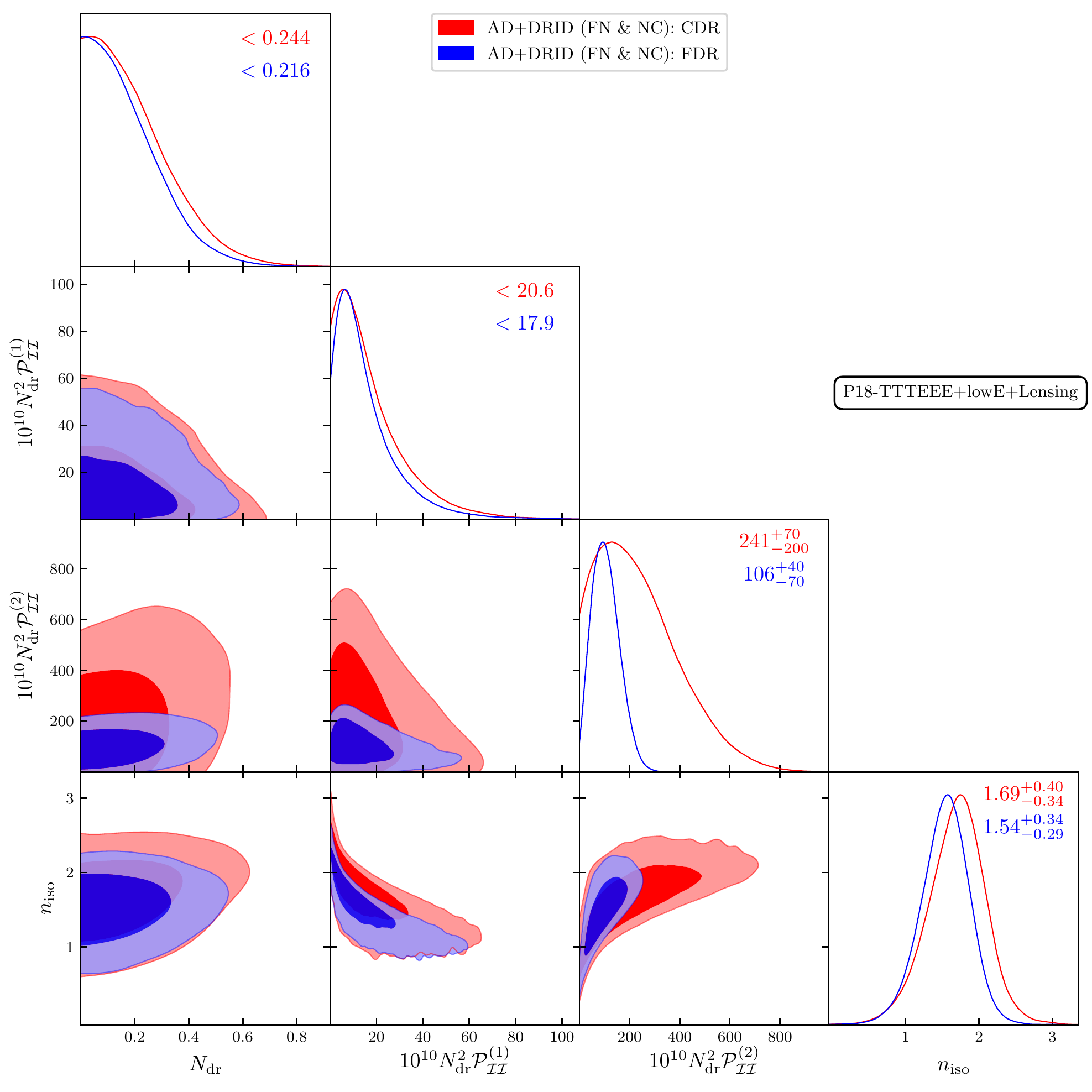}
		\caption{Triangle plot for isocurvature parameters and $ N_{\rm dr} $ for fixed $N_{\rm ur}$ analysis for P18-TTTEEE+lowE+lensing dataset. The constraints on individual parameters are mentioned on the diagonal 1-D posteriors with corresponding colors. The errors represent $1\sigma$ errorbar and the limits are at 68\% confidence level (C.L.). Here and in the following triangle plots, the inner and outer contours respectively denote $1\sigma$ and $2\sigma$ constraints.}
		\label{fig:is0-param-tri-fn}
	\end{figure}

First we analyse the case where $N_{\rm ur}$ is kept fixed to its standard value of $3.046$. This scenario represents the minimalist scenario where the curvaton decays exclusively to dark sector and thus does not affect the neutrino abundance. In Table~\ref{tab:param-fdr-fn} and Table~\ref{tab:param-cdr-fn}, we present the summary constraints for the uncorrelated FDR-DRID and CDR-DRID analysis, respectively.
In Fig.~\ref{fig:is0-param-tri-fn} we show the constraints on the isocurvature parameters along with $N_{\rm dr}$ for P18-TTTEEE+lowE+lensing dataset. Additionally, we
include the constraint on the derived parameter $n_\iso$ which is the spectral index of the DRID perturbation.
Among these two types of fluid DRID, the value of \neff{} is higher in CDR compared to FDR.
	
\begin{figure}[htb!]
	\centering
		\includegraphics[width=0.85\linewidth]{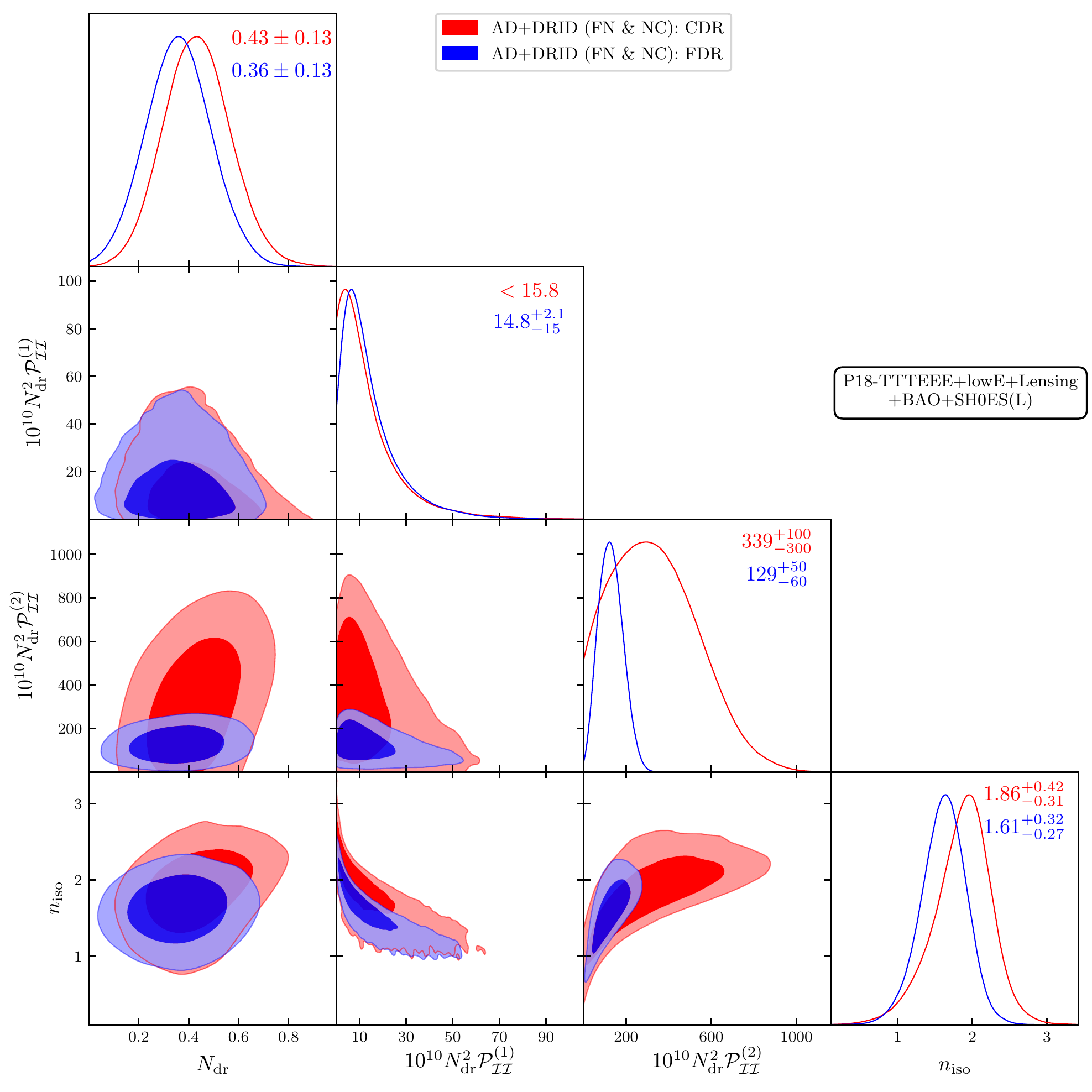}
		\caption{Triangle plot Isocurvature parameters and $ N_{\rm dr} $ for fixed $N_{\rm ur}$ analysis for P18-TTTEEE+lowE+lensing+lensing+SH0ES(L) dataset. The constraints on individual parameters are mentioned on the 1D diagonal posteriors with corresponding colors. The errors represent $1\sigma$ errorbar and the limits are at 68\% C.L.}
		\label{fig:is0-param-tri-wsh0es-fn}
	\end{figure}

For both these types of DRID perturbation, the amplitude of the physical isocurvature perturbation at small scale (\phpiib) is higher than the corresponding large scale value (\phpiia). Thus, the isocurvature initial spectrum is \emph{blue} tilted $(n_\iso > 1)$ contrary to the dominant red-tilted $(n_s < 1)$ adiabatic perturbation.
The magnitudes of the perturbation amplitudes at both scales are higher in case of CDR compared to FDR. Thus, $f_\iso$ is higher in CDR which means that CDR DRID allows for a larger isocurvature perturbation. The value of \phpiia{} for CDR is roughly factor of $1.2$  higher with respect to the FDR value. Whereas, the value at the small scale \phpiib{} is approximately $2.5$ times higher compared to FDR. 
These translate to the fact that the $n_\iso$ is higher for CDR compared to FDR. From the 2D posteriors we see that \ndr{}  and $n_\iso$ are positively correlated, which is especially apparent for the CDR case. This suggests that the effects of these parameters on the CMB spectrum compensate each other. Therefore, a bigger blue tilt of the initial DRID spectrum would prefer a higher value of \ndr{}. In the following subsections, where we study the effects of isocurvature parameters on the CMB spectrum, we will explain all these features discussed here. Note that, the bounds on $f_{\rm iso}$ in FDR and CDR are both quite weak. This is due to the degeneracy of the CMB spectrum between $\mathcal{P}_\mathcal{II}$ and $N_{\rm dr}$ (or \neff{}) as explained earlier. The high values of $f_{\rm iso}$, which occurs due to the high values of $\mathcal{P}_\mathcal{II}$,  corresponds to the small values of $N_{\rm dr}$.

In Fig.~\ref{fig:is0-param-tri-wsh0es-fn} we show the corresponding isocurvature parameter plots for P18-TTTEEE
+lowE+lensing+BAO+SH0ES dataset. All the qualitative features of the results from Planck only data set analysis are present here. Due to the inclusion of the SH0ES measurement, higher \ndr{} (correlated with higher $H_0$) is preferred in both FDR and CDR cases. The values of \phpiib{} for both cases are significantly higher compared to that of the Planck-only analysis. Therefore, $n_\iso$ for both FDR and CDR are larger for this dataset. This is expected since \ndr{} is higher for this dataset and $n_{\rm iso}$ and \ndr{} are positively correlated for isocurvature analysis. In summary, inclusion of BAO and SH0ES dataset results in a higher value of \ndr{} and a more blue tilted isocurvature spectrum compared to the Planck only analysis. 
In Appendix~\ref{sec:app:tri}, we show more detailed triangle plots of the isocurvature parameters along with the all other \lcdm{} parameters for all the datasets used in this paper.

\subsubsection{with varying $N_{\rm ur}$ (VN)}

\renewcommand{\arraystretch}{1.4}
\begin{table}[htb!]
    \centering
    \begin{tabular}{|l|c| c| }
\hline
FDR (VN \& NC)& \makecell[c]{P18-TTTEEE\\+lowE+lensing}& \makecell[c]{P18-TTTEEE+lowE+ \\lensing+BAO+SH0ES(L)}\\
\hline
$10^2 \omega_{b}$& $ 2.252\pm 0.025$& $ 2.282\pm 0.015$\\
\hline
$\omega_{cdm }$& $ 0.1200\pm 0.0031$& $ 0.1248\pm 0.0025$\\
\hline
$100\theta_{s }$& $ 1.04241\pm 0.00052$& $ 1.04189\pm 0.00047$\\
\hline
$\tau_{reio }$& $ 0.0554\pm 0.0077$& $ 0.0560\pm 0.0072$\\
\hline
$10^{10}\mathcal{P}_\mathcal{RR}^{(1)}$& $ 23.32\pm 0.55$& $ 22.80\pm 0.46$\\
\hline
$10^{10}\mathcal{P}_\mathcal{RR}^{(2)}$& $ 20.37\pm 0.44$& $ 20.73\pm 0.36$\\
\hline
$ 10^{10} N_{\rm dr}^2 \mathcal{P}_\mathcal{II}^{(1)}$& $< 13.0$& $< 13.3$\\
\hline
$10^{10} N_{\rm dr}^2\mathcal{P}_\mathcal{II}^{(2)}$& $ 74^{+20}_{-60}$& $ 107^{+40}_{-70}$\\
\hline
$N_{ur }$& $ 2.06^{+1.0}_{-0.50}$& $ 2.29^{+1.1}_{-0.49}$\\
\hline
$N_{\rm dr}$& $< 1.32$& $< 1.44$\\
\hline
\hline$H_0 ({\rm km/s/Mpc})$& $ 68.8\pm 1.6$& $ 71.04\pm 0.81$\\
\hline
$\sigma_8$& $ 0.820\pm 0.010$& $ 0.8318\pm 0.0086$\\
\hline
$10^9 A_s$& $ 2.086\pm 0.037$& $ 2.108\pm 0.032$\\
\hline
$n_{s }$& $ 0.9655\pm 0.0090$& $ 0.9756\pm 0.0062$\\
\hline
$n_{\rm iso}$& $ 1.51^{+0.34}_{-0.30}$& $ 1.62^{+0.32}_{-0.28}$\\
\hline
$f_{\rm iso}$& $ 14.8^{+7.8}_{-14}$& $ 15.0^{+7.1}_{-14}$\\
\hline
$N_{\rm tot}$& $ 3.09\pm 0.21$& $ 3.42\pm 0.13$\\
\hline
$f_{\rm dr}$& $ 0.33^{+0.14}_{-0.32}$& $ 0.33^{+0.14}_{-0.32}$\\
\hline
\hline
$ \chi^2 - \chi^2_{\Lambda \rm{CDM}}$& $-3.92$& $-14.08$\\
\hline
\end{tabular}
    \caption{Mean and $1\sigma$ error of parameters for FDR-DRID (uncorrelated and varying $N_{\rm ur}$) for the corresponding datasets. The limits are at 68\% C.L. The constraints on the primary parameters and the derived parameters are shown in two separate blocks. The $\chi^2$ difference with respect to the \lcdm{} (fixed \neff{}) model for the corresponding data-set is shown on the last line.}
    \label{tab:param-fdr-vn}
\end{table}

\begin{table}[htb!]
    \centering
    \begin{tabular}{|l|c| c| }
\hline
CDR (VN \& NC)& \makecell[c]{P18-TTTEEE\\+lowE+lensing}& \makecell[c]{P18-TTTEEE+lowE+ \\lensing+BAO+SH0ES(L)}\\
\hline
$10^2 \omega_{b}$& $ 2.257\pm 0.026$& $ 2.287\pm 0.016$\\
\hline
$\omega_{cdm }$& $ 0.1220^{+0.0033}_{-0.0038}$& $ 0.1276^{+0.0028}_{-0.0032}$\\
\hline
$100\theta_{s }$& $ 1.04258^{+0.00061}_{-0.00073}$& $ 1.04226^{+0.00063}_{-0.00081}$\\
\hline
$\tau_{reio }$& $ 0.0561^{+0.0071}_{-0.0081}$& $ 0.0565\pm 0.0072$\\
\hline
$10^{10}\mathcal{P}_\mathcal{RR}^{(1)}$& $ 23.45\pm 0.55$& $ 23.11\pm 0.50$\\
\hline
$10^{10}\mathcal{P}_\mathcal{RR}^{(2)}$& $ 20.19\pm 0.46$& $ 20.35\pm 0.45$\\
\hline
$10^{10} N_{\rm dr}^2 \mathcal{P}_\mathcal{II}^{(1)}$& $< 18.9$& $< 17.6$\\
\hline
$10^{10} N_{\rm dr}^2 \mathcal{P}_\mathcal{II}^{(2)}$& $< 264$& $ 407^{+200}_{-300}$\\
\hline
$N_{ur }$& $ 2.94\pm 0.25$& $ 3.18^{+0.27}_{-0.22}$\\
\hline
$N_{\rm dr}$& $< 0.304$& $ 0.35^{+0.15}_{-0.27}$\\
\hline
\hline$H_0 ({\rm km/s/Mpc})$& $ 69.2^{+1.6}_{-1.8}$& $ 71.46\pm 0.87$\\
\hline
$\sigma_8$& $ 0.820\pm 0.011$& $ 0.8304\pm 0.0092$\\
\hline
$10^9 A_s$& $ 2.073\pm 0.039$& $ 2.081\pm 0.038$\\
\hline
$n_{s }$& $ 0.9617\pm 0.0090$& $ 0.9675^{+0.0086}_{-0.0075}$\\
\hline
$n_{\rm iso}$& $ 1.66^{+0.43}_{-0.35}$& $ 1.87^{+0.39}_{-0.28}$\\
\hline
$f_{\rm iso}$& $ 58^{+22}_{-53}$& $ 31.7^{+6.7}_{-27}$\\
\hline
$N_{\rm tot}$& $ 3.18^{+0.21}_{-0.25}$& $ 3.53\pm 0.15$\\
\hline
$f_{\rm dr}$& $ 0.076^{+0.031}_{-0.068}$& $ 0.098^{+0.041}_{-0.074}$\\
\hline
\hline
$ \chi^2 - \chi^2_{\Lambda \rm{CDM}}$& $0.8$& $-11.68$\\
\hline
\end{tabular}
    \caption{Mean and $1\sigma$ error of parameters for CDR-DRID (uncorrelated and varying $N_{\rm ur}$)  for the corresponding datasets. The limits are at 68\% C.L. The constraints on the primary parameters and the derived parameters are shown in two separate blocks. The $\chi^2$ difference with respect to the \lcdm{} (fixed \neff{}) model for the corresponding datasets is shown on the last line.}
    \label{tab:param-cdr-vn}
\end{table}

\begin{figure}[htb!]
	\centering
		\includegraphics[width=0.85\linewidth]{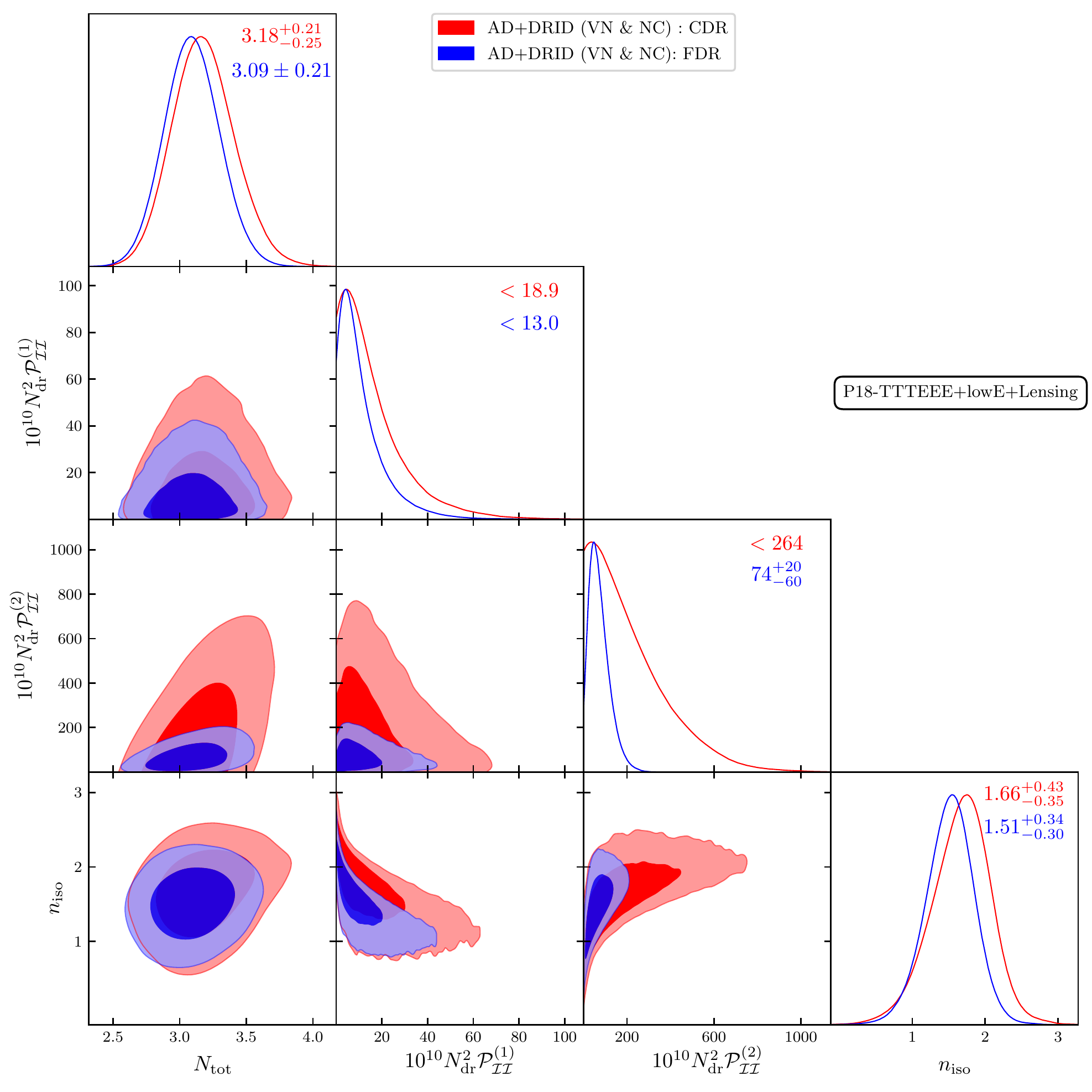}
		\caption{Triangle plot for isocurvature parameters and $ N_{\rm dr} $ for varying $N_{\rm ur}$ analysis for P18-TTTEEE+lowE+lensing dataset. The constraints on individual parameters are mentioned on the diagonal 1-D posteriors with corresponding colors. The errors represent $1\sigma$ errorbar and the limits are at 68\% confidence level (C.L.).}
		\label{fig:is0-param-tri-vn}
	\end{figure}

\begin{figure}[htb!]
	\centering
		\includegraphics[width=0.85\linewidth]{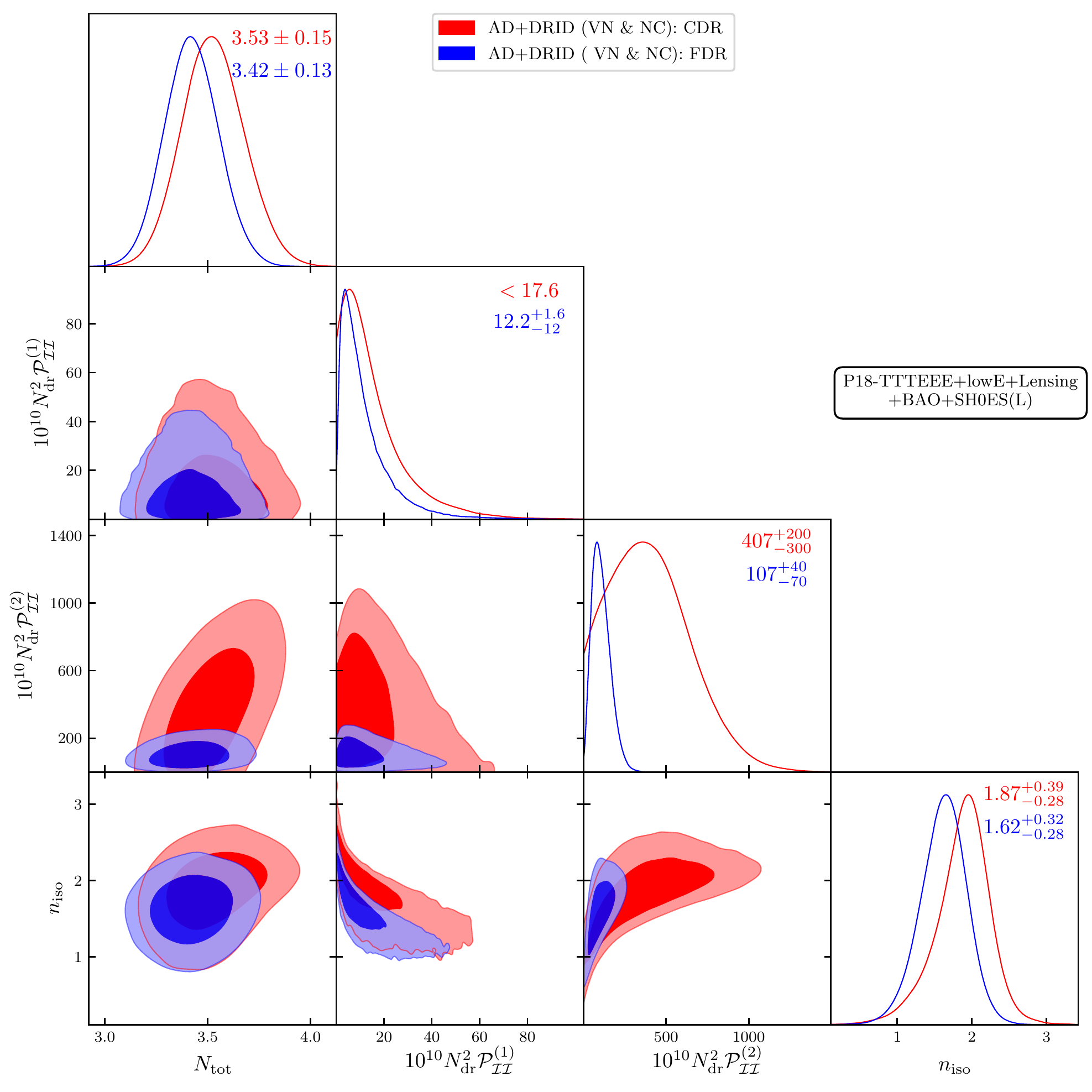}
		\caption{Triangle plot Isocurvature parameters and $ N_{\rm dr} $ for varying $N_{\rm ur}$ analysis for P18-TTTEEE+lowE+lensing+lensing+SH0ES(L) dataset. The constraints on individual parameters are mentioned on the 1D diagonal posteriors with corresponding colors. The errors represent $1\sigma$ errorbar and the limits are at 68\% C.L.}
		\label{fig:is0-param-tri-wsh0es-vn}
	\end{figure}

In the second scenario, we study the more general case where we also let the neutrino contribution to the energy density $N_{\rm ur}$ vary. This captures models where there are extra relativistic neutrinos or the presence of additional cooling or heating mechanism for SM neutrinos. In our setup, species contributing to $N_{\rm ur}$ only carry adiabatic perturbations. In Table.~\ref{tab:param-fdr-vn} and Table.~\ref{tab:param-cdr-vn}, we
present the summary of the uncorrelated FDR-DRID and CDR-DRID analysis for varying $N_{\rm ur}$ scenario, respectively.\footnote{ Note that the $N_{\rm dr}$ for FDR and $N_{ur}$ are highly degenerate as they give same signatures with adiabatic perturbation. Therefore, due to this strong degeneracy, it is hard to find the true minima in the $\chi^2$ plane. Because of this artifact, in some cases, the $\chi^2 $ difference (with respect to $\Lambda$CDM) for the FDR-VN is larger than FDR-FN scenario (where $N_{ur}$ is not varied). This artifact can be removed with more MCMC samples and also does not significantly affect the parameter estimation.} In Fig.~\ref{fig:is0-param-tri-vn} and \ref{fig:is0-param-tri-wsh0es-vn} we show the constraints on the isocurvature parameters along with \neff{} for P18-TTTEEE+lowE+lensing and P18-TTTEEE+lowE+lensing+BAO+SH0ES(L) datasets, respectively. The plots shows similar features to the fixed $N_{\rm ur}$ analysis. We find $n_{\rm iso}$ to be positively correlated with \neff{}, and CDR allows for larger amplitudes of isocurvature perturbations. The 1-D posteriors of \neff{} indicate that both FDR and CDR DRID prefer higher value of \neff{} compared the standard \lcdm{} value $N_{\rm tot} = 3.046$.

\subsection{Constraint on $\Delta N_{\rm tot}$ and application to the $H_0$ tension}
\begin{figure}[htb!]
	\centering
	\begin{subfigure}{0.49\linewidth}
		\includegraphics[width=\linewidth]{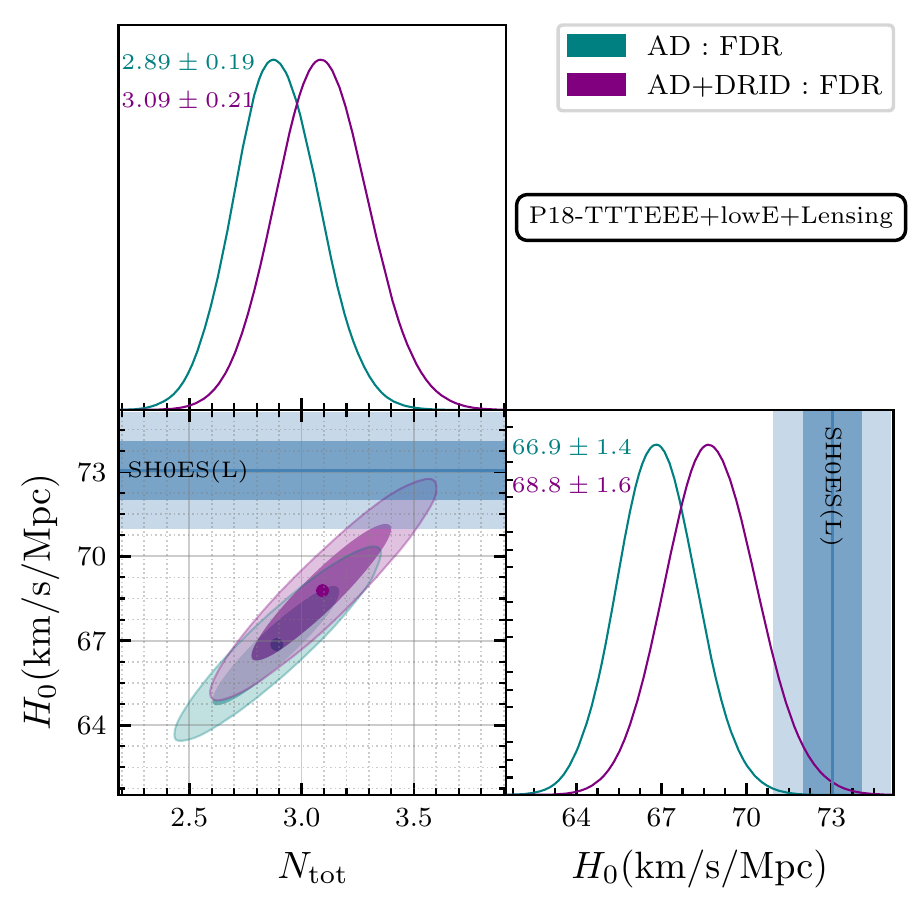}
		\subcaption{FDR}
	\end{subfigure}
    	\begin{subfigure}{0.49\linewidth}
    	\includegraphics[width=\linewidth]{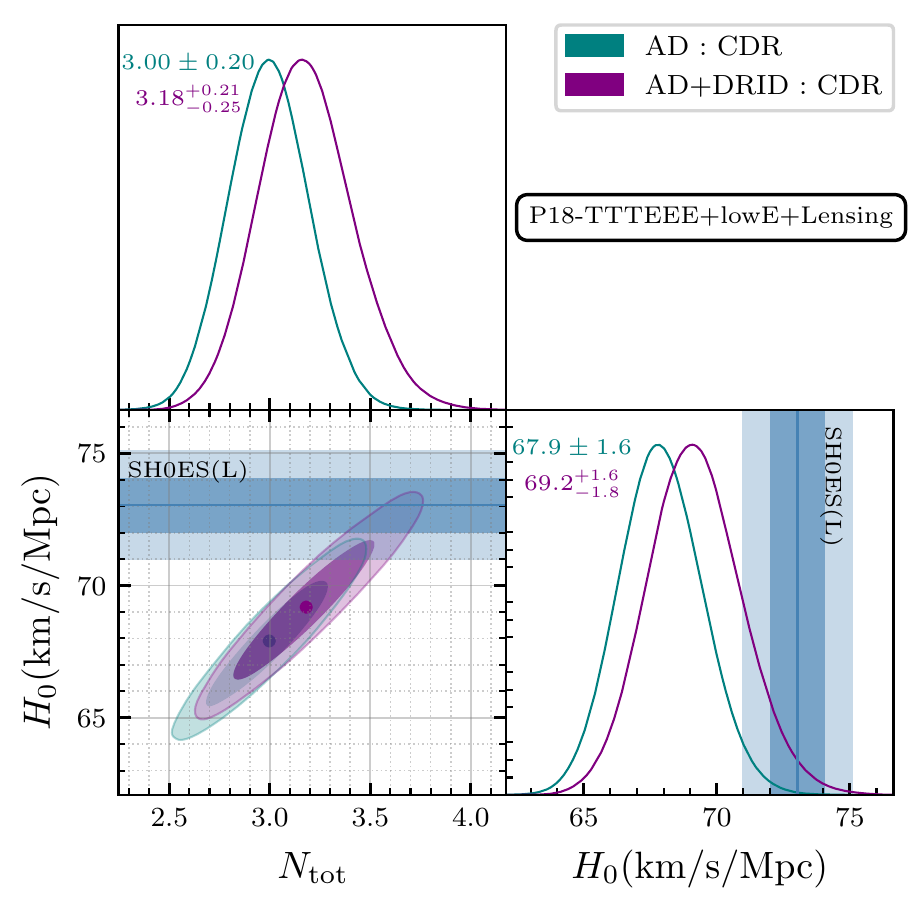}
    	\subcaption{CDR}
    \end{subfigure}
\caption{1D and 2D marginalised posteriors of $ H_0 $ and $ N_{\rm tot} $ in different models for P18-TTTEEE+lowE+lensing dataset for varying neutrino analysis (VN). The dots in the 2D contours represent the best-fit points. FDR (CDR) are the free streaming (coupled) dark radiation that either carries only adiabatic (AD) perturbation or additional isocurvature (DRID) perturbation.}
\label{fig:H0-vs-Neff-wo-sh0es}
\end{figure}
In this subsection, we compare the values of \neff{} and $H_0$ between AD+DRID
and AD only scenario for the corresponding datasets. First we analyse the varying $N_{\rm ur}$ (VN) scenario.
In Fig.~\ref{fig:H0-vs-Neff-wo-sh0es} we show the constraint on $N_{\rm tot}$ and $H_0$ for DRID with a triangle plot for P18-TTTEEE+lowE+lensing dataset. The blue bands depict the $1\sigma$ and $2\sigma$ value of the Hubble constant measured by the SH0ES collaboration~\cite{Riess:2021jrx}. In each plot we compare DRID results against the results from pure adiabatic perturbation to distinguish the effects of isocurvature perturbation.
In the left panel, we compare the FDR-DRID posteriors against \lcdm{} model with varying \neff{} because massless neutrinos and free-streaming DR are completely equivalent in the presence of pure adiabatic perturbation. These two species have identical Boltzmann evolution and initial conditions therefore affect the CMB spectrum identically. Thus, only by varying \neff{} via $N_{\rm ur}$ we are able simulate the effect of FDR for adiabatic case. However, in presence of DRID, these two species evolve differently since they have different initial condition as shown in Tables~\ref{tab:fdric}~and~\ref{tab:cdric}.

As we see from the left panel of the figure, FDR-DRID prefers a higher value of \neff{} compared to \lcdm{} (with varying \neff{}). As a result, it also accommodates a comparatively higher Hubble constant. The enhancement of CMB spectrum in the presence of DRID compensates for the additional Silk damping due to higher \neff{}. Consequently, the Hubble tension for `VN' scenario is reduced to  $\sim 2.2\sigma$ in this scenario from the $\Lambda\text{CDM}$ result. Note that, the reduction of the tension is primary due to the enlarged errorbar of $H_0$ for the DRID scenarios. 

In the right panel we compare results for the CDR scenario where the DR is treated as a perfect fluid which does not free-stream. The CDR being a perfect fluid, the higher moments of its Boltzmann evolution are identically zero which distinguishes it from a free-streaming radiation like neutrinos. Note that in this scenario, we have an admixture CDR and free-streaming neutrinos in the cosmological models.
The results for adiabatic initial condition is shown in the cyan contours which itself prefers a higher \neff{} compared to the \lcdm{}. Due to its fluid like nature, the perturbations in CDR do not get leaked to higher multipole. Therefore, compared to FDR, the density perturbation in CDR is larger which boost the potential and results in the enhancement of the CMB spectrum at small scale. To compensate for this effect, the CDR scenario allows for large \neff{} which sources greater Silk damping of the CMB tail. 

The addition of DRID in the CDR case further enhances the \neff{}. Similar to the FDR case, the presence of CDR-DRID boost the CMB spectrum which helps to accommodate even larger \neff{} to compensate for that through Silk damping. As a result, CDR-DRID accommodate the largest Hubble constant among all the scenarios discussed so far. With Planck data alone, the value of the $H_0$ is pushed to $69.2^{+1.6}_{-1.8}$~(km/s/Mpc) and the tension with the local measurement is reduced to $\sim 2.0\sigma$.

\begin{figure}[htb!]
	\centering
	\begin{subfigure}{0.49\linewidth}
		\includegraphics[width=\linewidth]{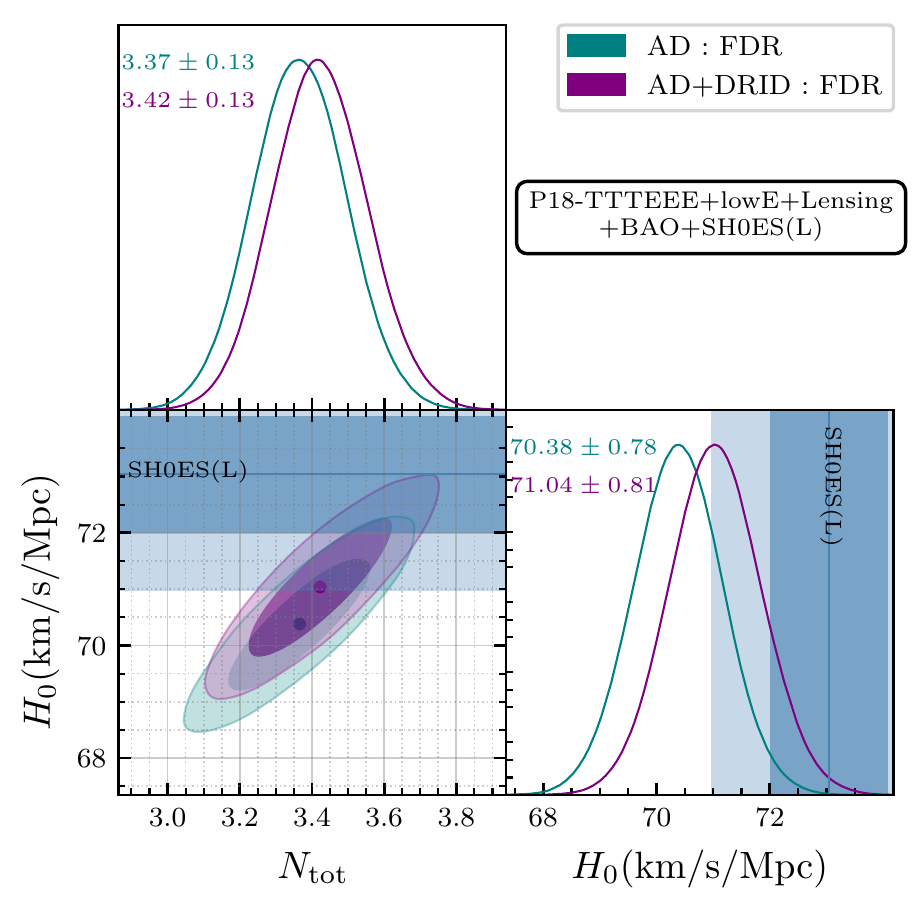}
		\subcaption{FDR}
	\end{subfigure}
	\begin{subfigure}{0.49\linewidth}
		\includegraphics[width=\linewidth]{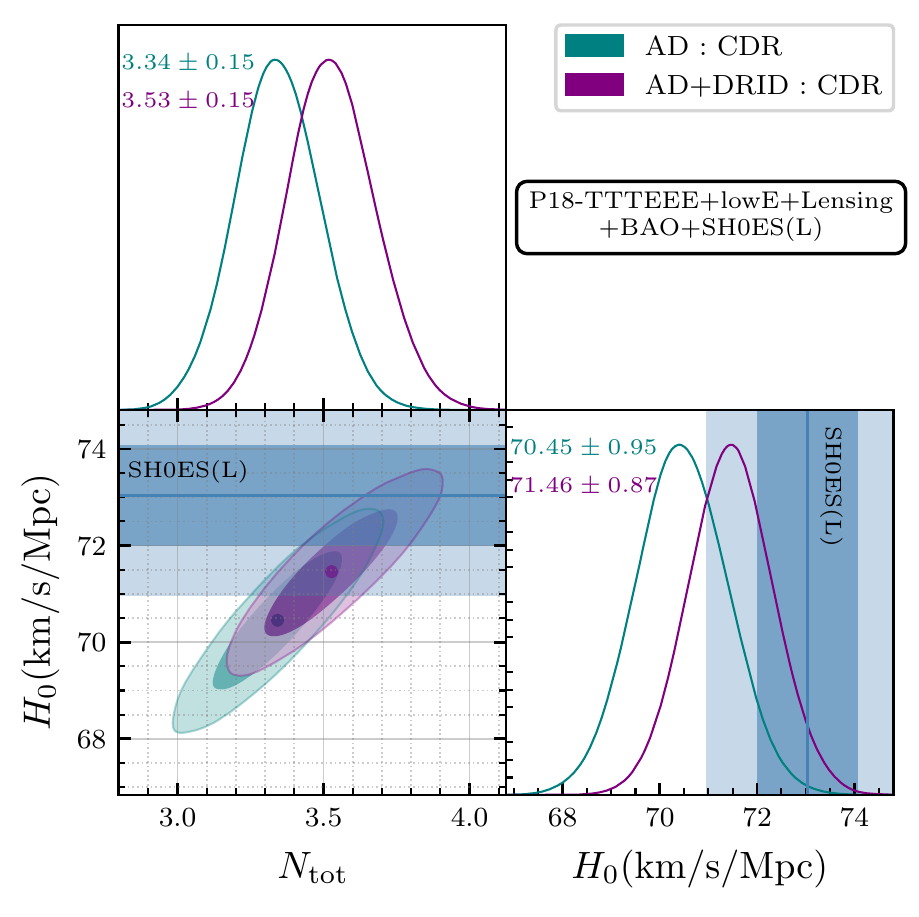}
		\subcaption{CDR}
	\end{subfigure}
	\caption{1D and 2D marginalised posteriors of $ H_0 $ and $ N_{\rm tot} $ in different models for P18-TTTEEE+lowE+lensing+BAO+SH0ES(L) dataset for varying neutrino analysis (VN). The dots in the 2D contours represent the best-fit points. FDR (CDR) are the free streaming (coupled) dark radiation that either carries only adiabatic (AD) perturbation or additional isocurvature (DRID) perturbation.}
	\label{fig:H0-vs-Neff-w-sh0es}
\end{figure}

In Fig.~\ref{fig:H0-vs-Neff-w-sh0es}, we show the same constraints with the inclusion of BAO and SH0ES datasets along with the Planck data. As the SH0ES data prefers higher $H_0$, the value of both \neff{} and $H_0$ get boosted compared to Planck only analysis across all models. As a result, in case of FDR-DRID, the disparity with the local Hubble measurement is relaxed to $\sim 1.5\sigma$. The scenario with CDR-DRID is particularly noteworthy which accommodate the largest Hubble constant for this dataset: $H_0 = 71.46\pm 0.87$~(km/s/Mpc). The discrepancy with the local Hubble measurement in this scenario is reduced to approximately $ 1.2\sigma$.

We also compare the values of the Hubble tension for fixed $N_{\rm ur}$ analysis (FN) which can be computed using Table~\ref{tab:param-fdr-fn} and \ref{tab:param-cdr-fn}. For P18-TTTEEE+lowE+lensing dataset the tension with the SH0ES measurement is $\sim 2.5\sigma$ for both FDR and CDR DRID. In addition when the BAO and SH0ES datas are included the discrepancy reduces to $1.6 \sigma$ and $1.3 \sigma$, respectively.

\subsection{Changes to the CMB spectra}

\begin{figure}[b!]
    \centering
    \includegraphics[width = 0.49\linewidth]{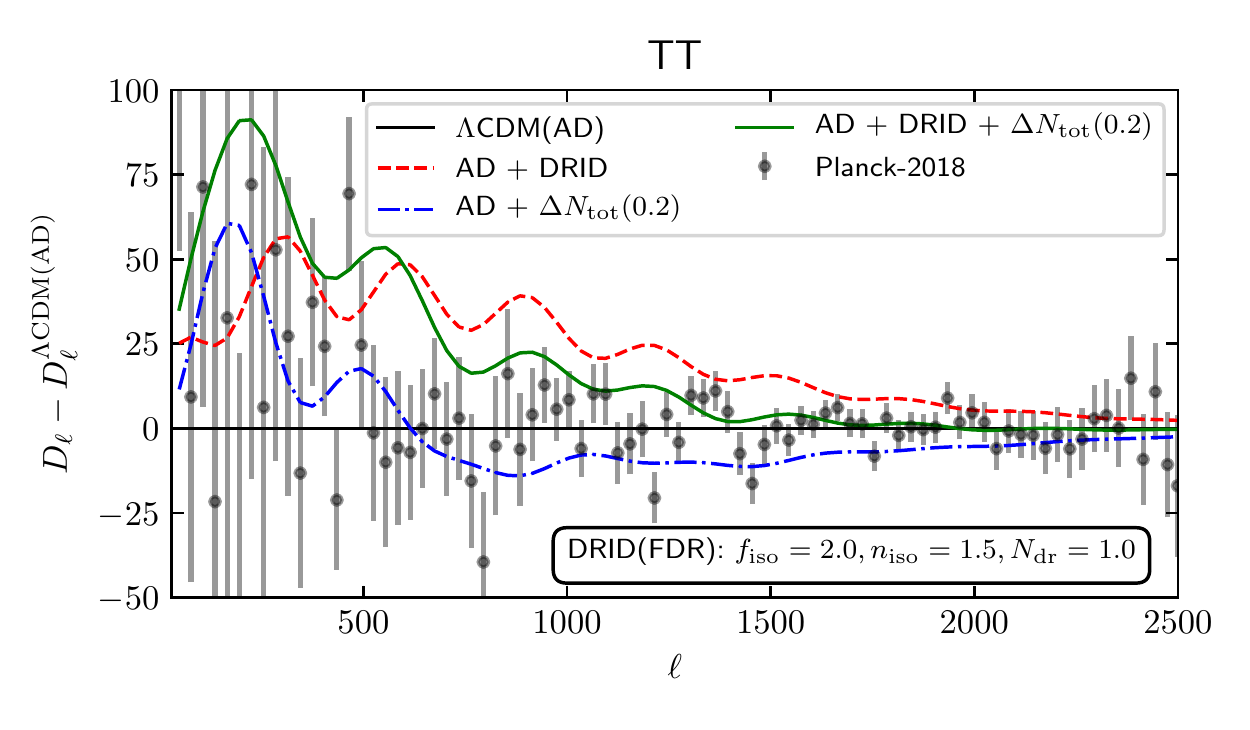}
    \includegraphics[width = 0.49\linewidth]{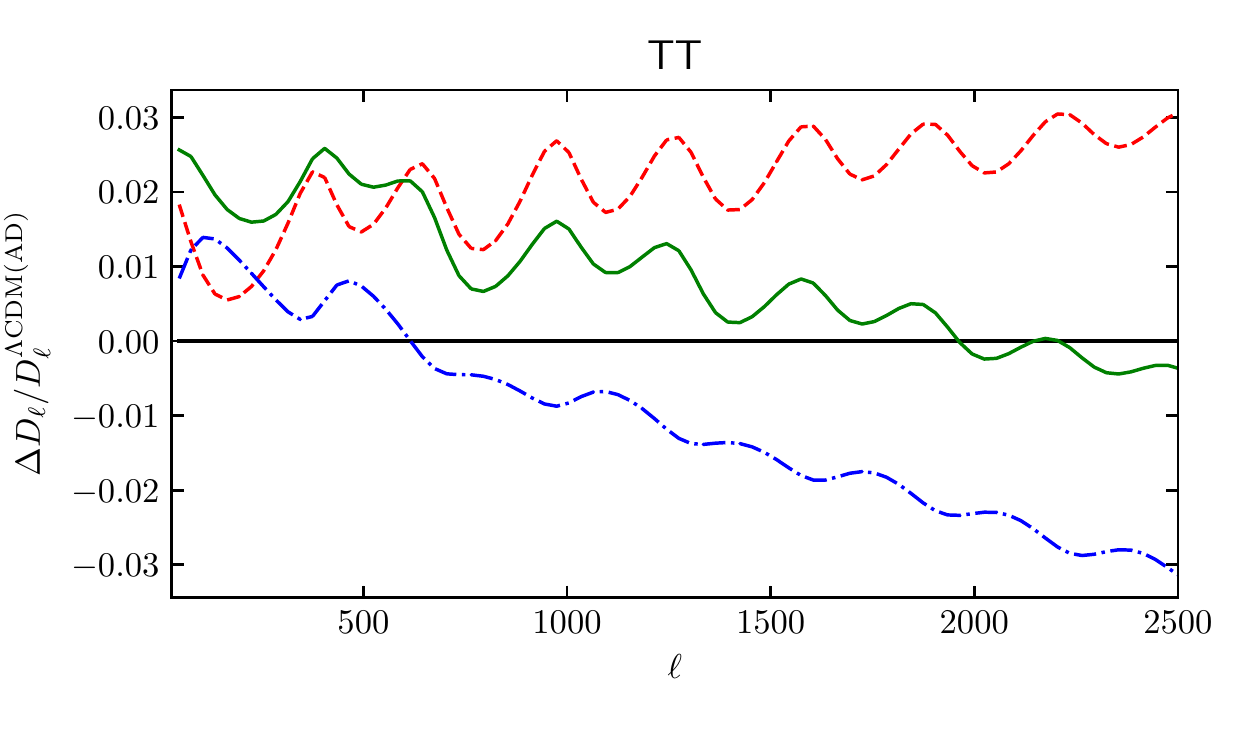}
    \includegraphics[width = 0.49\linewidth]{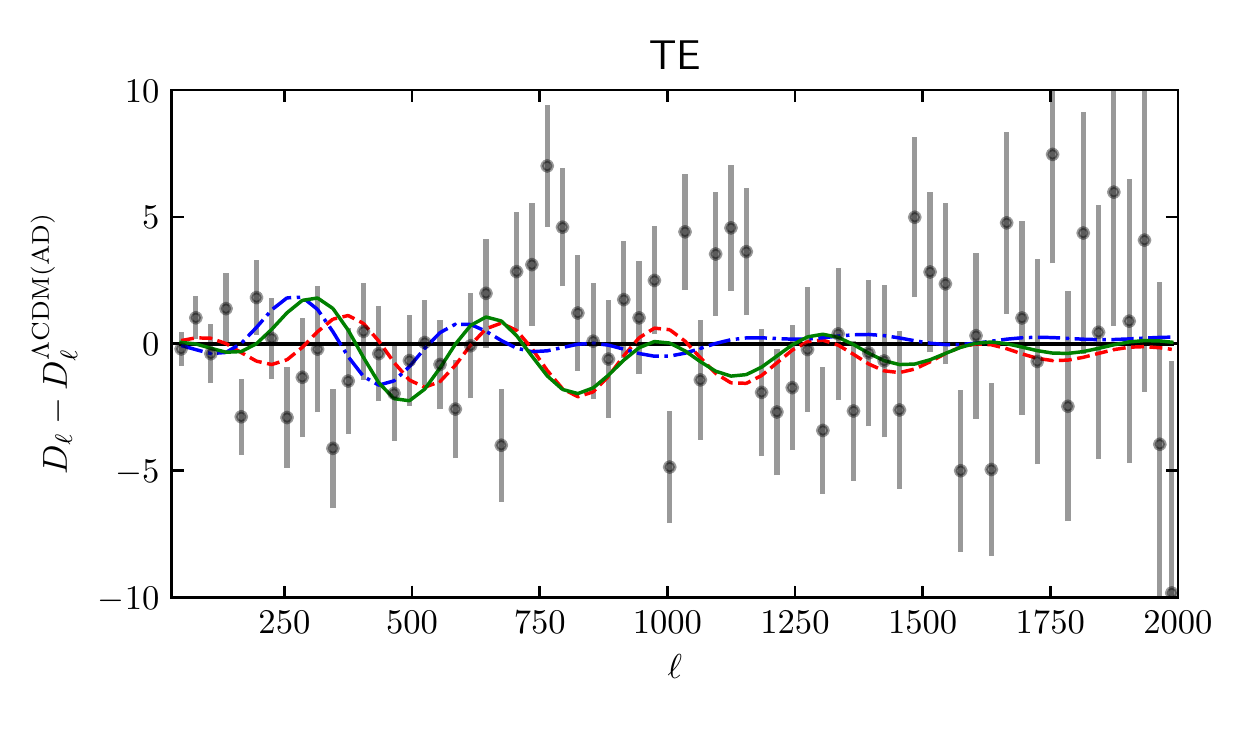}
    \includegraphics[width = 0.48\linewidth]{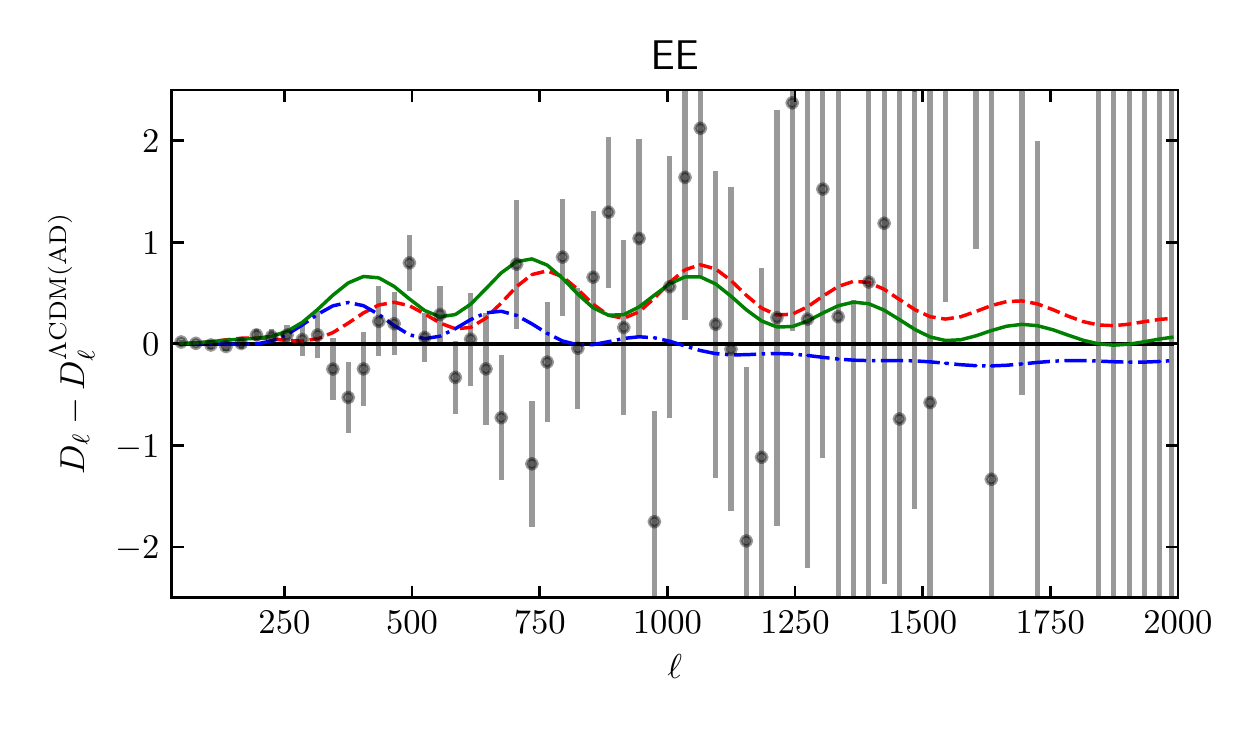}
    \caption{Relative changes to the CMB TT,TE and EE spectrum compared to \lcdm{} for different models in the context of FDR. The top-right plot shows the fractional changes for the TT spectrum. \neff{} is fixed to $3.046$ for the \lcdm{} model. For the DRID parameters we have chosen: $f_{\iso} = 2.0$ and $n_{\iso} = 1.5$ with $N_{\rm dr} = 1.0 $. The changes in the $ N_{\rm tot}$ $(\Delta N_{\rm tot})$, mentioned in the legends, are introduced by changing $N_{\rm ur}$.}
    \label{fig:fdr-spectra-w-data}
\end{figure}

In Fig.~\ref{fig:fdr-spectra-w-data}, we compare the DR isocurvature spectrum with Planck CMB data and demonstrate how it accommodates a larger \neff{}. The plots show the difference of the CMB spectra for TT, TE and EE mode for different models with respect to the bestfit \lcdm{} spectra having $N_{\rm tot} = 3.046$. The residual for the Planck 2018 data is also shown with the error-bar. In addition, for the TT spectra, we also show the relative changes of the spectra in the top-right panel. 

In this work, we have so far considered \emph{uncorrelated} DR isocurvature spectrum. In the absence of any correlation with the adiabatic spectrum, the additional contribution due to DR isocurvature is always positive across all CMB multipoles for the TT spectrum as shown in the top-left panel of Fig.~\ref{fig:fdr-spectra-w-data}. Therefore, it amounts to an enhancement of the CMB spectrum across all scales. In the red dashed curve , we show an example of the deviation due to FDR DR isocurvature with $f_{\iso} = 2.0$ and $n_{\iso} = 1.5$ for $N_{\rm dr} = 1.0 $ and $N_{\rm tot} = 3.046$. For this blue tilted isocurvature spectrum, which has excess power at small scales, the relative enhancement increases with higher multipole as can be seen in top right panel of Fig~\ref{fig:fdr-spectra-w-data}. A higher value of $n_\iso$ gives a stronger scale dependence of the enhancement. This enhancement due to isocurvature enables this scenario to accommodate a larger \neff{}, hence a higher $H_0$. Increase of \neff{} results in greater Silk damping which is the suppression the CMB spectra at higher multipole in the TT spectra due to enhanced diffusion damping of the photon perturbations~\cite{2013PhRvD..87h3008H}. The blue lines shows the effects of increasing \neff{} $(\Delta N_{\rm tot} = 0.2)$ in the \lcdm{} with only adiabatic perturbations and keeping all other parameter fixed. Since, Silk damping has an exponential dependence of the scale, the damping increases with higher multipole. Therefore, the damping has similar scale dependence with the enhancement arising from a blue tilted DR isocurvature spectrum. So, in case of DR isocurvature with varying \neff{}, these two effects can partially compensate each other, specifically at smaller scale, to allow for a larger \neff{}. We demonstrate this through the green curves which has the following DR parameters: $f_{\iso} = 2.0$ and $n_{\iso} = 1.5$ for $N_{\rm dr} = 1.0 $ and $ N_{\rm tot} = 3.246$. The resulting spectrum agrees quite well with the data for higher multipoles $(\ell \gtrsim 1000) $ for the TT spectrum.

The compensation mechanism however does not work for smaller multipoles, where higher \neff{} results in an enhancement of the CMB spectrum due to larger metric potential. Interestingly, the error of the CMB TT dataset is higher in this region comparing to the high-$\ell$ modes due to the large magnitude of the signal around the first acoustic peak. Therefore, the CMB TT dataset alone has comparatively limited constraining power to probe these changes at smaller multipole. Thus, TT dataset allows for the large compensation between isocurvature spectrum and \neff{} which results in a large degeneracy between \neff{} and $n_\iso$ (and $f_\iso$). This is the reason why the constraints from TT+lowE dataset are rather weak and permit large values of the isocurvature parameters and \neff{}, and thereby the $H_0$.

The scenario is modified significantly when CMB polarization data is added. Polarization measurements at large scale are more precise compared to the temperature data. So, TE and EE spectra have much smaller error bars at the large scale than the TT spectra. Therefore, the inclusion of the polarization data effectively constraints the large scale deviations introduced by the DR isocurvature setup. Thus, when we consider the full Planck CMB spectrum: TTTEEE+lensing dataset, the constraints on the parameters become much stricter and the allowed values of the \neff{} is much smaller compared to the TT-only analysis.  

Note that, addition of isocurvature spectrum introduces phase shift of the CMB spectrum compared to \lcdm{}~\cite{Baumann:2015rya}. This 
results in a change of the Hubble constant even when \neff{} is kept fixed~\cite{Planck:2018jri}.\footnote{As can be seen from Fig.[41] from Ref~\cite{Planck:2018jri}, the change in $H_0$ for Neutrino density isocurvature is $\approx +0.7 ~{\rm km/s/Mpc}$ which is smaller than the increase in $H_0$ in the presence of DRID with positive $\Delta N_{\rm tot}$.} However, we found out that this effect is subdominant and the bulk of the change in $H_0$ comes from the compensation effect which gives higher \neff{}, as described above. Although, we have only discussed FDR isocurvature in this context, the above characterization also holds true for CDR isocurvature.



\section{Analytical comparison between CDR and FDR}\label{sec.CDRFDR}
In this section, our goal is to analytically understand the difference between CDR and FDR in the presence of DRID, and show that isocurvature perturbations of FDR generate a larger contribution to $C^{\rm TT}_{\ell}$ than those of CDR. Before going into a detailed derivation, we can understand this intuitively as follows.

For the sake of simplicity, let us neglect the presence of neutrinos so that during radiation domination only $\gamma$ and DR contribute dominantly to the energy density. Now let us first consider CDR for which both $\gamma$ and DR behave as coupled fluids. For DRID initial conditions in the synchronous gauge, we start with vanishing total density perturbation in a patch along with vanishing metric perturbations. As time progresses, acoustic oscillations get set up in both $\gamma$ and DR. However, since we start with (say) overdensity in DR and compensating underdensity in $\gamma$, the acoustic oscillations progress in a way such that anytime DR gets under(over)dense, $\gamma$ gets over(under)dense. Therefore, metric perturbations do not get a chance to develop significantly and starting with an underdense region in $\gamma$ leads to a negative CMB temperature fluctuation without SW redshifting.

The situation changes when we have FDR instead of CDR. Again we start with vanishing metric perturbation in the superhorizon limit and an over(under)dense region in DR($\gamma$). Now due to anisotropic stress, DR starts developing a diffusion-like effect and no longer behaves as a perfect fluid. Thus DR acoustic oscillations are no longer synchronized with that in $\gamma$ to maintain a vanishing metric perturbation. That is, whereas for CDR there is an efficient outflow of DR from an overdense region, for FDR this outflow of DR is hindered by the random walk induced by the anisotropic stress.
As a result, DR does not move efficiently out of the overdense regions. Photons however do not experience this, and they continue their inflow into the photon-underdense region. As a result a net inflow of energy takes place which leads to formation of gravitational wells.\footnote{For the adiabatic case, diffusion leads to a decay of gravitational wells. However, here for isocurvature initial conditions there is no potential well to begin with, and diffusion-induced suppression of DR outflow leads to potential well formation.} This implies when the photons come out of those wells, they lose more energy. Hence, the photons coming from the underdense region appear even colder and in terms of the absolute value, leads to more CMB anisotropy. Similarly when starting with an under(over)dense patch in DR($\gamma$), free-streaming prevents a fluid-like inflow of DR and hence a potential ``hill'' develops. Correspondingly, when the already hot photons come out of such regions, they get an additional kick which makes them even hotter.


To show this more explicitly, we start with a discussion of isocurvature initial conditions in the synchronous gauge defined through eqs.~\eqref{eq.synch1}~and~\eqref{eq.synch2}. We will mostly follow the notations in Ma and Bertschinger~\cite{Ma:1995ey}. We will also track the contribution of $\nu$ unless mentioned otherwise.

\subsection{Properties of shear}
We expect the primary difference between CDR and FDR to come from the fact that due to its free-streaming nature, FDR would contribute to the total anisotropic stress, whereas CDR would not. Therefore, it is useful to consider the equation relating the metric perturbations $h,\eta$ to the shear for one $k-$mode~\cite{Ma:1995ey},
\begin{align}\label{eq.sheareq}
\Ddot{h}+6\Ddot{\eta}+2\frac{\dot{a}}{a}(\dot{h}+6\dot{\eta})-2k^2\eta=-24\pi G a^2(\Bar{\rho}+\Bar{P})\sigma.    
\end{align}
Here and in the following, the dots denote derivative with respect to conformal time $\tau$. The shear $\sigma$ is defined as $(\bar{\rho}+\bar{P})\sigma\equiv -\left(\hat{k}_i\hat{k}_j-\frac{1}{3}\delta_{ij}\right)\Sigma^i_j$ in terms of the traceless part of the energy-momentum tensor, $\Sigma^i_j\equiv T^i_j-\delta^i_j T^k_k/3$.
Assuming radiation domination and using $\Bar{\rho}=3H^2/(8\pi G)$, the RHS of eq.~\eqref{eq.sheareq} can be simplified to $-12a^2 H^2\sigma$.
Now for isocurvature initial conditions at $\tau\rightarrow 0$, $\sum_{\text{i=radiation}}\delta\rho_i=0$ (see eq.~\eqref{eq.sumrad}) at very early times and on superhorizon scales. Because of this fact, $h$ gets dominantly sourced by the matter fluctuations, and in particular by baryon fluctuations $\delta_b$.\footnote{The CDM fluctuation $\delta_c$ is subdominant compared to $\delta_b$ in synchronous gauge for this initial condition.} Using this fact, one can show $h\sim(k\tau)^2\omega\tau$ with 
$\omega\equiv a(\tau_i)\bar{\rho}_m(\tau_i)/(\sqrt{3\bar{\rho}_{r}(\tau_i)}M_{\rm pl})$
and $\omega\tau\ll 1$. However, $\eta\sim(k\tau)^2$ and hence is bigger than $h$ as we now check in a self-consistent manner.

Assuming terms involving $h$ can be neglected compared to $\eta$ in eq.~\eqref{eq.sheareq}, we get the leading order expansion in terms of $(k\tau)$, 
\begin{align}\label{eq.simpleshear}
\Ddot{\eta}+2\frac{\dot{a}}{a}\dot{\eta}=-2a^2H^2\sigma.    
\end{align}
This implies a going rate, $\sigma\sim\eta$, i.e., they are of the same order in $k\tau$ (using the fact that $aH\sim1/\tau$). Now using the fact that $\theta_{\text{FS}}\sim k^2\tau$ and $\dot\sigma\sim\theta_{\text{FS}}+\cdots$, we see $\sigma\sim(k\tau)^2$ implying $\eta\sim(k\tau)^2\gg h$ as well. Here `FS' refers to any free-streaming species, such as $\nu$ or FDR. Our full results in Tables~\ref{tab:fdric} and \ref{tab:cdric} which were obtained by solving the coupled set of equations governing metric and matter/radiation fluctuations confirm this parametric behavior.

Therefore we parametrize the leading dependence of $\eta$ as, $\eta=A(k\tau)^2$, where the precise values of $A$ are given in Tables~\ref{tab:fdric} and \ref{tab:cdric}. Then eq.~\eqref{eq.simpleshear} implies $\sigma=-3A (k\tau)^2$, i.e., for superhorizon modes there is a simple relation,
\begin{align}\label{eq.sigma_eta}
\sigma\approx-3\eta.    
\end{align}
To use the above relation, we first note that for any free-streaming species $i$, 
\begin{align}
\dot{\delta}_i=&-\frac{4}{3}\theta_i-\frac{2}{3}\dot{h},\\  
\dot{\sigma}_i=&\frac{4}{15}\theta_i-\frac{3}{10}k F_{i3}+\frac{2}{15}\dot{h}+\frac{4}{5}\dot{\eta}.
\end{align}
These can be combined to give,
\begin{align}
\dot{\sigma}_i=-\frac{1}{5}\dot{\delta}_i-\frac{3}{10}k F_{i3}+\frac{4}{5}\dot{\eta}.    
\end{align}
On superhorizon scales, the third moment $F_{i3}$ is suppressed compared to the other terms. Therefore, summing over all free-streaming (FS) perturbations,
\begin{align}
\dot{\sigma}\equiv\sum_{i=\rm{FS}} R_i\dot{\sigma}_i= -\frac{1}{5}\sum_{i=\rm{FS}}R_i\dot{\delta}_i +\frac{4}{5}R_{\rm FS}\dot{\eta},  
\end{align}
where $R_i=\bar{\rho}_i/\bar{\rho}_{\rm tot}$ is the energy fraction in species $i$, and $R_{\rm FS}=\sum_{i=\rm{FS}} R_i$ measures the energy fraction in FS radiation. Using Eq.~\eqref{eq.sigma_eta} we then get,
\begin{align}
\left(1+\frac{4}{15}R_{\rm FS}\right)\dot{\sigma}=-\frac{1}{5}\sum_{i=\rm{FS}} R_i\dot{\delta}_i.    
\end{align}
We further note that on superhorizon scales, 
\begin{align}
\Ddot{\delta}_i=-\frac{k^2}{3}\delta_i+\frac{4}{3}k^2\sigma_i-\frac{2}{3}\Ddot{h}\approx-\frac{k^2}{3}\delta_i,    
\end{align} 
implying,\footnote{Since $\theta_i\rightarrow 0$ as $\tau\rightarrow 0$, there is no term linear in $k\tau$ in $\delta_i$.} 
\begin{align}
\delta_i=\delta_i^{(0)} -\frac{1}{6}(k\tau)^2\delta_i^{(0)},  
\end{align} 
where $\delta_i^{(0)}$ is the time-independent piece of $\delta_i$.
Thus we can finally write,
\begin{align}\label{eq.sigmatot}
\sigma=\frac{1}{2(15+4R_{\rm FS})} (k\tau)^2\delta_{\rm FS}^{(0)},   
\end{align}
with $\delta_{\rm FS}=\sum_i R_i\delta_i$ is the weighted density perturbation of the free-streaming species. Now for DRID initial conditions, the total density perturbation, $R_\gamma\delta_\gamma+R_\nu\delta_\nu+R_{\rm DR}\delta_{\rm DR}=0$ and $\delta_\nu=\delta_\gamma$ (to ensure there is no neutrino density isocurvature). This means with a normalization choice $\delta_{\rm DR}=1$,
the initial density perturbations can be written as,
\begin{align}
\delta_{\rm DR}=1;~~\delta_\nu=\delta_\gamma=-\frac{R_{\rm DR}}{1-R_{\rm DR}},
\end{align} implying,
\begin{align}
\delta_{\rm FS}^{(0)}=R_\nu\delta_\nu=-\frac{R_{\rm DR} R_\nu}{1-R_{\rm DR}}\Rightarrow \sigma<0~~\text{for CDR},\\
\delta_{\rm FS}^{(0)}=R_\nu\delta_\nu+R_{\rm DR}\delta_{\rm DR}=\frac{R_{\rm DR}-R_{\rm DR}^2-R_\nu R_{\rm DR}}{1-R_{\rm DR}}\Rightarrow \sigma>0~~\text{for FDR}.
\end{align}
Crucially therefore, the sign of $\sigma$ depends on the nature of DR. To summarize this intuitively, we consider a patch with an overdensity in DR. For isocurvature initial conditions, this implies an underdensity in $\nu$. Since shear is determined by the fluctuations in free-streaming radiation and in scenarios with CDR only $\nu$ free streams, we effectively have an \textit{under}dense region contributing to shear. On the other hand, for FDR both DR and $\nu$ free stream, and due to the overdesity in DR, the patch effectively has an \textit{over}density in free-streaming radiation---flipping the sign of $\sigma$, as seen from eq.~\eqref{eq.sigmatot}. Since $\sigma$ is a gauge invariant quantity, this conclusion remains true in conformal Newtonian gauge as well. To see how this sign affects CMB it is more convenient to gauge transform to conformal Newtonian gauge.

\subsection{Effects of shear}
\begin{figure}[h]
    \centering
    \includegraphics[width=10cm]{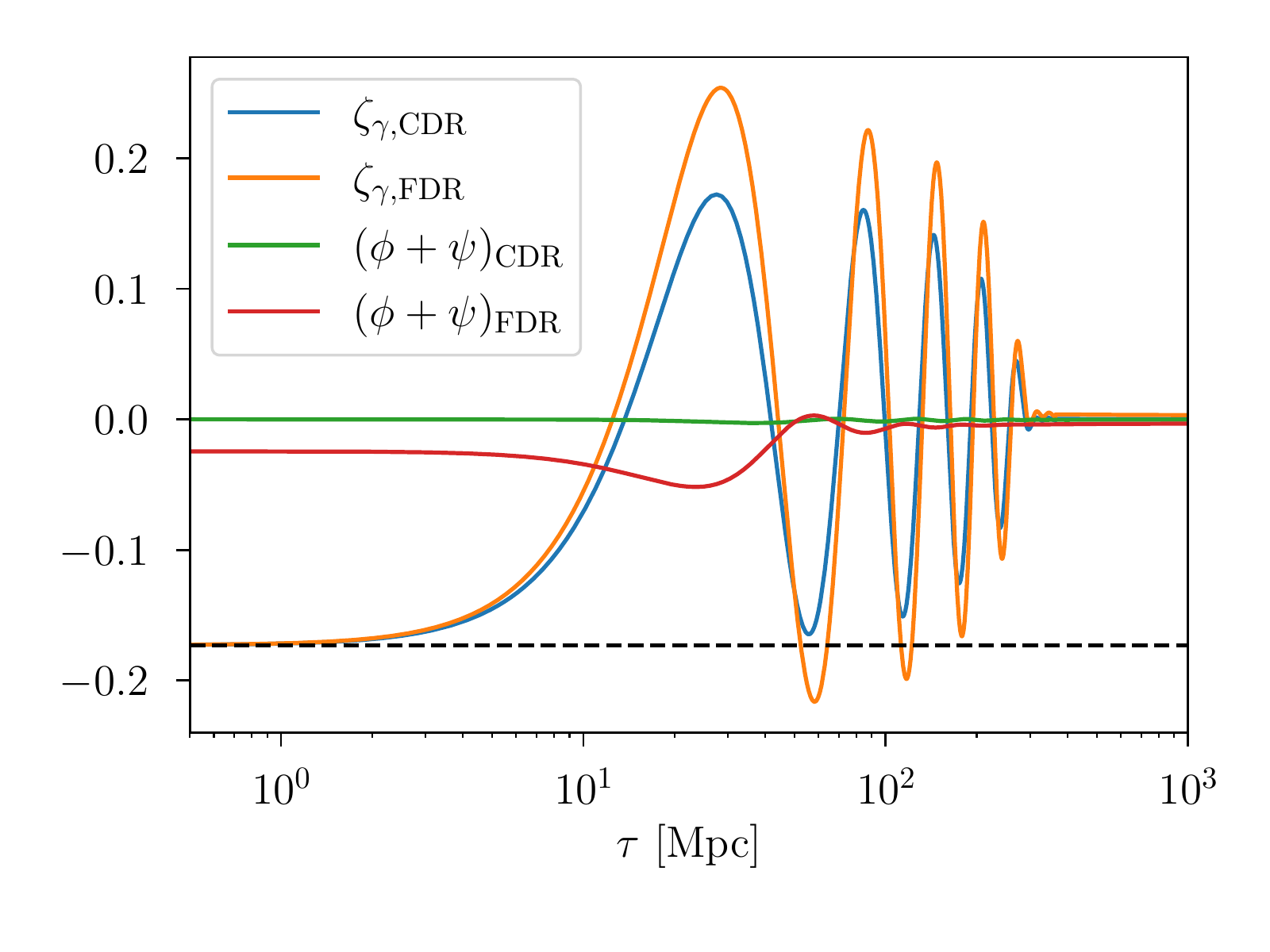}
    \caption{Time evolution of $\zeta_\gamma$ and $\phi+\psi$ for FDR and CDR with isocurvature initial condition for $k=0.2~\text{Mpc}^{-1}$. The dashed black line shows that for both FDR and CDR, $\zeta_\gamma$ has the same superhorizon value given by eq.~\eqref{eq.zetasuper}. These quantities are computed using our modified version of \texttt{CLASS} using $N_{\mathrm{ur}}=0.0,~N_{\mathrm{dr}}=3.046$.}
    \label{fig:mode_evolution}
\end{figure}

The conformal Newtonian gauge is parametrized as,
\begin{align}
ds^2=a^2(\tau)\left[-(1+2\psi)d\tau^2+(1-2\phi)d\Vec{x}^2\right].    
\end{align}
To go from the synchronous gauge to the conformal Newtonian gauge, we use the transformation~\cite{Ma:1995ey}
\begin{align}
\psi=\frac{1}{2k^2}\left(\ddot{h}+6\ddot{\eta}+\frac{\dot{a}}{a}\left(\dot{h}+6\dot{\eta}\right)\right),\label{eq.gauge1} \\
\phi=\eta-\frac{1}{2k^2}\frac{\dot{a}}{a}\left(\dot{h}+6\dot{\eta}\right).\label{eq.gauge2}
\end{align}
Once again, to get the leading behavior in the superhorizon limit, we approximate $\eta\gg h$,
\begin{align}
\psi\approx \frac{3}{k^2}\left(\ddot{\eta}+\frac{\dot{a}}{a}\dot{\eta}\right),\\
\phi\approx -\frac{3}{k^2}\frac{\dot{a}}{a}\dot{\eta},
\end{align}
implying
\begin{align}
\phi+\psi\approx\frac{3}{k^2}\ddot{\eta}\approx -\frac{2\sigma}{(k\tau)^2}.    
\end{align}
Since we showed above that the sign of $\sigma (<0~\text{or}~>0)$ depends upon the nature of DR (CDR or FDR), and we also showed $\sigma\approx-3\eta$ on superhorizon scales, we see that $\phi+\psi>0~(<0)$ for CDR (FDR). To see the implication of this, we consider the photon-baryon fluid in the tight-coupling regime where it obeys the equations,
\begin{align}
\Dot{\delta}_\gamma^{\rm con}=-\frac{4}{3}\theta_\gamma^{\rm con}+4\dot{\phi},\nonumber\\
\dot{\theta}_\gamma^{\rm con}\approx\frac{1}{4}k^2\delta_\gamma^{\rm con}+k^2\psi.
\end{align}
Here and below the superscript ``con'' denotes that the relevant quantities are evaluated in the conformal Newtonian gauge. By defining the gauge invariant perturbation for the photons~\cite{Malik:2008im}, $\zeta_\gamma=-\phi+\frac{1}{4}\delta_\gamma^{\rm con}$ we can rewrite the above as,
\begin{align}
\Ddot{\zeta}_\gamma\approx -\frac{1}{3}k^2\left(\zeta_\gamma+\phi+\psi\right). 
\end{align}
Now our goal is to study the solution of the above equations for times just after the horizon reentry for a given $k-$mode. To this end, we can ignore the time evolution of $\phi,\psi$ since they are subdominant on superhorizon scales compared to $\zeta_\gamma$ for isocurvature initial conditions, as can also be seen from Fig.~\ref{fig:mode_evolution}. 
Then we can further approximate the above equation as,
\begin{align}
\left(\ddot{\zeta_\gamma}+\ddot{\phi}+\ddot{\psi}\right)\approx -\frac{1}{3}k^2(\zeta_\gamma+\phi+\psi),
\end{align}
which has the solution,
\begin{align}
\zeta_\gamma+\phi+\psi\approx \mathcal{A}\cos(k\tau/\sqrt{3}).  
\end{align}
In the above, we have used the fact $\zeta_\gamma+\phi+\psi$ remains a non-zero constant on superhorizon scales. Conveniently, $\zeta_\gamma+\phi+\psi=\frac{1}{4}\delta_\gamma^{\rm con}+\psi$, i.e., the Sachs-Wolfe (SW) contribution. Now to compare CDR and FDR,  we note that for both CDR and FDR, $\zeta_\gamma$ is identical and negative at the time of horizon reentry, and can be written in terms of the density perturbation in the synchronous gauge,
\begin{align}\label{eq.zetasuper}
\zeta_\gamma=\frac{1}{4}\delta_\gamma-\eta\approx-\frac{1}{4}\frac{R_{\rm DR}}{1-R_{\rm DR}}<0.~~\text{(superhorizon initial condition)} 
\end{align}
However, as we have seen above, $\left(\phi+\psi\right)(\tau\rightarrow 0)$ is $>0~(<0)$ for CDR (FDR). Therefore, $\mathcal{A}=\left(\zeta_\gamma+\phi+\psi\right)(\tau\rightarrow0)$ is bigger in magnitude for FDR compared to CDR. Thus, as the modes reenter the horizon, they start a bigger amplitude of oscillations for the case of FDR. This then translates into a larger value of the SW term at recombination $\tau_*$,
$\left(\frac{1}{4}\delta_\gamma^{\rm con}+\psi\right)(\tau_*)$, leading to more SW anisotropy. 
To obtain the specific values of $\frac{1}{4}\delta_\gamma^{\rm con}+\psi$ for CDR and FDR, we can use gauge transformations~\eqref{eq.gauge1}~and~\eqref{eq.gauge2} from the conformal Newtonian gauge to the synchronous gauge to write in the superhorizon limit,
\begin{align}\label{eq.SWic}
\frac{1}{4}\delta_\gamma^{\rm con}+\psi \approx \frac{1}{4} \delta_\gamma+\frac{3}{k^2}\ddot{\eta}\approx
\begin{dcases} 
  -\dfrac{1}{4}\dfrac{R_{\rm DR}}{1-R_{\rm DR}}+\dfrac{R_{\rm DR}R_\nu}{(1-R_{\rm DR})(15+4R_\nu)} & \text{CDR}\\
  -\dfrac{1}{4}\dfrac{R_{\rm DR}}{1-R_{\rm DR}}-\dfrac{R_{\rm DR}-R_{\rm DR}R_\nu-R_{\rm DR}^2}{(1-R_{\rm DR})(15+4R_\nu+4R_{\rm DR})} & \text{FDR}
\end{dcases}.
\end{align}
Here we have used the superhorizon values of $\delta_\gamma,\eta$ from Tables~\ref{tab:fdric}~and~\ref{tab:cdric}. In particular, it is instructive to consider a parametric limit, $R_{\rm DR}\ll 1, R_{\nu}=0$, for which we get the leading order result,
\begin{align}\label{eq.SWicParam}
\frac{1}{4}\delta_\gamma^{\rm con}+\psi\approx
\begin{dcases} 
  -\dfrac{1}{4}R_{\rm DR} & \text{CDR}\\
  -\dfrac{19}{60}R_{\rm DR} & \text{FDR}
\end{dcases},
\end{align}
and now the $\mathcal{O}(1)$ difference between CDR and FDR is manifest.

\begin{figure}[h]
    \centering
    \begin{subfigure}[b]{0.47\textwidth}
        \includegraphics[width=\textwidth]{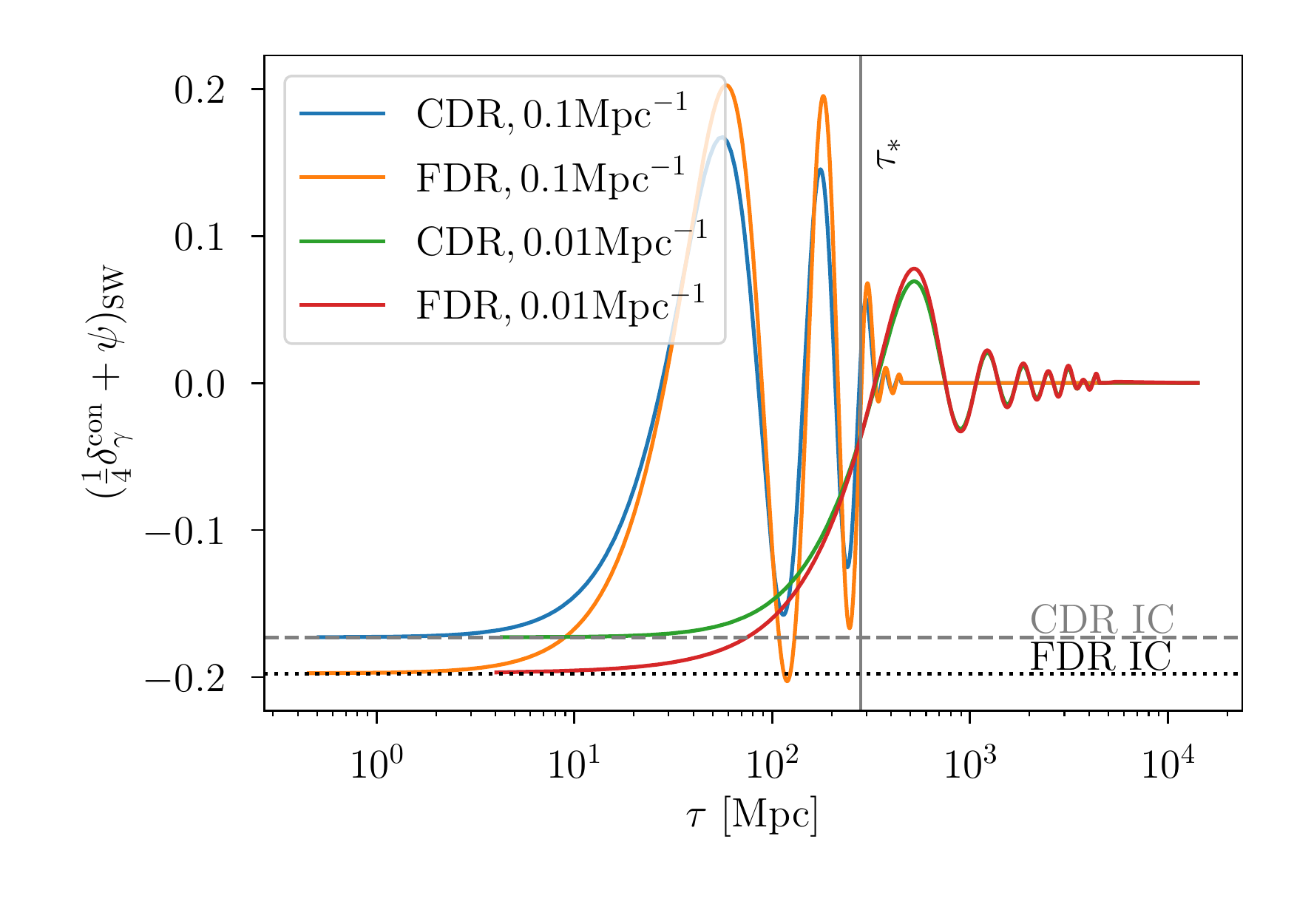}
        \caption{Evolution of the Sachs-Wolfe term}
        \label{fig:deltag}
    \end{subfigure}
    \begin{subfigure}[b]{0.47\textwidth}
        \includegraphics[width=\textwidth]{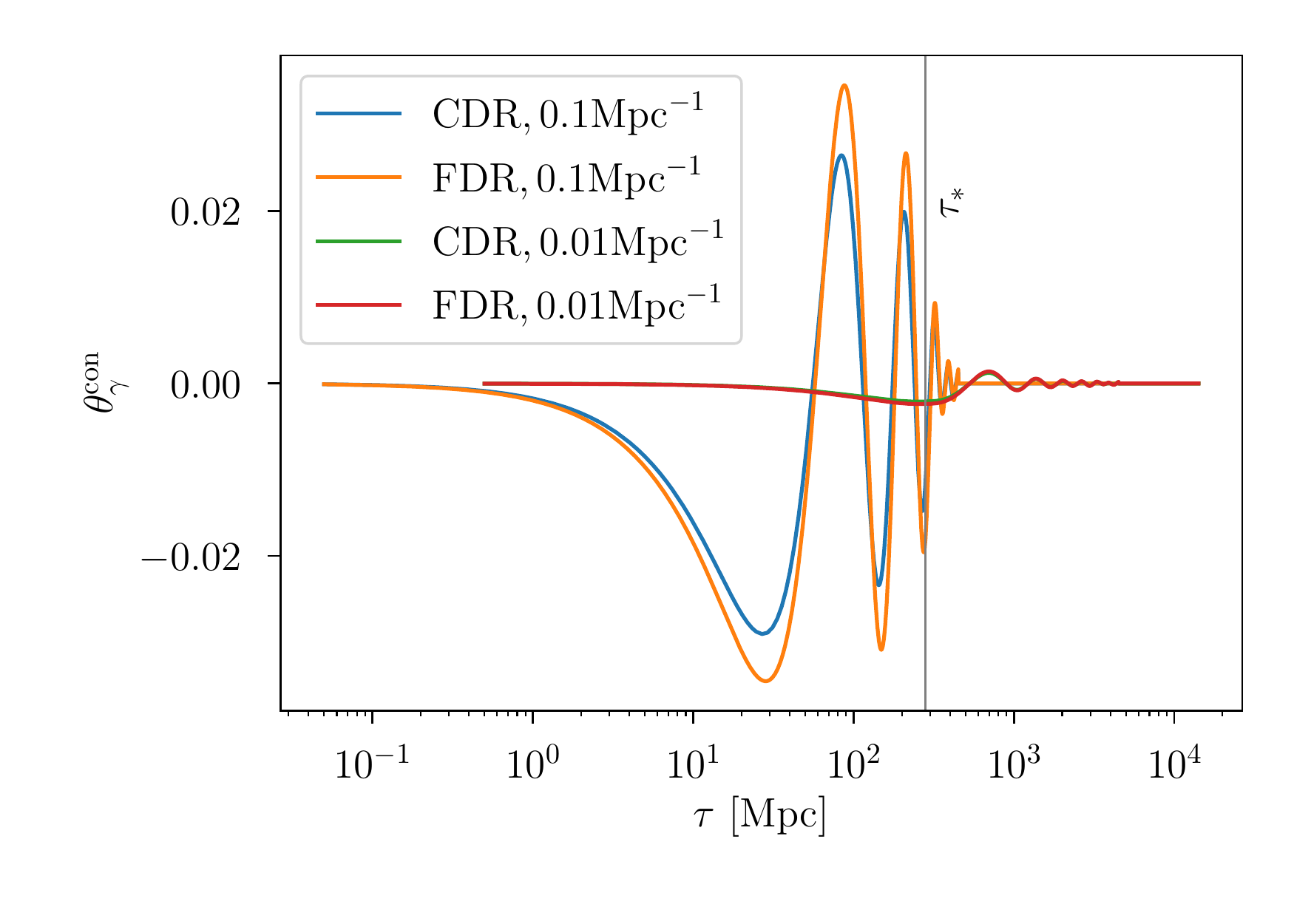}
        \caption{Evolution of the Doppler term}
        \label{fig:iso}
    \end{subfigure}
    \caption{Comparison of time evolution of two different $k$-modes for isocurvature initial conditions (IC). For $k=0.1~\text{Mpc}^{-1}$, the mode reenters the horizon earlier and get a chance to oscillate multiple times before recombination at $\tau_*$. Therefore generically, FDR exhibits larger anisotropies as argued in the text. For $k=0.01~\text{Mpc}^{-1}$, the mode enters later and happens to go near its minimum around $\tau_*$, and therefore the distinction between CDR and FDR is less significant. In panel (a), the dashed gray and dotted black lines show the superhorizon values derived above for CDR and FDR respectively, using eq.~\eqref{eq.SWic} for $R_\nu=0$. These are exactly matched by the numerical initial values computed using our modified version of \texttt{CLASS} using $N_{\mathrm{ur}}=0.0,~N_{\mathrm{dr}}=3.046$. In panel (b), we show the evolution of photon velocity where similar enhancement for FDR is observed.}
    \label{fig:evolution}
\end{figure}

The same conclusion is also true for the Doppler contribution to CMB anisotropy. To see this, we can approximate the photon velocity equation as, 
\begin{align}
\theta_\gamma^{\rm con}\approx-\frac{3}{4}\dot{\delta}_\gamma^{\rm con}\approx-3\dot{\zeta}_\gamma\approx\sqrt{3}\mathcal{A}k\sin\left(k\tau/\sqrt{3}\right).   
\end{align}
Therefore, once again for FDR, $\theta_\gamma$ oscillations have a bigger amplitude compared to CDR and this leads to larger $\theta_\gamma^{\rm con}(\tau_*)$. In Fig.~\ref{fig:evolution}, we numerically evaluate the evolution of both SW term (controlled by $\left(\frac{1}{4}\delta_\gamma^{\rm con}+\psi\right)$),  and the Doppler term (controlled by $\theta_\gamma^{\rm con}$) for two different $k$-modes where enhanced oscillations for FDR is apparent. 

\begin{figure}[h]
    \centering
    \begin{subfigure}[b]{0.45\textwidth}
        \includegraphics[width=\textwidth]{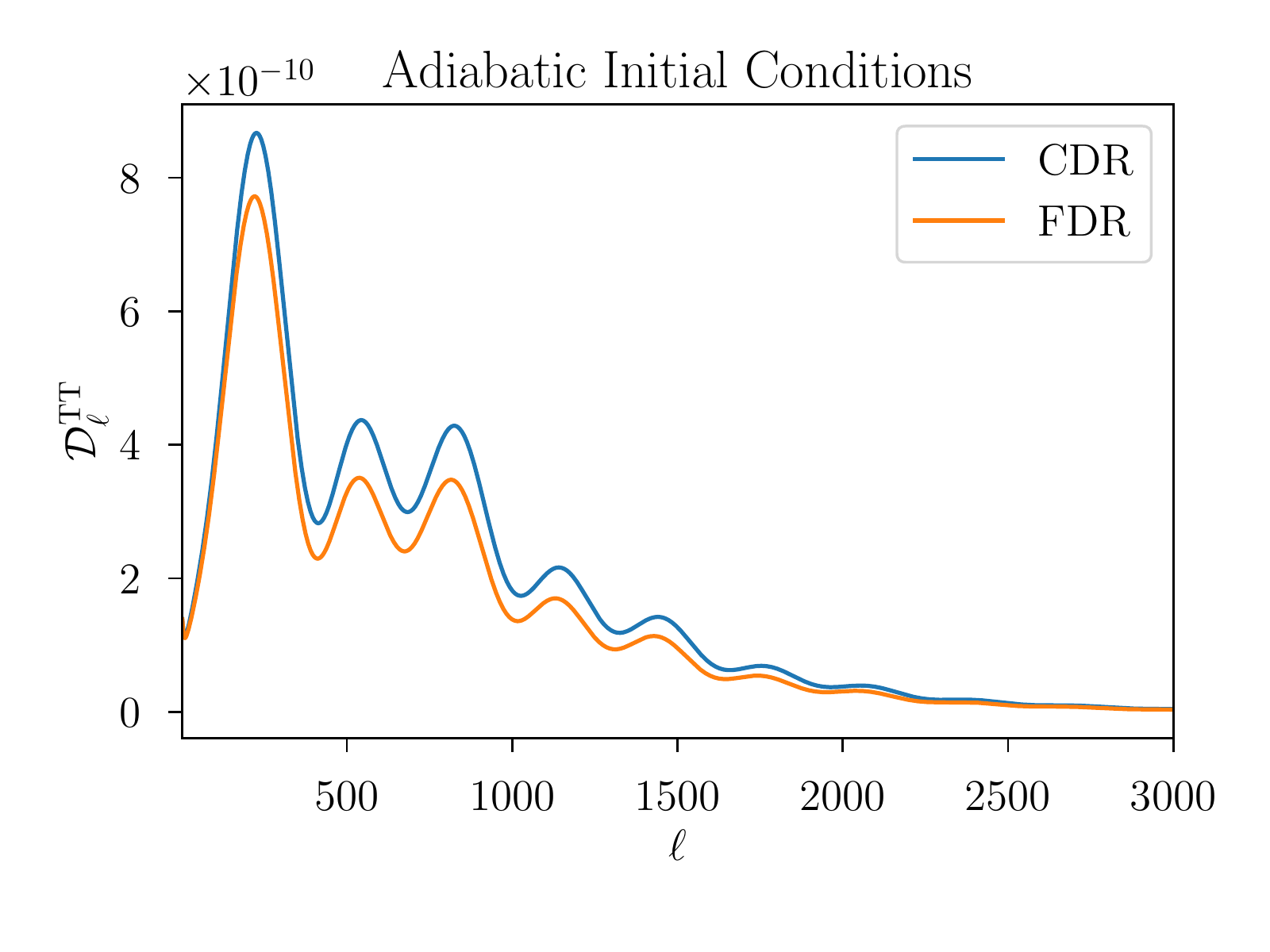}
        \caption{}
        \label{fig:adia}
    \end{subfigure}
    \begin{subfigure}[b]{0.45\textwidth}
        \includegraphics[width=\textwidth]{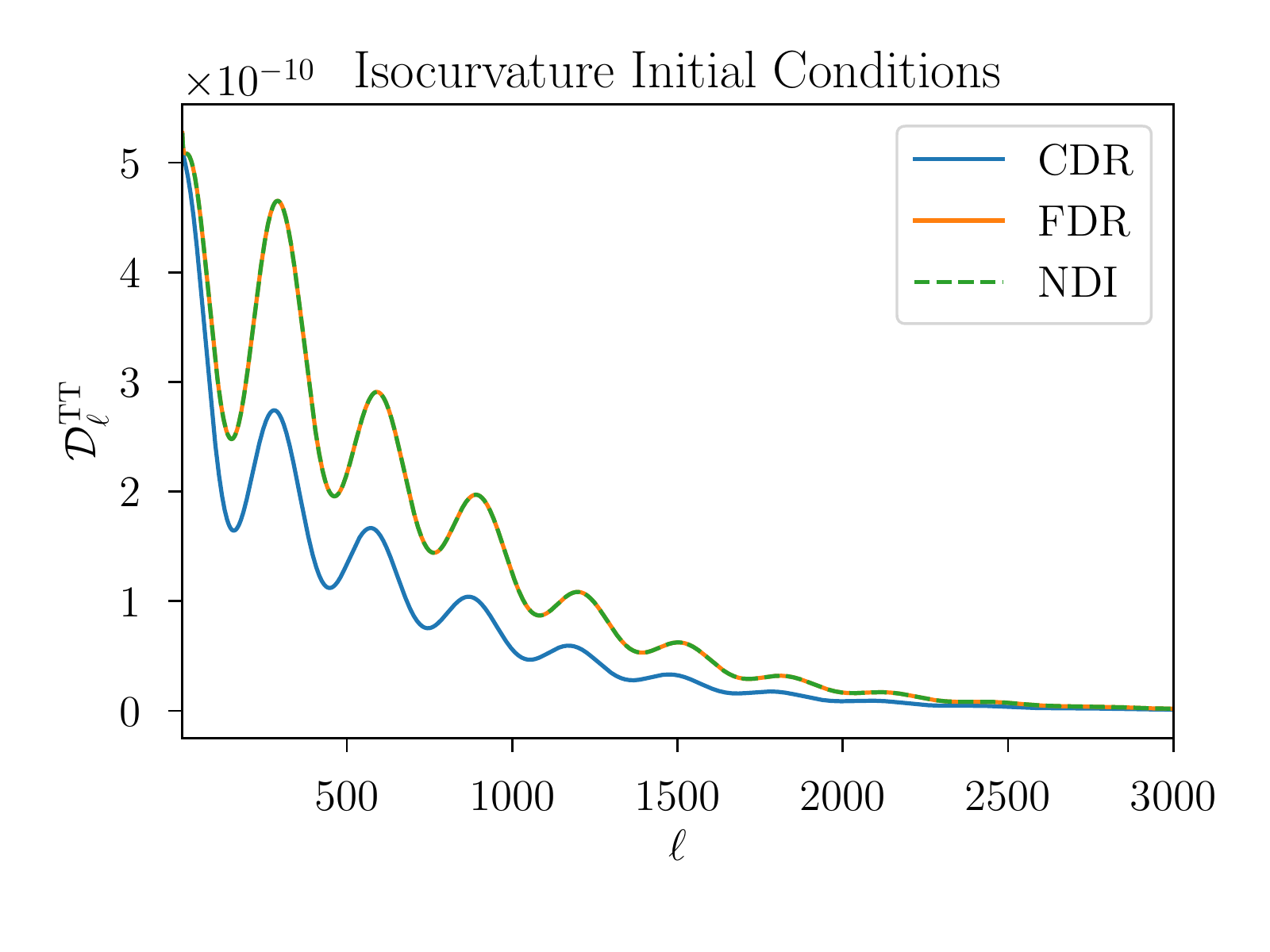}
        \caption{}
        \label{fig:compare_NDI}
    \end{subfigure}
    \caption{Comparison of $\mathcal{D}_\ell^{\mathrm{TT}}\equiv\left[\ell(\ell+1)/(2\pi)\right]C_\ell^{\rm TT}$ for various initial conditions. In panel (a) for adiabatic initial conditions, and $N_{\mathrm{ur}}=0.0,~N_{\mathrm{dr}}=3.046$, we see the well known fact that CDR gives a more enhanced contribution to $\mathcal{D}_\ell^{\mathrm{TT}}$ than FDR. In panel (b) for the same choice of $N_{\mathrm{ur}}$ and $N_{\mathrm{dr}}$ but with pure isocurvature initial conditions, and $f_{\mathrm{iso}}=3.0,~n_{\mathrm{iso}}=1.0$ we see that FDR instead gives a more enhanced contribution to $\mathcal{D}_\ell^{\mathrm{TT}}$ than CDR. From panel (b) we also see that FDR spectrum coincides with a pure NDI induced spectrum with $N_{\mathrm{ur}}=3.046,~N_{\mathrm{dr}}=0.0$, as expected. In these comparisons, all the other parameters are set to their $\Lambda$CDM values.}
    \label{fig:cDRvsfDR}
\end{figure}

Now let us translate this into our final expectations on CMB power spectrum. We have seen above that both for the SW term and the Doppler term, FDR is expected to exhibit more anisotropy for DRID initial conditions. The integrated Sachs-Wolfe (ISW) term gets its primary contribution around the time of matter-radiation equality and the onset of dark energy domination. Therefore at least for large enough $\ell$, where SW and Doppler effects dominate, we expect FDR to exhibit more anisotropy compared to CDR. This is indeed what is seen in panel~(b) of Fig.~\ref{fig:cDRvsfDR}.

The disparity in the sizes of the anisotropy perturbations for CDR and FDR is also reflected in the DRID constraints shown in Fig.~\ref{fig:is0-param-tri-fn} and \ref{fig:is0-param-tri-wsh0es-fn} (or in Fig.~\ref{fig:is0-param-tri-vn} and \ref{fig:is0-param-tri-wsh0es-vn}). It can be seen from those figures that the sizes of $N_{\rm dr}^2\mathcal{P}_\mathcal{II}^{(1)}$ and $N_{\rm dr}^2\mathcal{P}_\mathcal{II}^{(2)}$ are systematically higher on CDR compared to FDR. To induce similar sized modification in the CMB spectrum, the amplitude of the initial DRID perturbation must be higher for CDR compared to the FDR scenario to compensate for the suppression of CDR spectrum as shown in Fig.~\ref{fig:cDRvsfDR}~(b).
\paragraph{Comparison with adiabatic initial conditions.}
One can repeat the above exercise for the case of adiabatic initial conditions. In particular, the primary difference with the DRID scenario is that, for adiabatic IC, the sign of the total shear perturbations $\sigma$ does not change between CDR and FDR. For CDR, the shear perturbation is identical to the \lcdm{} result since DR does not contribute to it. For FDR, on the other hand, $\sigma_{\rm DR}=\sigma_{\nu}$ and they are given by changing $R_\nu\rightarrow R_\nu+R_{\rm DR}$ from the \lcdm{} result.
Furthermore, it can be derived that the analog of eq.~\eqref{eq.SWic} is given by,
\begin{align}\label{eq.SWicad}
\frac{1}{4}\delta_\gamma^{\rm con}+\psi\bigg\rvert_{\tau\rightarrow 0} \approx
\begin{dcases} 
  \dfrac{5}{15+4R_\nu} & \text{CDR}\\
  \dfrac{5}{15+4R_\nu+4 R_{\rm DR}} & \text{FDR}
\end{dcases}.  
\end{align}
The difference in the initial condition is well studied in the literature~\cite{Bashinsky:2003tk,Baumann:2015rya,Blinov:2020hmc}, which results in larger CMB anisotropies for the case of CDR than FDR.

\section{Conclusion}\label{sec.con}

In this paper, we show how the current CMB data alone and also in combination with BAO and SH0ES data, constrain the DR isocurvature perturbation and how the existence of DR isocurvature density perturbation (DRID) helps to relax the $H_0$ tension. We start the discussion with a simple curvaton model to explain how the curvaton fluctuations can source DRID perturbations with different sizes and tilts from the adiabatic perturbations. We then turn to a more general study of the DRID cosmology by first systematically deriving the initial conditions for both the FDR-DRID and CDR-DRID scenarios. Using a MCMC study with current cosmological datasets, we then calculate bounds on the size ($\mathcal{P_{II}}$) and the tilt ($n_{\rm iso}$) of the DRID perturbations, and find constraints on the energy density of the DR ($N_{\rm dr}$). We showed two sets of analyses where the neutrino effective number $N_{\rm ur}$ is kept fixed to $3.046$ in one case and varied in the other. The constraints derived in these two cases are similar. While we considered only uncorrelated DRID perturbations in the main text, we show the constraints for correlated DRID scenario in the Appendix.

The FDR-DRID constraint from Planck 2018 dataset for fixed $N_{\rm ur}$ (for varying $N_{\rm ur}$) analysis is summarized by the blue contours in Fig.~\ref{fig:is0-param-tri-fn} (Fig.~\ref{fig:is0-param-tri-vn}) and the numbers in Table~\ref{tab:param-fdr-fn} (Table~\ref{tab:param-fdr-vn}) with additional contours shown in Appendix~\ref{sec.triangle}. The CMB spectrum induced by DRID perturbation has an approximate degeneracy in the form of $C_{\ell,\rm DRID}\propto N^2_{\rm dr} \mathcal{P_{II}}$. If the curvaton $\chi$ produces FDR with $N_{\rm dr}=0.4$, the current data at $2 \sigma$ allows a blue-tilted FDR-DRID with $10^{10}N^2_{\rm dr}\mathcal{P_{II}}^{(2)}\leq 200$ (TTTEEE+lowE+lensing), and $10^{10}N^2_{\rm dr}\mathcal{P_{II}}^{(2)}\leq 220$ (TTTEEE+lowE+lensing+BAO+SH0ES(L)), where $\mathcal{P_{II}}^{(2)}$ is the magnitude of isocurvature power spectrum at $k_2=0.1\,{\rm Mpc}^{-1}$. This means that current Planck data constrains the curvaton perturbation to be $\delta\sigma/\sigma\lsim 2\times10^{-4}$ around comoving time $\sim k_2^{-1}$ if $\sigma$ decays into FDR.
We also present the first cosmological bounds on the CDR-DRID, which can easily exist if the curvaton decays into DR that interact with each other. For $N_{\rm dr}=0.4$, the $2\sigma$ upper bound on the CDR-DRID is $10^{10}N^2_{\rm dr}\mathcal{P_{II}}^{(2)}\leq 600$ (TTTEEE+lowE+lensing), and $10^{10}N^2_{\rm dr}\mathcal{P_{II}}^{(2)}\leq 1000$ (TTTEEE+lowE+lensing+SH0ES(L)). The bound on the curvaton perturbation from the Planck data is $\delta\sigma/\sigma\lsim 5\times 10^{-4}$ around comoving time $\sim k_2^{-1}$ if $\sigma$ decays into CDR. The above numbers are in regards to the varying $N_{\rm ur}$ analysis and similar sized bounds are derived for the fixed $N_{\rm ur}$ case.

The weaker bounds on $N_{\rm dr}^2\mathcal{P_{II}}$ for CDR shows an intriguing feature of the DRID scenario very different from the adiabatic case---the FDR-DRID enhances the CMB TT spectrum more than the CDR-DRID, opposite to the relative FDR versus CDR contribution to the TT spectrum in the adiabatic case. We give an analytical explanation of the behavior, which comes from the nature of the isocurvature perturbations existing on a manifold with constant total radiation density. Because of this, the metric perturbation mainly comes from the shear of neutrino (for CDR-DRID) or the DR (for FDR-DRID), which results in different SW contributions and a larger TT spectrum from the DRID in the FDR case. 

Besides studying the DRID constraint, we also check if the presence of DRID helps to reconcile the tension between the CMB and the local $H_0$ measurements. As summarized in Fig.~\ref{fig:H0-vs-Neff-wo-sh0es}~and~\ref{fig:H0-vs-Neff-w-sh0es}, the presence of isocurvature perturbation does indeed help to reduce the tension better than the adiabatic DR. The FDR-DRID with varying $N_{\rm ur}$ (with fixed $N_{\rm ur}$) setup reduce the tension between the Planck and SH0ES measurements to $\sim 2.2\sigma$ ($\sim 2.5\sigma$), while the CDR-DRID scenario further reduces the tension to $\sim 2.0\sigma$ ($\sim 2.5\sigma$). As a comparison, the same estimate of the tension with adiabatic CDR studied in~\cite{Blinov:2020hmc} is $\sim 3.1\sigma$. Therefore the DRID perturbation does help to further suppress the tension. If including both the Planck and SH0ES data, the fit of $H_0$ from the CDR-DRID (FDR-DRID) scenarios has a $\sim 1.2\sigma\,(1.5\sigma)$ discrepancy to the SH0ES measurement for the varying $N_{\rm ur}$ scenario and $\sim 1.3\sigma\,(1.6\sigma)$ discrepancy for the fixed $N_{\rm ur}$ scenario. As is demonstrated in Fig.~\ref{fig:fdr-spectra-w-data}, the blue-tilted DRID perturbation helps to compensate for the suppressed $C_{\ell}^{\rm TT}$ at higher multipoles due to the larger Silk damping caused by the energy density in DR. This allows the existence of larger DR energy density that enhances the $H_0$ from fitting the Planck data. However, the enhancement is still limited by the TE and EE data that sets strong constraints on the perturbations of lower-$\ell$ modes.  

Free streaming and coupled DR are plausible forms of energy in the early Universe motivated by many BSM scenarios, and depending on the inflationary and reheating process, DR can easily carry isocurvature perturbations uncorrelated or correlated to the adiabatic perturbations. Our work shows that besides being sensitive to an additional source of perturbations and energy density in the invisible radiation, CMB measurements can further probe the interactions of DR in the early Universe. 
Since we have seen relatively large, blue-tilted isocurvature perturbation from DR is allowed by the present data, and that affects the tail of the CMB spectrum, future small scale CMB measurements will be able to put stronger constraints or 
discover 
isocurvature perturbations in the Universe.

\begin{acknowledgments}
We thank Tommi Tenkanen and Jussi Väliviita for the helpful discussion in the beginning of the project. We also thank Nikita Blinov, Gustavo Marques Tavares and Jussi Väliviita for useful comments on the draft. 
The research of SG and YT is supported by the NSF grant PHY-2014165. SK is supported by the NSF grant PHY-1915314 and the U.S. DOE Contract DE-AC02-05CH11231. SG  acknowledges  the  support  from  Department  of  Atomic  Energy, Government of India during the initial stage of this project.

\end{acknowledgments}
\appendix

\section{Constraint on DR isocurvature with correlation}\label{sec.corr}
In the main text, we have discussed and derived the constraints on \emph{uncorrelated} dark radiation isocurvature perturbations by setting the correlations between the adiabatic and DRID perturbation at both scales (in the two-scale parametrization) $P_\mathcal{RI}^{(1)} = P_\mathcal{RI}^{(2)} = 0$. In this section, we relax this restriction and consider non-zero correlation between these two types of perturbations.

The most general parametrization of the correlation will require independent values of both $P_\mathcal{RI}^{(1)} $ and $ P_\mathcal{RI}^{(2)}$ which can be translated into the amplitude and tilt of the correlation power spectrum. Further, the positive definiteness of the initial condition matrix requires, 
\begin{equation}\label{eq:defcorr}
    \pow_\mathcal{RR}(k)\pow_\mathcal{II}(k) \geq \left[\pow_\mathcal{RI}(k)\right]^2
\end{equation}
for all $k$ modes of interest. However, the positive definiteness of perturbations at scales $k_1$ and $k_2$ only guarantee positive definiteness for $k$-modes lying within the range $k_1<k<k_2$. Using the definition of power spectrum in Eq.~\eqref{eq:twoscaledef}
\begin{align}\label{eq:genpos}
2\ln \powri(k) &= 2\ln \powri^{(1)} \ {\ln k - \ln k_2 \over \ln k_1 - \ln k_2 } +2 \ln\powri^{(2)} \ {\ln k - \ln k_1 \over \ln k_2 - \ln k_1 }\;.
\end{align}
The positive definiteness of the input power spectrum means,
\begin{align}\label{eq:posdefin1}
    2\ln \powri^{(1)} \leq \ln \powrr^{(1)}+ \ln \powii^{(1)}\;,\\\label{eq:posdefin2}
    2\ln \powri^{(2)} \leq \ln \powrr^{(2)}+ \ln \powii^{(2)}\;.
\end{align}
Using Eq.~\eqref{eq:posdefin1} and Eq.~\eqref{eq:posdefin2} we can derive from Eq.~\eqref{eq:defcorr} the following condition:
\begin{align}
    2\ln \powri(k) \leq &( \ln \powrr^{(1)}+ \ln \powii^{(1)}) {\ln k - \ln k_2 \over \ln k_1 - \ln k_2 } + ( \ln \powrr^{(2)}+ \ln \powii^{(2)}) {\ln k - \ln k_1 \over \ln k_2 - \ln k_1 }\\
    \leq& \ln \powrr(k) + \ln \powii(k)
\end{align}
for positive definiteness of a general $k $ mode if and only if
\begin{equation}
    {\ln k - \ln k_2 \over \ln k_1 - \ln k_2 } \geq 0 \quad\text{and}\quad {\ln k - \ln k_1 \over \ln k_2 - \ln k_1 } \geq 0 \quad\Rightarrow \quad k_1 < k< k_2\;.
\end{equation}
Therefore, for modes outside $[k_1,k_2]$ positive definiteness of initial condition matrix in not guaranteed. 

To remedy this problem, following isocurvature studies by Planck collaboration~\cite{Ade:2015lrj,Planck:2018jri}, we vary only $\powri^{(1)}$ as an independent parameter and $\powri^{(2)}$ is determined as,
\begin{equation}\label{eq:defptab}
\powri^{(2)} = \powri^{(1)} \sqrt{\powrr^{(2)}\powii^{(2)} \over \powrr^{(1)}\powii^{(1)}}\;,
\end{equation}
Using the above definition we can show that for any $k$ mode
\begin{multline}
    2\ln \powri(k) - \left(\ln \powrr(k) + \ln \powii(k)\right) = 2\ln \powri^{(1)} - \left(\ln \powrr^{(1)}+\ln \powii^{(1)}\right) \leq 0 \\\text{[following Eq.~\eqref{eq:posdefin1}]}
\end{multline}
Thus, in this scheme, the positive definiteness of \emph{all modes} are guaranteed as long as the initial spectrum at $k_1$ is positive definite. It can easily be shown that Eq.~\eqref{eq:defptab} amounts to setting the spectral index of the correlation power spectrum, which is defined as~\cite{lesgourgues2011cosmic}
\begin{equation}\label{eq:ncor}
n_\text{cor} \equiv {\ln \mathcal{P}_\mathcal{RI}^{(1)} -\ln \mathcal{P}_\mathcal{RI}^{(2)}  \over \ln k_1 -\ln k_2}- {1 \over 2}\left(n_s+n_\text{iso} - 2\right)
\end{equation}
to zero i.e., $n_{\rm cor} = 0$.\footnote{The definition of $n_\text{cor}$ is taken from \texttt{CLASS}~\cite{lesgourgues2011cosmic}. Note the difference between the definitions of $n_\text{cor}$ and $n_\mathcal{RI} \equiv {(\ln \mathcal{P}_\mathcal{RI}^{(1)} -\ln \mathcal{P}_\mathcal{RI}^{(2)}) /( \ln k_1 -\ln k_2)} $ which is used in the analysis of isocurvature perturbation by the Planck collaboration~\cite{Ade:2015lrj,Planck:2018jri}. } Therefore, effectively we are assuming a scale invariant correlation spectrum in this analysis. Eq.~\eqref{eq:defptab} also dictates that correlation fraction $\cos \Delta$ which is defined as,
\begin{equation}
    \label{eq:cosD}
    \cos \Delta = \dfrac{\mathcal{P}_\mathcal{RI}(k)}{\sqrt{\mathcal{P}_\mathcal{RR}(k)\mathcal{P}_\mathcal{II}(k)}}
\end{equation}
is constant for all $k$ modes.

\subsection{MCMC results}

We performed MCMC analysis of correlated DRID model with the same dataset described section~\ref{sec.sig}. We redo the analysis as done in that section with the addition of one new parameter for correlation: $N_{\rm dr} \powri^{(1)}$. Note the different $N_{\rm dr}$ scaling in the correlation parameter compared to the isocurvature parameters: $N_{\rm dr}^2 \powii^{(1)}$ and $N_{\rm dr}^2 \powii^{(2)}$. Since the correlation is the interference between adiabatic and isocurvature perturbations, it is linear in isocurvature perturbation. Therefore, we only scale it by a single power of $N_{\rm dr}$ to take care of the generic convergence issue of DR isocurvature discussed in section~\ref{sec.sig}. Thus, we vary the `physical' correlation parameter $N_{\rm dr} \powri^{(1)}$ as a primary parameter and $\powri^{(1)}$ is recovered as a derived parameter by diving the former by $N_{\rm dr}$. 

\renewcommand{\arraystretch}{1.4}
\begin{table}[h!]
    \centering
    \begin{tabular}{|l|c| c| }
\hline
FDR (FN \& WC)& \makecell[c]{P18-TTTEEE\\+lowE+lensing}& \makecell[c]{P18-TTTEEE+lowE+ \\lensing+BAO+SH0ES(L)}\\
\hline
$10^2 \omega_{b}$& $ 2.264^{+0.018}_{-0.022}$& $ 2.281\pm 0.015$\\
\hline
$\omega_{cdm }$& $ 0.1217^{+0.0016}_{-0.0025}$& $ 0.1247\pm 0.0026$\\
\hline
$100\theta_{s }$& $ 1.04219^{+0.00055}_{-0.00069}$& $ 1.04206^{+0.00063}_{-0.00075}$\\
\hline
$\tau_{reio }$& $ 0.0564^{+0.0071}_{-0.0081}$& $ 0.0564\pm 0.0072$\\
\hline
$10^{10}\mathcal{P}_\mathcal{RR}^{(1)}$& $ 23.10\pm 0.50$& $ 22.79\pm 0.47$\\
\hline
$10^{10}\mathcal{P}_\mathcal{RR}^{(2)}$& $ 20.55\pm 0.40$& $ 20.68\pm 0.42$\\
\hline
$ 10^{10} N_{\rm dr}^2 \mathcal{P}_\mathcal{II}^{(1)}$& $ 19.8^{+3.0}_{-20}$& $< 19.1$\\
\hline
$10^{10} N_{\rm dr}^2\mathcal{P}_\mathcal{II}^{(2)}$& $ 99^{+40}_{-70}$& $ 118^{+50}_{-70}$\\
\hline
$10^{10} N_{\rm dr} \mathcal{P}_\mathcal{RI}^{(1)}$& $ -0.18^{+1.4}_{-0.99}$& $ 0.24^{+1.2}_{-0.95}$\\
\hline
$N_{\rm dr}$& $< 0.208$& $ 0.37\pm 0.13$\\
\hline
\hline$H_0 ({\rm km/s/Mpc})$& $ 69.66^{+0.82}_{-1.4}$& $ 71.01\pm 0.83$\\
\hline
$\sigma_8$& $ 0.8250^{+0.0075}_{-0.0087}$& $ 0.8318\pm 0.0087$\\
\hline
$10^9 A_s$& $ 2.098\pm 0.035$& $ 2.104\pm 0.037$\\
\hline
$n_{s }$& $ 0.9702^{+0.0063}_{-0.0075}$& $ 0.9752^{+0.0077}_{-0.0061}$\\
\hline
$n_{\rm iso}$& $ 1.45^{+0.37}_{-0.28}$& $ 1.55^{+0.37}_{-0.28}$\\
\hline
$f_{\rm iso}$& $ 57^{+15}_{-51}$& $ 7.0^{+1.1}_{-3.5}$\\
\hline
$\cos \Delta$& $ 0.003^{+0.058}_{-0.075}$& $ 0.028^{+0.055}_{-0.082}$\\
\hline
$N_{\rm tot}$& $< 3.25$& $ 3.41\pm 0.13$\\
\hline
$f_{\rm dr}$& $ 0.051^{+0.020}_{-0.047}$& $ 0.106^{+0.036}_{-0.032}$\\
\hline
\hline
$ \chi^2 - \chi^2_{\Lambda \rm{CDM}}$& $-1.9$& $-14.62$\\
\hline
\end{tabular}
    \caption{Mean and $1\sigma$ error of parameters for FDR-DRID \emph{with correlation} for fixed $N_{\rm ur}$ scenario for the corresponding datasets. The limits are at 68\% C.L. The constraints on the primary parameters and the derived parameters are shown in two separate blocks. The $\chi^2$ difference with respect to the \lcdm{} (fixed \neff{}) model for the corresponding data-set is shown on the last line.}
    \label{tab:param-fdr-wcor-fn}
\end{table}

\renewcommand{\arraystretch}{1.4}
\begin{table}[h!]
    \centering
    \begin{tabular}{|l|c| c| }
\hline
CDR (FN \& WC)& \makecell[c]{P18-TTTEEE\\+lowE+lensing}& \makecell[c]{P18-TTTEEE+lowE+ \\lensing+BAO+SH0ES(L)}\\
\hline
$10^2 \omega_{b}$& $ 2.262^{+0.018}_{-0.025}$& $ 2.286\pm 0.016$\\
\hline
$\omega_{cdm }$& $ 0.1229^{+0.0018}_{-0.0030}$& $ 0.1267\pm 0.0028$\\
\hline
$100\theta_{s }$& $ 1.04221^{+0.00035}_{-0.00040}$& $ 1.04250\pm 0.00035$\\
\hline
$\tau_{reio }$& $ 0.0556^{+0.0070}_{-0.0079}$& $ 0.0559\pm 0.0074$\\
\hline
$10^{10}\mathcal{P}_\mathcal{RR}^{(1)}$& $ 23.29\pm 0.45$& $ 23.11\pm 0.45$\\
\hline
$10^{10}\mathcal{P}_\mathcal{RR}^{(2)}$& $ 20.49\pm 0.36$& $ 20.40\pm 0.37$\\
\hline
$ 10^{10} N_{\rm dr}^2 \mathcal{P}_\mathcal{II}^{(1)}$& $ 28^{+6}_{-30}$& $ 23^{+4}_{-20}$\\
\hline
$10^{10} N_{\rm dr}^2\mathcal{P}_\mathcal{II}^{(2)}$& $< 386$& $< 603$\\
\hline
$10^{10} N_{\rm dr} \mathcal{P}_\mathcal{RI}^{(1)}$& $ -1.3^{+1.5}_{-1.3}$& $ -1.3^{+1.4}_{-1.0}$\\
\hline
$N_{\rm dr}$& $< 0.253$& $ 0.43\pm 0.13$\\
\hline
\hline$H_0 ({\rm km/s/Mpc})$& $ 69.62^{+0.90}_{-1.6}$& $ 71.28\pm 0.86$\\
\hline
$\sigma_8$& $ 0.8285\pm 0.0081$& $ 0.8324\pm 0.0084$\\
\hline
$10^9 A_s$& $ 2.096\pm 0.033$& $ 2.085\pm 0.034$\\
\hline
$n_{s }$& $ 0.9673\pm 0.0057$& $ 0.9681^{+0.0065}_{-0.0048}$\\
\hline
$n_{\rm iso}$& $ 1.59^{+0.44}_{-0.31}$& $ 1.77^{+0.43}_{-0.27}$\\
\hline
$f_{\rm iso}$& $ 66^{+22}_{-59}$& $ 8.5^{+3.5}_{-4.8}$\\
\hline
$\cos \Delta$& $ -0.050^{+0.041}_{-0.066}$& $ -0.055^{+0.040}_{-0.061}$\\
\hline
$N_{\rm tot}$& $< 3.30$& $ 3.48\pm 0.13$\\
\hline
$f_{
m dr}$& $ 0.061^{+0.025}_{-0.053}$& $ 0.123\pm 0.034$\\
\hline
\hline
$ \chi^2 - \chi^2_{\Lambda \rm{CDM}}$& $0.48$& $-11.62$\\
\hline
\end{tabular}
    \caption{Mean and $1\sigma$ error of parameters for CDR-DRID \emph{with correlation} for fixed $N_{\rm ur}$ scenario for the corresponding datasets. The limits are at 68\% C.L. The constraints on the primary parameters and the derived parameters are shown in two separate blocks. The $\chi^2$ difference with respect to the \lcdm{} (fixed \neff{}) model for the corresponding data-set is shown on the last line.}
    \label{tab:param-cdr-wcor-fn}
\end{table}

\renewcommand{\arraystretch}{1.4}
\begin{table}[h!]
    \centering
    \begin{tabular}{|l|c| c| }
\hline
FDR (VN \& WC)& \makecell[c]{P18-TTTEEE\\+lowE+lensing}& \makecell[c]{P18-TTTEEE+lowE+ \\lensing+BAO+SH0ES(L)}\\
\hline
$10^2 \omega_{b}$& $ 2.248\pm 0.026$& $ 2.282\pm 0.015$\\
\hline
$\omega_{cdm }$& $ 0.1193\pm 0.0032$& $ 0.1249\pm 0.0026$\\
\hline
$100\theta_{s }$& $ 1.04225\pm 0.00061$& $ 1.04200\pm 0.00067$\\
\hline
$\tau_{reio }$& $ 0.0548^{+0.0070}_{-0.0082}$& $ 0.0564\pm 0.0073$\\
\hline
$10^{10}\mathcal{P}_\mathcal{RR}^{(1)}$& $ 23.39\pm 0.60$& $ 22.76\pm 0.48$\\
\hline
$10^{10}\mathcal{P}_\mathcal{RR}^{(2)}$& $ 20.39\pm 0.44$& $ 20.70\pm 0.41$\\
\hline
$ 10^{10} N_{\rm dr}^2 \mathcal{P}_\mathcal{II}^{(1)}$& $< 18.8$& $< 16.2$\\
\hline
$10^{10} N_{\rm dr}^2\mathcal{P}_\mathcal{II}^{(2)}$& $ 65^{+20}_{-60}$& $ 99^{+40}_{-70}$\\
\hline
$10^{10} N_{\rm dr} \mathcal{P}_\mathcal{RI}^{(1)}$& $ -0.59^{+1.4}_{-0.94}$& $ 0.21^{+1.0}_{-0.86}$\\
\hline
$N_{ur }$& $ 2.08^{+1.0}_{-0.46}$& $ 2.40^{+1.0}_{-0.44}$\\
\hline
$N_{\rm dr}$& $< 1.23$& $< 1.29$\\
\hline
\hline$H_0 ({\rm km/s/Mpc})$& $ 68.4\pm 1.7$& $ 71.06\pm 0.83$\\
\hline
$\sigma_8$& $ 0.819\pm 0.011$& $ 0.8324\pm 0.0087$\\
\hline
$10^9 A_s$& $ 2.089\pm 0.037$& $ 2.105\pm 0.036$\\
\hline
$n_{s }$& $ 0.9650\pm 0.0093$& $ 0.9758^{+0.0072}_{-0.0064}$\\
\hline
$n_{\rm iso}$& $ 1.38^{+0.36}_{-0.31}$& $ 1.54^{+0.36}_{-0.28}$\\
\hline
$f_{\rm iso}$& $< 3.35$& $ 14.7^{+5.9}_{-14}$\\
\hline
$\cos \Delta$& $ -0.027\pm 0.072$& $ 0.025^{+0.053}_{-0.076}$\\
\hline
$N_{\rm tot}$& $ 3.05\pm 0.22$& $ 3.43\pm 0.13$\\
\hline
$f_{\rm dr}$& $ 0.32^{+0.13}_{-0.31}$& $ 0.30^{+0.12}_{-0.30}$\\
\hline
\hline
$ \chi^2 - \chi^2_{\Lambda \rm{CDM}}$& $-2.64$& $-14.68$\\
\hline
\end{tabular}
    \caption{Mean and $1\sigma$ error of parameters for FDR-DRID \emph{with correlation} for varying $N_{\rm ur}$ scenario for the corresponding datasets. The limits are at 68\% C.L. The constraints on the primary parameters and the derived parameters are shown in two separate blocks. The $\chi^2$ difference with respect to the \lcdm{} (fixed \neff{}) model for the corresponding data-set is shown on the last line.}
    \label{tab:param-fdr-wcor-vn}
\end{table}

\renewcommand{\arraystretch}{1.4}
\begin{table}[h!]
    \centering
    \begin{tabular}{|l|c| c| }
\hline
CDR (VN \& WC)& \makecell[c]{P18-TTTEEE\\+lowE+lensing}& \makecell[c]{P18-TTTEEE+lowE+ \\lensing+BAO+SH0ES(L)}\\
\hline
$10^2 \omega_{b}$& $ 2.249\pm 0.028$& $ 2.286\pm 0.016$\\
\hline
$\omega_{cdm }$& $ 0.1205^{+0.0032}_{-0.0039}$& $ 0.1267\pm 0.0029$\\
\hline
$100\theta_{s }$& $ 1.04295^{+0.00069}_{-0.00090}$& $ 1.04253^{+0.00075}_{-0.00097}$\\
\hline
$\tau_{reio }$& $ 0.0546\pm 0.0077$& $ 0.0563\pm 0.0074$\\
\hline
$10^{10}\mathcal{P}_\mathcal{RR}^{(1)}$& $ 23.67^{+0.57}_{-0.64}$& $ 23.13\pm 0.49$\\
\hline
$10^{10}\mathcal{P}_\mathcal{RR}^{(2)}$& $ 20.15\pm 0.48$& $ 20.40\pm 0.44$\\
\hline
$ 10^{10} N_{\rm dr}^2 \mathcal{P}_\mathcal{II}^{(1)}$& $ 26^{+5}_{-30}$& $ 25^{+5}_{-20}$\\
\hline
$10^{10} N_{\rm dr}^2\mathcal{P}_\mathcal{II}^{(2)}$& $< 310$& $< 568$\\
\hline
$10^{10} N_{\rm dr} \mathcal{P}_\mathcal{RI}^{(1)}$& $ -2.4^{+2.3}_{-1.2}$& $ -1.5^{+1.7}_{-1.1}$\\
\hline
$N_{ur }$& $ 2.73\pm 0.30$& $ 3.03^{+0.35}_{-0.27}$\\
\hline
$N_{\rm dr}$& $< 0.436$& $ 0.45^{+0.20}_{-0.33}$\\
\hline
\hline$H_0 ({\rm km/s/Mpc})$& $ 68.6^{+1.7}_{-2.0}$& $ 71.30\pm 0.84$\\
\hline
$\sigma_8$& $ 0.821^{+0.010}_{-0.012}$& $ 0.8325\pm 0.0096$\\
\hline
$10^9 A_s$& $ 2.073\pm 0.039$& $ 2.086\pm 0.038$\\
\hline
$n_{s }$& $ 0.959\pm 0.010$& $ 0.9680\pm 0.0079$\\
\hline
$n_{\rm iso}$& $ 1.51^{+0.52}_{-0.34}$& $ 1.71^{+0.47}_{-0.27}$\\
\hline
$f_{\rm iso}$& $< 12.0$& $ 21.8^{+3.3}_{-18}$\\
\hline
$\cos \Delta$& $ -0.101\pm 0.088$& $ -0.061^{+0.055}_{-0.065}$\\
\hline
$N_{\rm tot}$& $ 3.08^{+0.21}_{-0.26}$& $ 3.48\pm 0.15$\\
\hline
$f_{\rm dr}$& $ 0.112^{+0.046}_{-0.095}$& $ 0.130^{+0.056}_{-0.093}$\\
\hline
\hline
$ \chi^2 - \chi^2_{\Lambda CDM}$& $-0.74$& $-12.26$\\
\hline
\end{tabular}
    \caption{Mean and $1\sigma$ error of parameters for CDR-DRID \emph{with correlation} for varying $N_{\rm ur}$ scenario for the corresponding datasets. The limits are at 68\% C.L. The constraints on the primary parameters and the derived parameters are shown in two separate blocks. The $\chi^2$ difference with respect to the \lcdm{} (fixed \neff{}) model for the corresponding data-set is shown on the last line.}
    \label{tab:param-cdr-wcor-vn}
\end{table}

\begin{figure}[htb!]
	\centering
		\includegraphics[width=\linewidth]{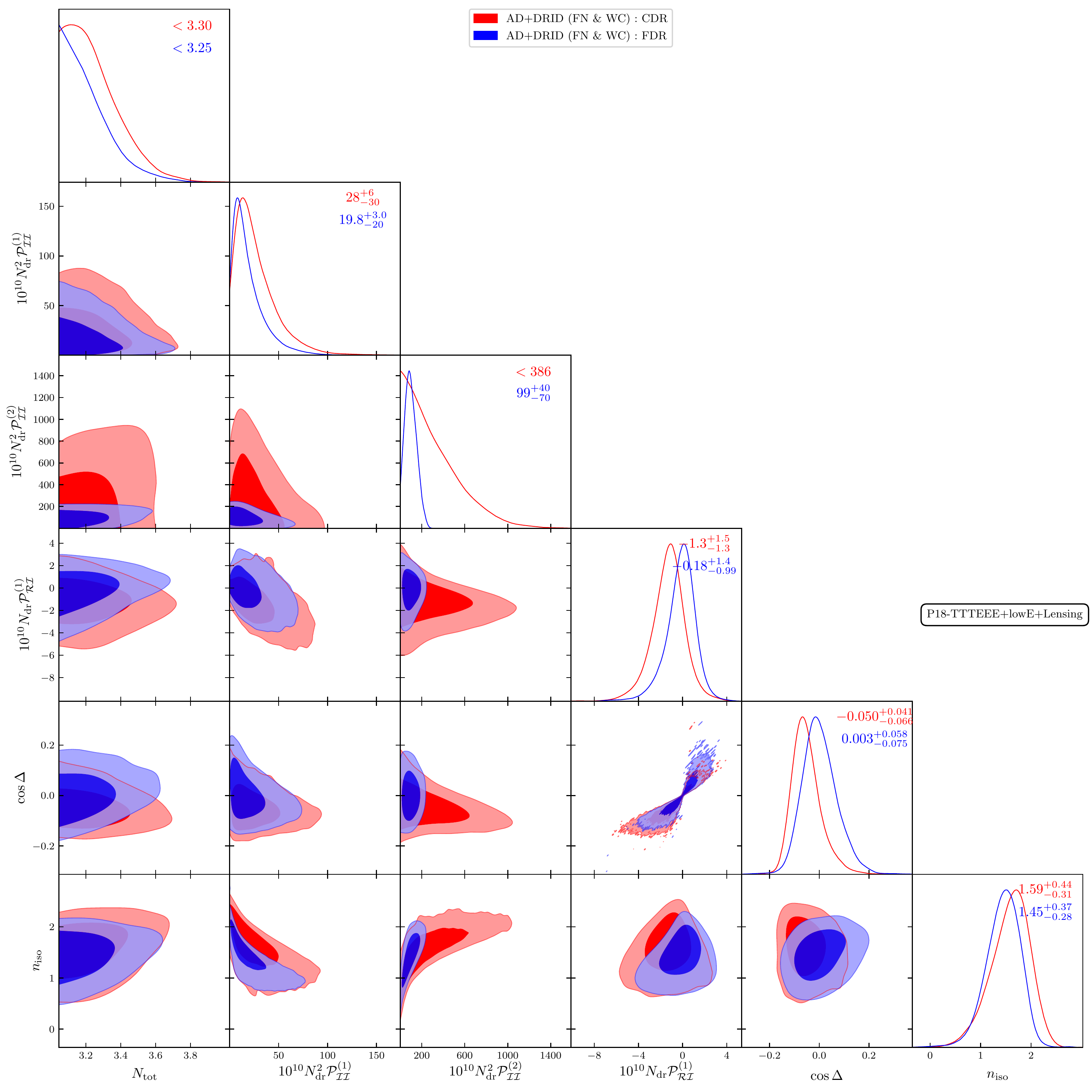}
		\caption{Triangle plot for DR isocurvature with correlation for fixed $N_{\rm ur}$ for P18-TTTEEE+lowE+lensing dataset. The constraints on individual parameters are mentioned on the diagonal 1-D posteriors with corresponding colors. The errors represent $1\sigma$ errorbar and the limits are at 68\% C.L.}
		\label{fig:is0-param-tri-wcor-fn}
	\end{figure}

\begin{figure}[htb!]
	\centering
		\includegraphics[width=\linewidth]{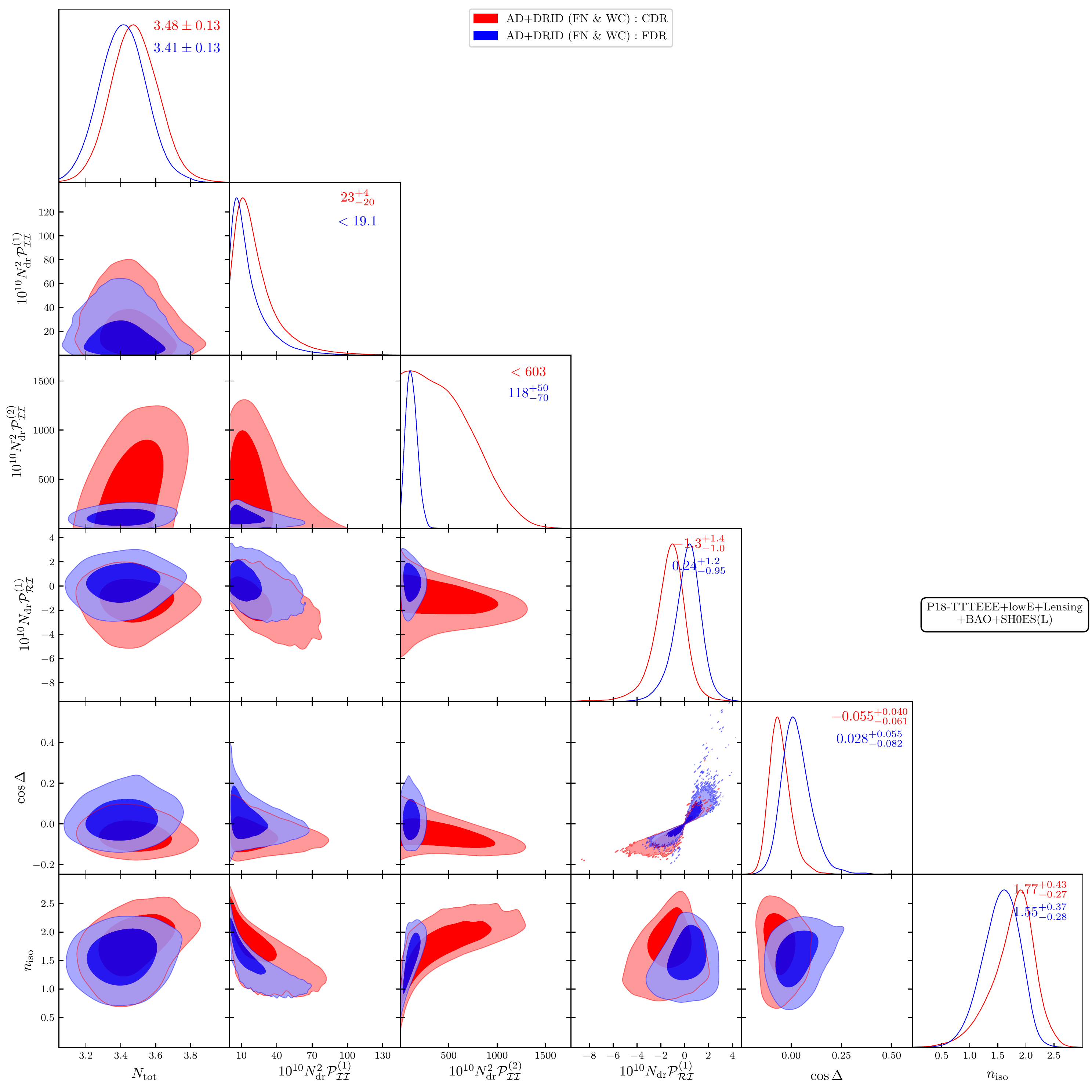}
		\caption{Triangle plot for DR isocurvature with correlation for fixed $N_{\rm ur}$ for P18-TTTEEE+lowE+lensing+lensing+SH0ES(L) dataset. The constraints on individual parameters are mentioned on the 1-D diagonal posteriors with corresponding colors. The errors represent $1\sigma$ errorbar and the limits are at 68\% C.L.}
		\label{fig:is0-param-tri-wsh0es-wcor-fn}
	\end{figure}

\begin{figure}[htb!]
	\centering
		\includegraphics[width=\linewidth]{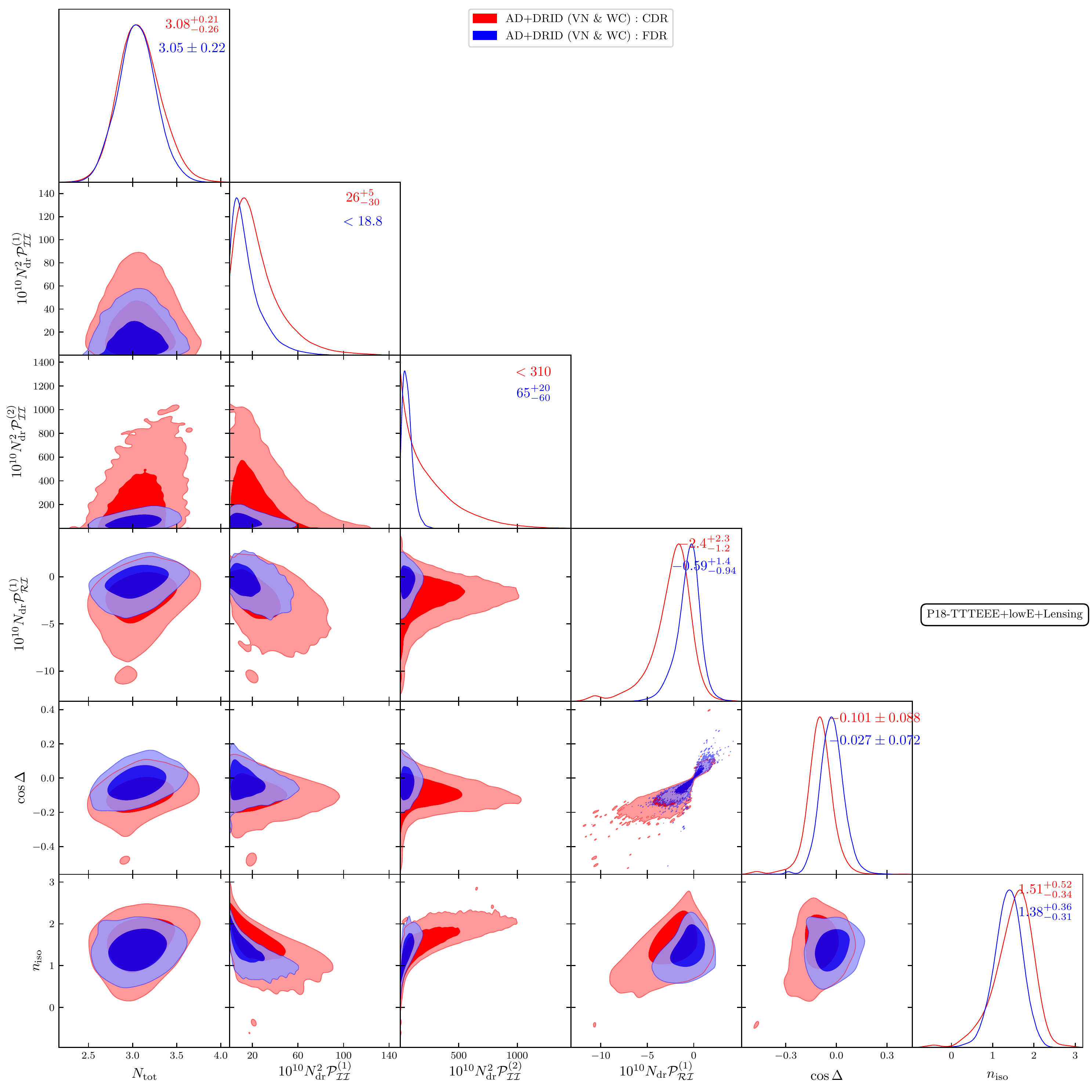}
		\caption{Triangle plot for DR isocurvature with correlation for varying $N_{\rm ur}$ for P18-TTTEEE+lowE+lensing dataset. The constraints on individual parameters are mentioned on the diagonal 1-D posteriors with corresponding colors. The errors represent $1\sigma$ errorbar and the limits are at 68\% C.L.}
		\label{fig:is0-param-tri-wcor-vn}
	\end{figure}

\begin{figure}[htb!]
	\centering
		\includegraphics[width=\linewidth]{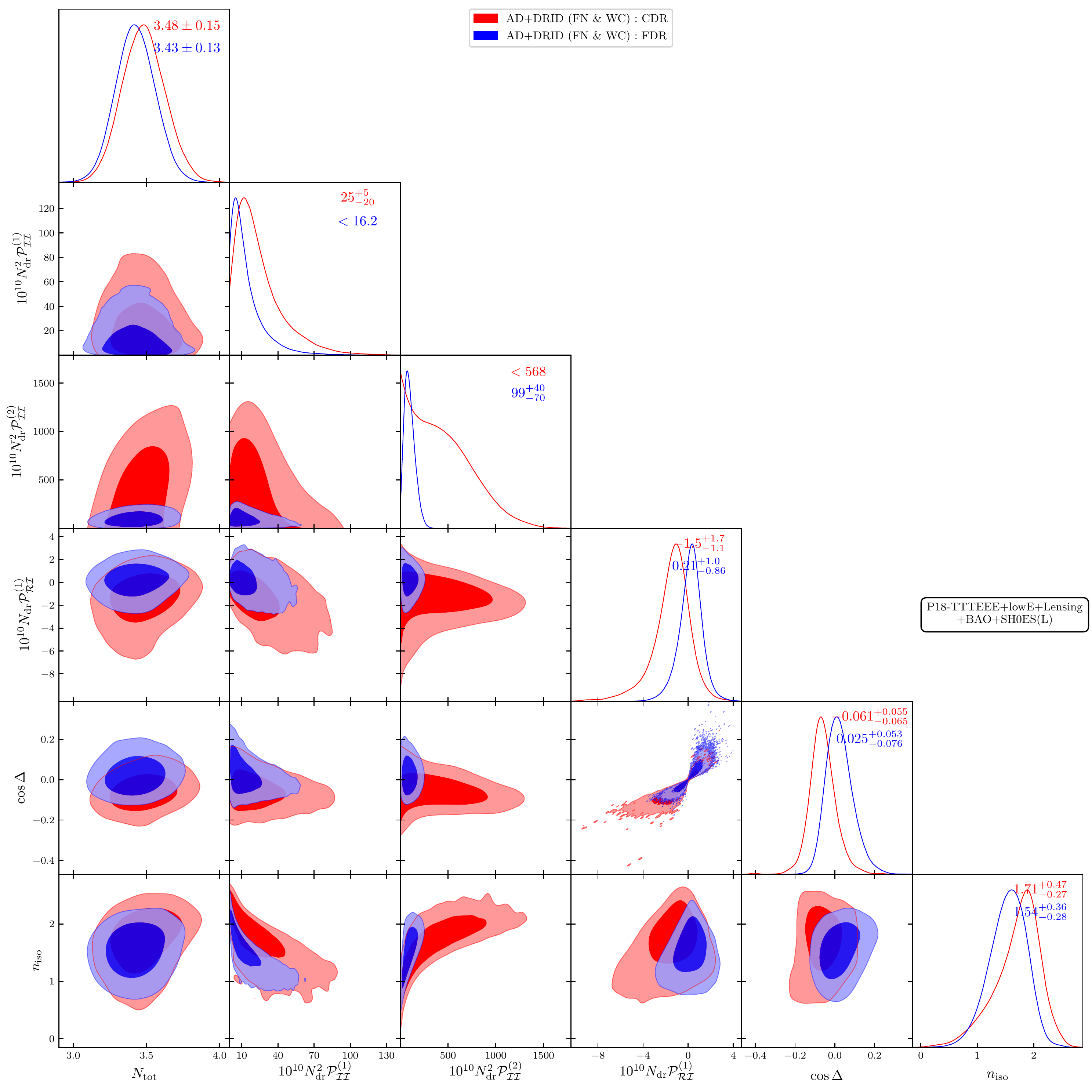}
		\caption{Triangle plot for DR isocurvature with correlation for varying $N_{\rm ur}$ for P18-TTTEEE+lowE+lensing+lensing+SH0ES(L) dataset. The constraints on individual parameters are mentioned on the 1-D diagonal posteriors with corrosponding colors. The errors represent $1\sigma$ errorbar and the limits are at 68\% C.L.}
		\label{fig:is0-param-tri-wsh0es-wcor-vn}
	\end{figure}

In Fig.~\ref{fig:is0-param-tri-wcor-fn},~\ref{fig:is0-param-tri-wsh0es-wcor-fn},~\ref{fig:is0-param-tri-wcor-vn} and ~\ref{fig:is0-param-tri-wsh0es-wcor-vn} we show the constraints on the isocurvature parameters which now include the correlation parameter and correlation fraction. The value correlation parameter in all our runs are consistent with zero. The other isocurvature parameters also exhibit similar degeneracies that was observed in the runs without correlation.

\section{Triangle plots for uncorrelated DRID}\label{sec.triangle}
\begin{figure}
    \centering
    \includegraphics[width=\textwidth]{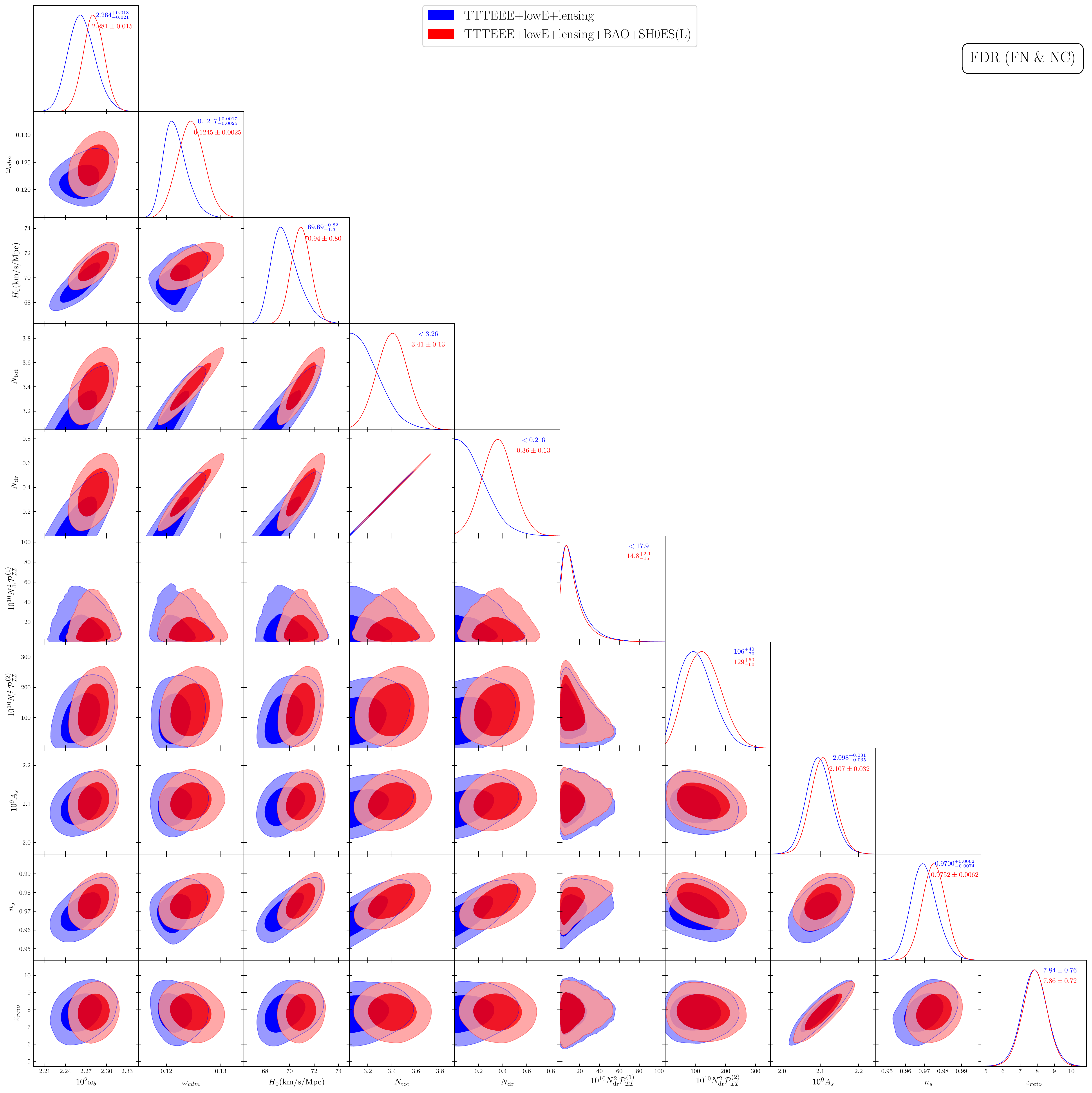}
    \caption{Triangle plot for FDR isocurvature without correlation for the fixed $N_{\rm ur}$ analysis for all the datasets. The constraints on individual parameters are mentioned on the 1-D diagonal posteriors with corresponding colors. The errors represent $1\sigma$ errorbar and the limits are at 68\% C.L.}
    \label{fig:tri-all-fDR-fn}
\end{figure}

\begin{figure}
    \centering
    \includegraphics[width=\textwidth]{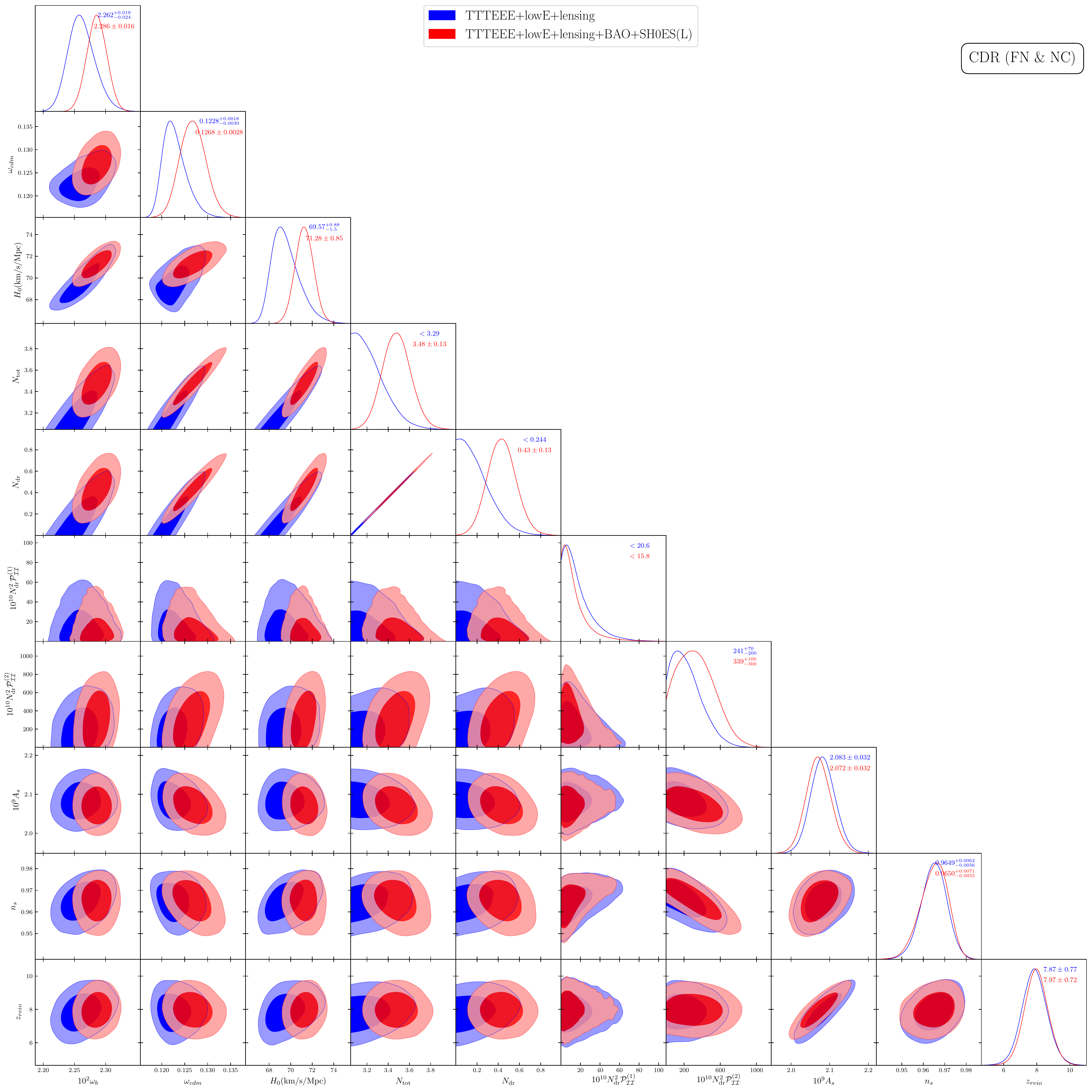}
    \caption{Triangle plot for CDR isocurvature without correlation for the fixed $N_{\rm ur}$ analysis for all the datasets. The constraints on individual parameters are mentioned on the 1-D diagonal posteriors with corresponding colors. The errors represent $1\sigma$ errorbar and the limits are at 68\% C.L.}
    \label{fig:tri-all-cDR-fn}
\end{figure}

\begin{figure}
    \centering
    \includegraphics[width=\textwidth]{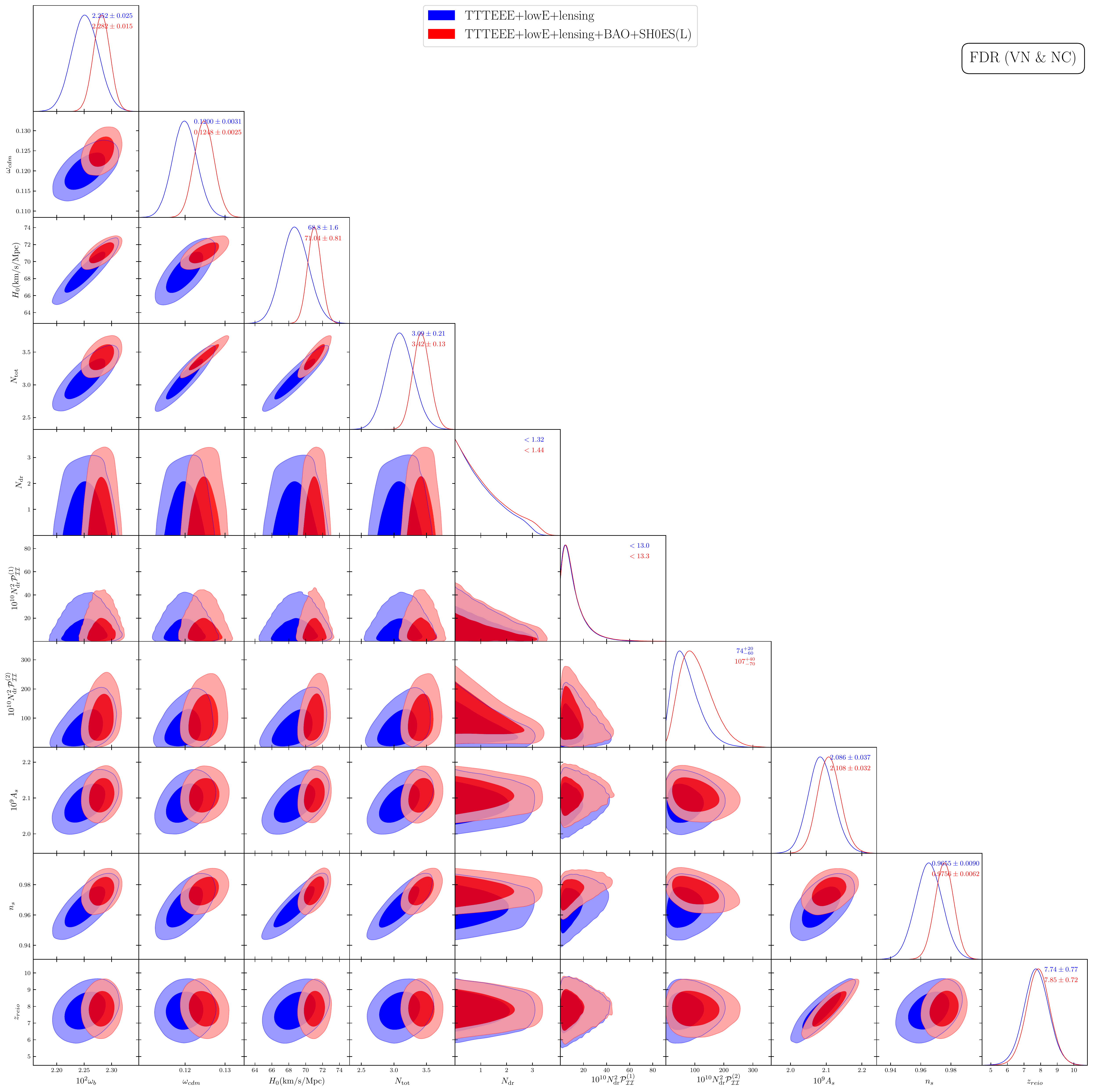}
    \caption{Triangle plot for FDR isocurvature without correlation for the varying $N_{\rm ur}$ analysis for all the datasets. The constraints on individual parameters are mentioned on the 1-D diagonal posteriors with corresponding colors. The errors represent $1\sigma$ errorbar and the limits are at 68\% C.L.}
    \label{fig:tri-all-fDR-vn}
\end{figure}

\begin{figure}
    \centering
    \includegraphics[width=\textwidth]{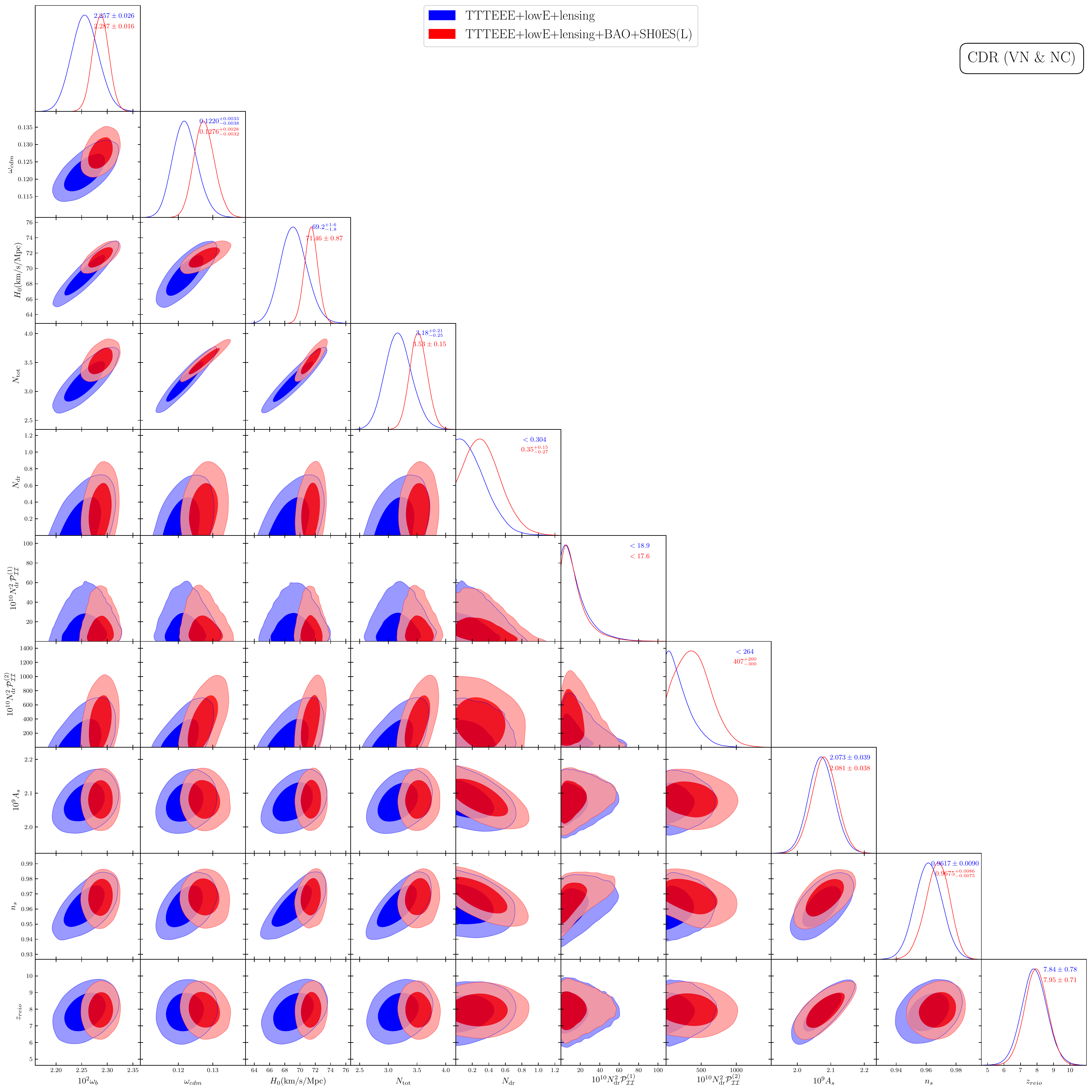}
    \caption{Triangle plot for CDR isocurvature without correlation for the varying $N_{\rm ur}$ analysis for all the datasets. The constraints on individual parameters are mentioned on the 1-D diagonal posteriors with corresponding colors. The errors represent $1\sigma$ errorbar and the limits are at 68\% C.L.}
    \label{fig:tri-all-cDR-vn}
\end{figure}

Fig.~\ref{fig:tri-all-fDR-fn}, \ref{fig:tri-all-cDR-fn}, \ref{fig:tri-all-fDR-vn} and \ref{fig:tri-all-cDR-vn} show the constraints on the adiabatic as well as the isocurvature parameters for uncorrelated FDR-DRID and uncorrelated CDR-DRID, respectively with both fixed and varying $N_{ur}$ for all the datasets used in this paper. 
\label{sec:app:tri}

\section{Triangle plots for correlated DRID }
\label{sec:app:tri-wcor}
\begin{figure}
    \centering
    \includegraphics[width=\textwidth]{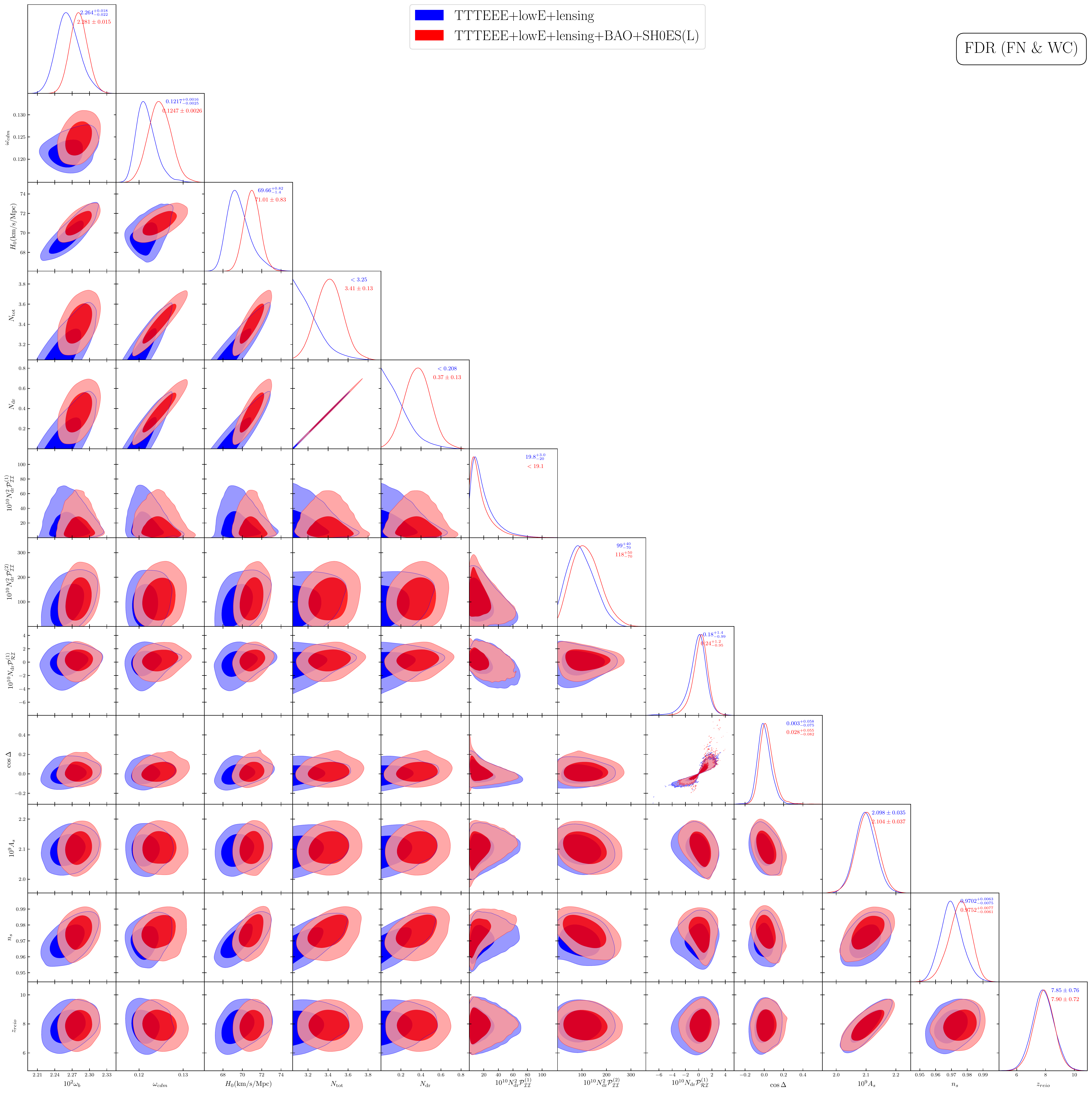}
   \caption{Triangle plot for FDR isocurvature with correlation for the fixed $N_{\rm ur}$ analysis. The constraints on individual parameters are mentioned on the 1-D diagonal posteriors with corresponding colors. The errors represent $1\sigma$ errorbar and the limits are at 68\% C.L.}
    \label{fig:tri-all-fDR-wcor-fn}
\end{figure}

\begin{figure}
    \centering
    \includegraphics[width=\textwidth]{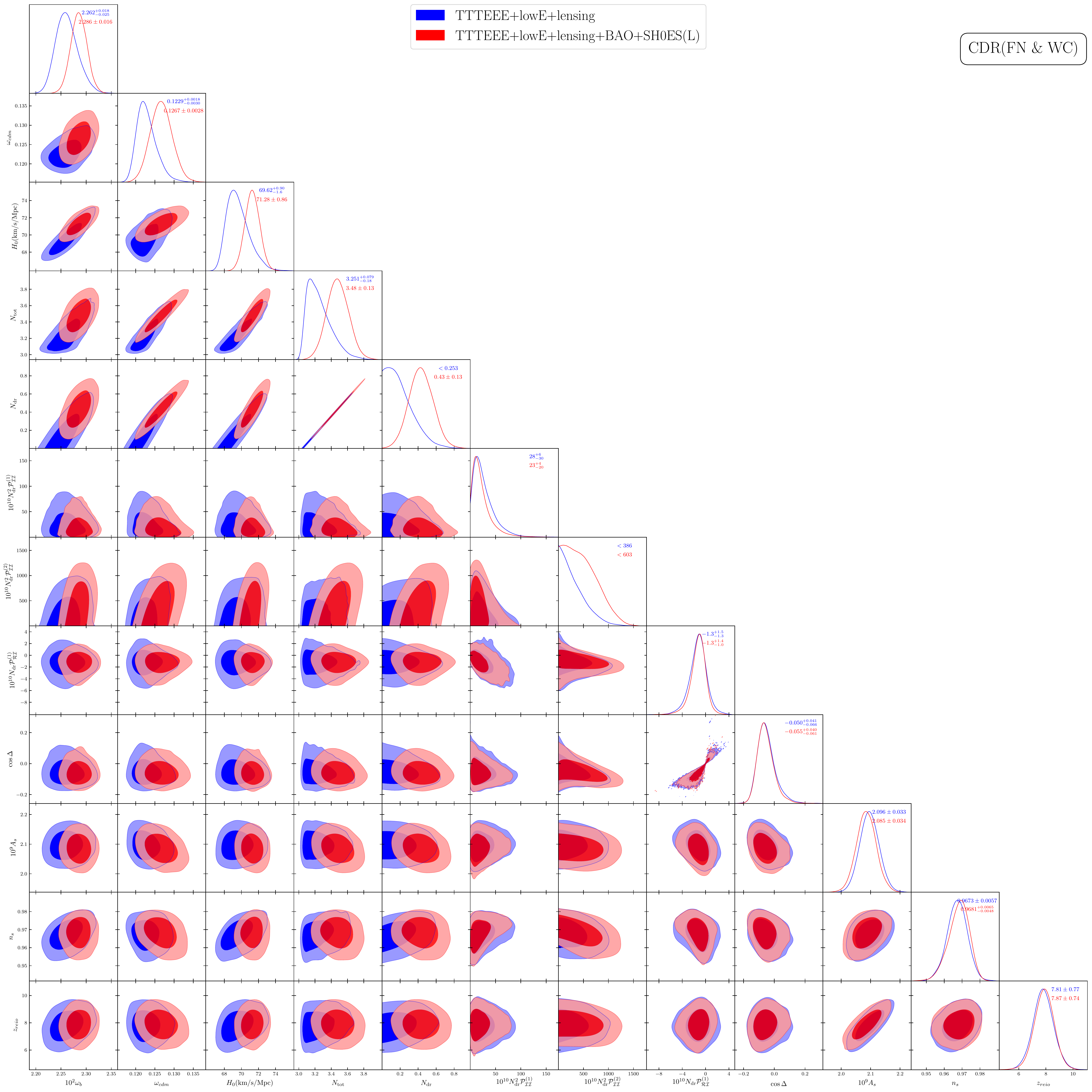}
    \caption{Triangle plot for CDR isocurvature with correlation for the fixed $N_{\rm ur}$ analysis. The constraints on individual parameters are mentioned on the 1-D diagonal posteriors with corresponding colors. The errors represent $1\sigma$ errorbar and the limits are at 68\% C.L.}
    \label{fig:tri-all-cDR-wcor-fn}
\end{figure}

\begin{figure}
    \centering
    \includegraphics[width=\textwidth]{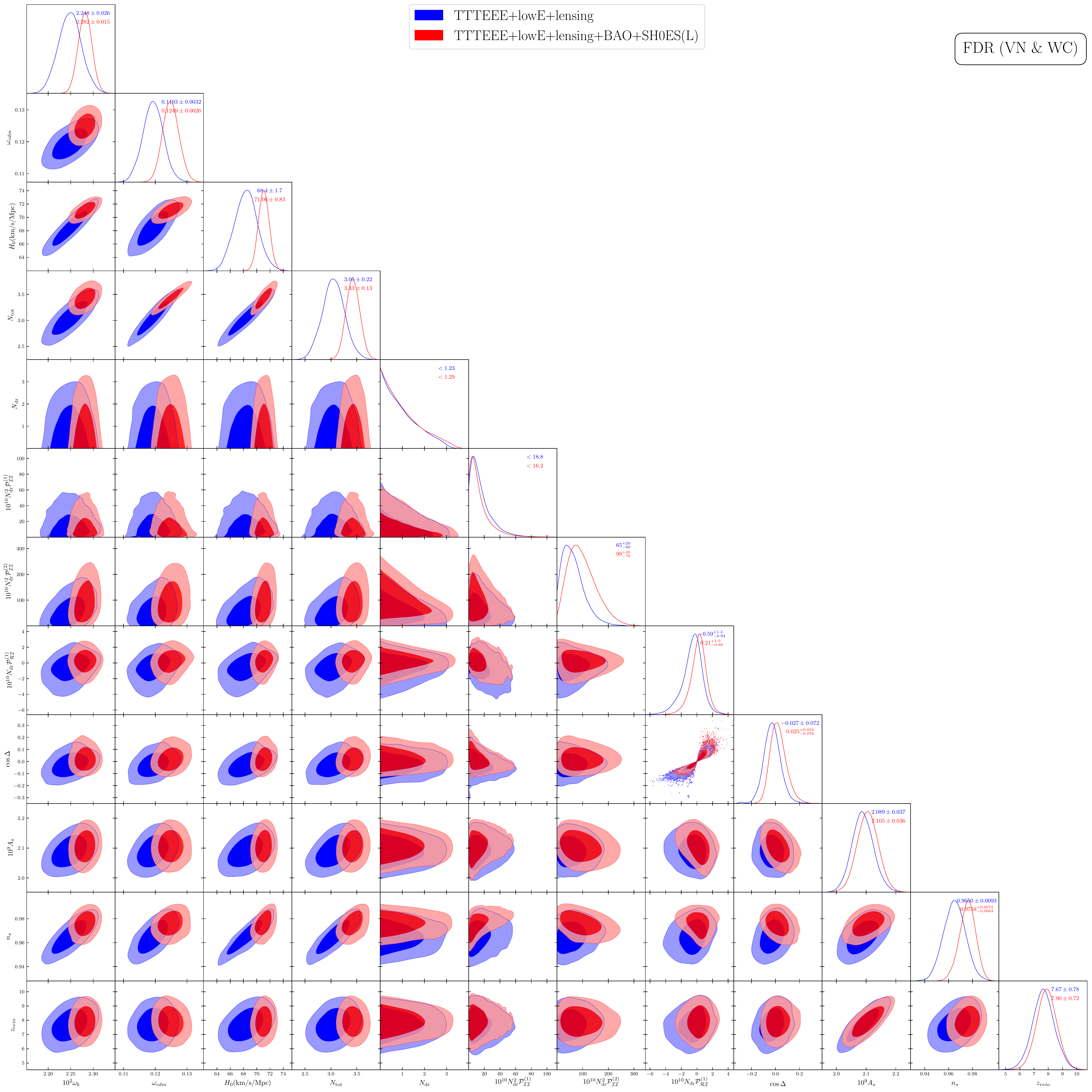}
   \caption{Triangle plot for FDR isocurvature with correlation for the varying $N_{\rm ur}$ analysis. The constraints on individual parameters are mentioned on the 1-D diagonal posteriors with corresponding colors. The errors represent $1\sigma$ errorbar and the limits are at 68\% C.L.}
    \label{fig:tri-all-fDR-wcor-vn}
\end{figure}

\begin{figure}
    \centering
    \includegraphics[width=\textwidth]{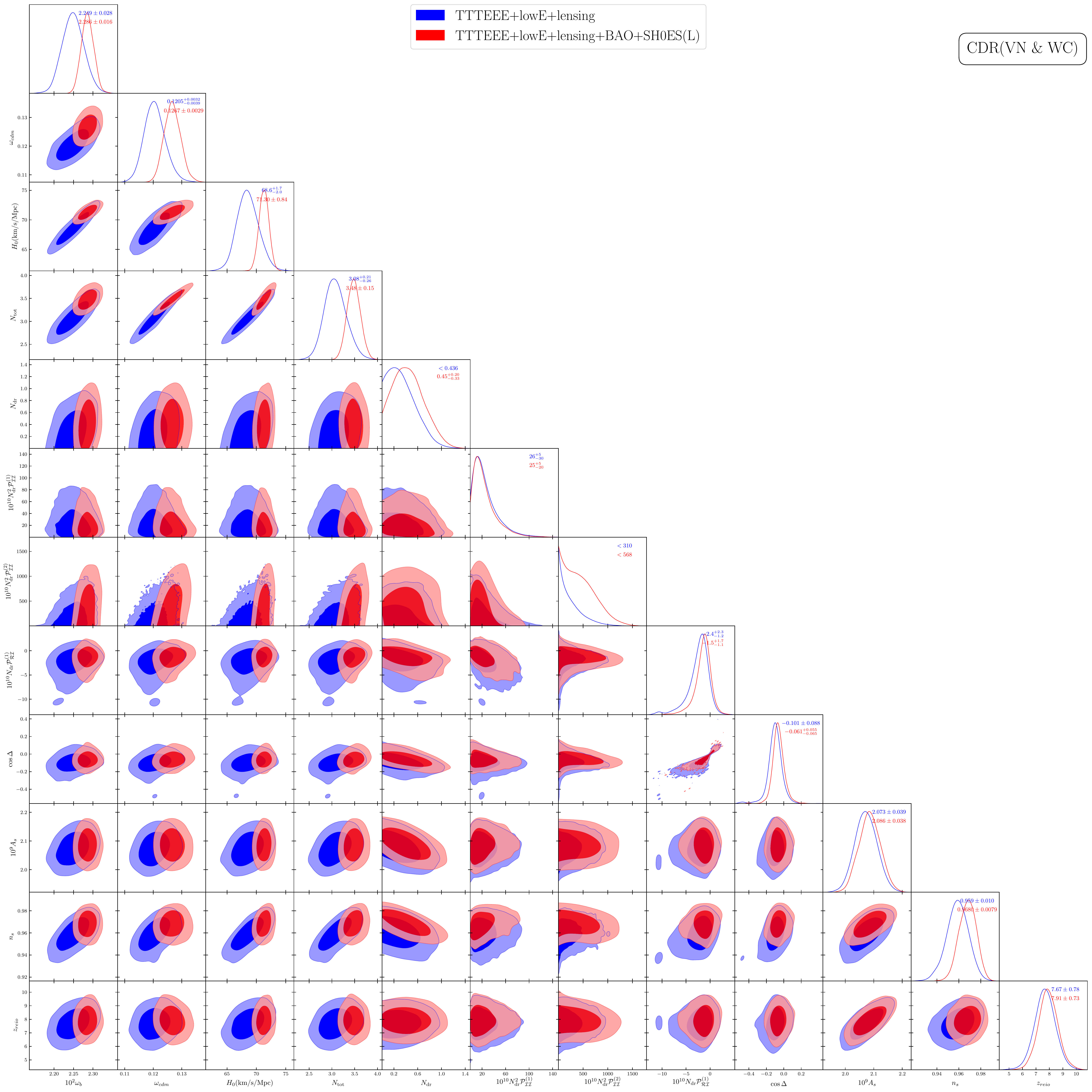}
    \caption{Triangle plot for CDR isocurvature with correlation for the varying $N_{\rm ur}$ analysis. The constraints on individual parameters are mentioned on the 1-D diagonal posteriors with corresponding colors. The errors represent $1\sigma$ errorbar and the limits are at 68\% C.L.}
    \label{fig:tri-all-cDR-wcor-vn}
\end{figure}

Fig.~\ref{fig:tri-all-fDR-wcor-fn}, \ref{fig:tri-all-cDR-wcor-fn}, \ref{fig:tri-all-fDR-wcor-vn} and \ref{fig:tri-all-cDR-wcor-vn} show the constraints on the adiabatic as well as the isocurvature parameters for FDR-DRID and CDR-DRID with correlation, respectively for both fixed and varying $N_{\rm ur}$. 


\bibliographystyle{JHEP}
\bibliography{references}
\end{document}